\documentclass[11pt]{article}
\usepackage{geometry}                
\pdfoutput=1

\usepackage{setspace}
\usepackage{graphicx}
\usepackage{amssymb}
\usepackage{bm}
\usepackage{epstopdf}
\newcommand{\ra}[1]{\renewcommand{\arraystretch}{#1}}

\usepackage[square,authoryear]{natbib}
\usepackage[titletoc,title]{appendix}
\usepackage[font=small,labelfont=bf]{caption}
\usepackage{placeins}

\title{Enceladus's crust as a non-uniform thin shell:\\ II tidal dissipation}
\author{Mikael Beuthe\\
\it Royal Observatory of Belgium,
\it Avenue Circulaire 3, 1180 Brussels, Belgium\\
\it E-mail: mikael.beuthe@observatoire.be}      
\date{}                                             

\begin{document}
\maketitle

\begin{abstract}

Tidal heating is the prime suspect behind Enceladus's south polar heating anomaly and global subsurface ocean.
No model of internal tidal dissipation, however, can explain at the same time the total heat budget and the focusing of the energy at the south pole.
I study here whether the non-uniform icy shell thickness can cause the north-south heating asymmetry by redistributing tidal heating either in the shell or in the core.
Starting from the non-uniform tidal thin shell equations, I compute the volumetric rate, surface flux, and total power generated by tidal dissipation in shell and core.
The micro approach is supplemented by a macro approach providing an independent determination of the core-shell partition of the total power.
Unless the shell is incompressible, the assumption of a uniform Poisson's ratio implies non-zero bulk dissipation.
If the shell is laterally uniform, the thin shell approach predicts shell dissipation with a few percent error while the error on core dissipation is negligible.
Variations in shell thickness strongly increase the shell dissipation flux where the shell is thinner.
For a hard shell with long-wavelength variations, the shell dissipation flux can be predicted by scaling with the inverse local thickness the flux for a laterally uniform shell.
If Enceladus's shell is in conductive thermal equilibrium with isostatic thickness variations,
the nominal shell dissipation flux at the south pole is about three times its value for a shell of uniform thickness, which remains negligible compared to the observed flux.
The shell dissipation rate should be ten times higher than nominal in order to account for the spatial variations of the observed flux.
Dissipation in an unconsolidated core can provide the missing power, but does not generate any significant heating asymmetry as long as the core is homogeneous.
Non-steady state models, though not investigated here, face similar difficulties in explaining the asymmetries of tidal heating and shell thickness.

\end{abstract}

\vspace{\stretch{1}}
\newpage

{\small
\tableofcontents
\listoffigures
\listoftables
}
\newpage

\section{Introduction}
\label{Introduction}

Despite being small and cooling fast, Enceladus is warm.
The first evidence for its present-day warm state was the detection by Cassini of a south polar hot spot \citep{spencer2006,howett2011},
with temperature reaching $200\rm\,K$ along the south polar faults \citep{goguen2013,spencer2018}.
A second line of evidence arises from the presence of underground liquid water, first inferred at the south pole from the composition of Saturn's E ring \citep{postberg2009} and Enceladus's plume \citep{waite2009,postberg2011}.
Geodesy measurements (gravity, topography, and libration) then proved beyond doubt the existence of a global ocean beneath a thin shell of non-uniform thickness \citep{iess2014,thomas2016,beuthe2016b}.

Tidal heating is the only source of energy sufficient to keep Enceladus warm \citep{squyres1983,ross1989}, and might be sufficient to keep it in a steady state \citep{fuller2016,nimmo2018}.
Energy conservation in the Saturn-Enceladus-Dione system, however, does not constrain where and how energy is dissipated inside Enceladus.
At the present time, viscoelastic dissipation within the shell is too small to produce the observed heat flux and to maintain thermal equilibrium \citep{squyres1983,roberts2008}, while dissipation within the ocean is completely negligible \citep{beuthe2016a,rovira2019}.
Theoretically, dissipation within an unconsolidated core can be large enough \citep{ross1989}, but it implies unverified assumptions about the viscoelastic behaviour of porous matter \citep{roberts2015,choblet2017}.
Instead of being in a steady state, Enceladus could oscillate in a thermal-orbital evolution cycle between phases of high and low dissipation.
Most models of this sort postulate that Enceladus is now in a short-lived stage of high dissipation, leading to the same problem as above regarding heat production, besides raising the question of Enceladus's special status at the present time (see review in \citet{nimmo2018}).
Alternatively, Enceladus could oscillate between two phases of nearly equal duration: the present-day phase of low dissipation in a medium-thick shell and another phase of high dissipation in a very thin shell \citep{luan2017}.

The south polar localization of the thermal anomaly and the spatial variations of shell thickness are often studied separately from the problem of the global heat budget.
This means throwing away a strong constraint on dissipation models.
For example, does a very dissipative core produce a north-south heating asymmetry if the shell is non-uniform, as suggested by \citet{choblet2017}?
Or else, is the non-steady state model compatible with the generation of shell thickness variations?
Before addressing these questions, we must be able to compute tidal dissipation in a body made of a viscoelastic core, a global ocean, and a viscoelastic shell of non-uniform thickness.
Here I will solve this problem with the non-uniform thin shell approach of \citet{beuthe2018} (hereafter named Paper~I).
The approach will be benchmarked against the solution for dissipation  in a laterally uniform thin shell, which has not appeared before in the literature (dissipation in a laterally uniform membrane was studied in \citet{beuthe2014}).
Besides being a non-trivial check of the consistency of the thin shell approach, this step provides an estimate of the error due to the thin shell approximation.

Without prejudging whether Enceladus is truly in a steady state, it seems natural to apply first the method to the simplest case, i.e.\ thermal equilibrium between tidal heating and conductive cooling.
As a concrete example, I will assume that Enceladus's shell varies in thickness according to the isostatic interpretation of gravity-topography data \citep{beuthe2016b}.
Given that previous studies made it clear that dissipation within the shell does not provide enough power at the present time, I will push the envelope of allowable parameters \citep{wolfe1979} in order to address the following questions:
\begin{itemize}
\item Does the variation of shell thickness increase the total dissipated power?
\item How much does shell thinning increase the dissipation flux at the south pole?
\item Is there a general relation between local shell thickness and local tidal dissipation?
\item Can we tune core and shell parameters so as to match both the total conductive flux and its spatial distribution, assuming thermal equilibrium?
\item Does the non-uniform shell induce a north-south asymmetry in core dissipation?
\end{itemize}

In a previous study of the topic, \citet{behounkova2017} used the finite-element method (FEM) of \citet{soucek2016} to solve for the deformations of an elastic shell with non-uniform thickness, and predicted several tens of GW of tidal heating by assuming a shell of uniform viscosity.
This last assumption is not realistic for a conductive shell because viscosity increases by orders of magnitude from the bottom of the shell to the surface; the total power dissipated in the shell is actually much lower.
Older 3D studies (reviewed in \citet{behounkova2017}) investigated the effect of laterally varying rheology on dissipation within a very thick convecting shell, now disfavoured by Cassini data.
\citet{soucek2019} recently improved the FEM approach by taking into account temperature-dependent viscosity and now predict less than $2\rm\,GW$ of dissipation in Enceladus's shell.
The FEM approach has the advantages of including faults (modelled as frictionless open slots) and of avoiding the errors intrinsic to the thin shell approach (mainly the thin shell approximation and the non-zero bulk dissipation).
Its disadvantages are that it does not include core dissipation; that it becomes unstable at high viscosity contrasts (need of viscosity cutoff); that it is not readily adaptable to non-Maxwell rheological models (e.g.\ the pseudo-Andrade rheology in \citet{soucek2019}); and that it neglects self-gravity (this approximation is justified for Enceladus but is not valid for large satellites).
Preliminary benchmarking against the thin shell approach shows that the two methods agree within a few percent error \citep{behounkova2018}.
Thus, the technical differences mentioned above are not problematic.
The choice of the method rather depends on whether one wants to include faulting or core dissipation in the model.

The rest of the paper is made of four parts:
(1) summary of the tidal thin shell equations from Paper~I,
(2) thin shell approach to dissipation in a non-uniform shell and in a spherically symmetric core,
(3) benchmark against a laterally uniform thick shell,
and (4) tidal heating in a body with a dissipative core and a non-uniform conductive shell in thermal equilibrium.

\section{Tidal thin shell equations}
\label{NonUniformThinShellTheory}

\subsection{Flexure equations}
\label{FlexureEquations}

In Paper~I, I used thin shell theory to compute the tidal deformations of an icy satellite with a viscoelastic core, a global ocean and a laterally non-uniform viscoelastic shell.
This approach rests on the following assumptions about the shell: thickness less than 10 to 20\% of the surface radius, uniform Poisson's ratio, homogeneous density, no density contrast with the top layer of the ocean, linear viscoelasticity.
Enceladus's internal structure approximately satisfies these criteria (Table~\ref{TableParam}).
In addition, the core structure must be spherically symmetric.

\begin{table}[ht]\centering
\ra{1.2}
\small
\caption[Physical parameters]
{Physical parameters used in this paper.}
\begin{tabular}{@{}llll@{}}
\hline
Parameter &  Symbol & Value & Unit \\
Mean eccentricity${}^a$ & $e$ & $0.0047$ & - \\
Rotation rate${}^a$  & $\omega$ & $5.307\times10^{-5}$ & $\rm\,s^{-1}$ \\
Bulk density${}^a$ & $\rho_b$ & $1610$ & $\rm kg \, m^{-3}$ \\
Surface gravity${}^a$ & $g$ & $0.1135$ & $\rm m \, s^{-2}$ \\
Surface radius${}^a$  & $R$ & $252.1$  & km \\
Reference thickness${}^b$ (if uniform shell) & $d$ & 23 & km \\
Core radius${}^b$ & $R_c$ & 192 & km \\
Density of ice and ocean & $\rho$ & 1000 & $\rm kg/m^3$ \\
Shear modulus of ice${}^c$ & $\mu_{\rm e}$ & 3.5 & GPa \\
Bulk modulus of ice${}^c$ & $K_{\rm e}$ & 9.13 & GPa  \\
Poisson's ratio of ice${}^c$ & $\nu_{\rm e}$ & 0.33 & - \\
Shear modulus of core (if non-porous) & $\mu_{\rm ce}$ & 40 & GPa \\
Conductivity of ice${}^d$ (if uniform shell) & $k_{ice}$ & $651/T$ & $\rm W m^{-1}K^{-1}$ \\
Activation energy & $E_a$ & $59.4$ & $\rm kJ\,mol^{-1}$ \\
Melting ice temperature & $T_{\rm m}$ & 273 & K \\
Viscosity of ice at melting point & $\eta_{\rm m}$ & $10^{13}-10^{15}$ & Pa.s \\
\hline
\multicolumn{4}{l}{\scriptsize ${}^a$ Source given in Table~1 of \citet{beuthe2016a}.}
\vspace{-1.5mm}\\
\multicolumn{4}{l}{\scriptsize ${}^b$ \citet{beuthe2016b}.}
\vspace{-1.5mm}\\
\multicolumn{4}{l}{\scriptsize ${}^c$ \citet{helgerud2009}.}
\vspace{-1.5mm}\\
\multicolumn{4}{l}{\scriptsize ${}^d$ \citet{petrenko1999} (Section~\ref{BenchmarkingU} only); Eq.~(\ref{condfit}) is used in Section~\ref{ThermalEquilibrium}.}
\end{tabular}
\label{TableParam}
\end{table}%

The thin shell approach to tidal deformations is neatly summarized by the \textit{tidal thin shell equations}.
These two differential equations govern the tidal deformations of a non-uniform thin spherical shell floating on a global ocean:
\begin{eqnarray}
{\cal C} ( D \,; w ) - (1-\nu) \, {\cal A}( D \,; w) + R^3 \, {\cal A}(\chi \,; F) 
&=& R^4 q \, ,
\nonumber \\
{\cal C} ( \alpha \,; F ) - (1+\nu) \, {\cal A}( \alpha \,; F) - R^{-1} \, {\cal A}(\chi \,; w)
&=& 0 \, ,
\label{TidalThinShellEq}
\end{eqnarray}
where
\begin{itemize}
\item $w$ is the radial displacement and $F$ is the auxiliary stress function;
\item  $\alpha$ and $D$ are the primary viscoelastic shell parameters (Table~\ref{TableVisco}); $\chi$ is a secondary viscoelastic shell parameter, close to one (Table~\ref{TableVisco});
\item ${\cal C}$ and ${\cal A}$ are spherical differential operators of order 4 (Appendix~\ref{SphericalOperators}).
\end{itemize}
In a spherical harmonic basis, the tidal thin shell equations become a system of coupled linear equations which can be solved for $(F,w)$ with standard matrix methods.
Other variables, such as tangential displacements, strains, and stresses, can be computed from $(F,w)$ with the pseudospectral transform method: angular derivatives are applied in the spectral domain whereas products of fields are computed in the spatial domain (see Paper~I).

\begin{table}[ht]\centering
\ra{1.2}
\small
\caption[Viscoelastic shell parameters]
{\small
Viscoelastic shell parameters.
}
\begin{tabular}{@{}lll@{}}
 \hline
Symbol & Name & Definition \\
\hline
\multicolumn{3}{l}{Depth-dependent} \\
$\zeta$ & radial shell coordinate & $r-R$ \\
$z$ & depth parameter & $\zeta/(R+\zeta) - \left( 1-\chi \right)$ \\
$\eta$ & viscosity of ice${}^a$ & $\eta_{\rm m} \, \exp(E_a(T^{-1}-T_{\rm m}^{-1})/R_g)$ \\
$\mu$ & complex shear modulus${}^b$ & $\mu_{\rm e}/(1-i\mu_e/\omega\eta)$ \\
\hline
\multicolumn{3}{l}{Depth-integrated ($\varepsilon=d/R$)} \\
$\mu_p$ & $p$th moment of $\mu$ & $(1/d)^{p+1} \int_d \mu \, \zeta^p \, d\zeta$ \\
$\mu_{\rm inv}$ & invariant second moment  &  $\mu_2 - (\mu_1)^2/\mu_0$ \\
$\chi$ & - & $(\mu_0 + \varepsilon \mu_1 )/(\mu_0 + 2 \varepsilon \mu_1 + \varepsilon^2 \mu_2 )$ \\
$\psi$ & - & $\mu_0/(\mu_0+\varepsilon \mu_1 )$ \\
$\alpha_{\rm inv}$ & invariant extensibility &  $\left( 2 (1+\nu ) \, \mu_0 d \, \right)^{-1}$ \\
$D_{\rm inv}$ & invariant bending rigidity &  $2\,\mu_{\rm inv} d^3/( 1-\nu )$ \\
$\alpha$ & extensibility & $\chi\psi \, \alpha_{\rm inv}$ \\
$D$ & bending rigidity & $\chi\psi \, D_{\rm inv}$ \\
\hline
\multicolumn{3}{l}{\scriptsize ${}^a$ $R_g=8.314\rm\,J\,K^{-1}mol^{-1}$ is the gas constant.}
\vspace{-1mm}\\
\multicolumn{3}{l}{\scriptsize ${}^b$ For Maxwell rheology, but other linear rheological models can be used.}
\end{tabular}
\label{TableVisco}
\end{table}%

\subsection{Tidal forcing}
\label{TidalLoad}

The RHS of Eq.~(\ref{TidalThinShellEq}) represents the tidal loading which can be expressed as a sum of spherical harmonics of degree $n$:
\begin{equation}
q = q_0 + \sum_{n\geq2} q_n \, ,
\end{equation}
where $q_0$ is the degree-0 load (see Section~3.4 of Paper~I) while the degree-1 load vanishes.
If $n\geq2$, the degree-$n$ load is given by
\begin{equation}
q_n = - \rho g \left(w_n - \Gamma_n/g \right) ,
\label{qn}
\end{equation}
where $\rho{}gw_n$ is the weight of the tidal bulge in the unperturbed gravity field.
The total perturbing potential $\Gamma_n$ depends on the primary tidal potential $U_n^T$ and on the secondary potential induced by the deformation of the whole body.
In the thin shell approach, it is given by
\begin{equation}
\Gamma_n = \upsilon_n \left( U_n^T + g \, \xi_n  w_n \right) ,
\label{Gamman}
\end{equation}
where $\xi_n $ is the degree-$n$ density ratio,
\begin{equation}
\xi_n = \frac{3}{2n+1} \, \frac{\rho}{\rho_b} \, .
\label{xin}
\end{equation}
The spherically symmetric structure below the shell (stratified ocean, viscoelastic core) enters into the problem via the nondimensional factor $\upsilon_n$ ($\upsilon_n\geq1$),
\begin{equation}
\upsilon_n = \frac{h_n^\circ}{1+\xi_n h_n^\circ} \, ,
\label{upsilon}
\end{equation}
where $h_n^\circ$ is the tidal radial Love number of the fluid-crust body (i.e.\ the same body except that the icy shell behaves as a fluid).
If the core is not deformable, $h_n^\circ=(1-\xi_n)^{-1}$ and $\upsilon_n=1$.
In that case, the term $\xi_nw_n$ in $\Gamma_n$ (Eq.~(\ref{Gamman})) can be interpreted as the geoid perturbation due to the tidal deformation (self-attraction or self-gravity).
In the simplest model with a deformable core,  the core is viscoelastic, homogeneous, and incompressible, and the ocean is homogeneous, in which case $h_n^\circ$ can be computed with Eq.~(\ref{hn0}).

Since Enceladus is in a synchronous orbit with negligible obliquity, tidal deformations are mainly due to eccentricity tides of degree~2.
These tides can be expressed as the sum of a radial tide (due to the varying distance to Saturn) and a librational tide (due to the optical libration, i.e.\ the varying direction of Saturn in the frame fixed to the satellite) \citep{murray1999}.
The latter is enhanced by the 1:1 forced (or physical) libration.
Including degree-2 eccentricity tides plus the forced libration (denoted $\gamma=-\gamma_0\sin \omega t$), I write the tidal potential as \citep{vanhoolst2013}
\begin{equation}
U(t,\theta,\varphi) = (\omega R)^2 \, e \left( - \frac{3}{2} \, P_{20} \cos \omega t +  P_{22} \left( \frac{3}{4} \cos2\varphi \cos \omega t + \left(1 + \frac{\gamma_0}{2e} \right) \sin2\varphi \sin \omega t \right)  \right) \, ,
\label{TidalPot}
\end{equation}
where $\gamma_0=0.12^\circ$ for Enceladus \citep{thomas2016}.
$P_{2m}$ are the associated Legendre functions of degree $2$ and order $m$ depending on $\cos\theta$ ($\theta$ is the colatitude and $\varphi$ is the longitude).
The amplitudes of the optical and forced librations must be added because the forced libration and the tidal torque are $180^\circ$ out of phase (e.g.\ Fig.~5 of \citet{hemingway2018}).
Thus, including the 1:1 forced libration increases tidal dissipation (by about 28\%, see Eq.~(\ref{U2sq})), as already shown for a homogeneous body \citep{wisdom2004}, instead of decreasing it as concluded by \citet{behounkova2017}.

Introducing the forced libration into the tidal potential, as above, makes sense for a completely solid body, in which the differential rotation between layers is negligible.
This procedure is however problematic if the shell and the solid core are decoupled by a global ocean.
In that case, the core has a smaller libration than the shell (by a factor of 10 in amplitude, \textit{A.~Trinh, private comm.}), so that the tidal potential (\ref{TidalPot}) is only valid for the shell.
Moreover, the differential rotation of the core and shell induces gravitational and pressure couplings between the shell and core which should be taken into account in an additional forcing term.
Here I assume that the shell and core have no differential rotation so that they are forced by the same tidal potential (\ref{TidalPot}).
For the shell, this approximation is reasonable as long as the forced libration is significantly smaller than the optical libration: $\gamma_0/2e=0.223$ implies that the neglected corrections are of order $(\gamma_0/2e)^2\ll1$.
For the core, overestimating dissipation does not pose a problem because core rheology is adjusted so that the total heat budget is satisfied.

\section{Dissipation in the thin shell approach}

In this section, I set forth the full methodology required to compute dissipation in the non-uniform shell and the internally spherically symmetric core.
First, I explain how the assumption of a uniform Poisson's ratio affects dissipation.
Next, I derive formulas for dissipation in the shell (rate, flux, and power) in terms of the basic thin shell variables $(F,w)$.
Finally, I show how to compute dissipation in the core by the way of the effective tidal potential.
This method is applied to partition the total power into core and shell contributions and the compute the spatially dependent core dissipation rate.

\subsection{Shear, bulk and Poisson dissipation}
\label{ShellDissipationRate}

In the micro approach, dissipation is computed locally by multiplying at each point the microscopic stress by the strain rate.
As Enceladus is rotating synchronously with its mean motion, tidal deformations due to the eccentric orbit are periodic with an angular frequency $\omega$ equal to the mean motion.
It is thus convenient to work with Fourier-transformed variables: $V(t) = {\rm Re}(\tilde V(\omega) \, e^{i \omega t })$.
The dissipation power per unit volume averaged over one orbital period $T$, in short the \textit{dissipation rate}, reads  (Appendix~A of \citet{beuthe2013})
\begin{equation}
P(r,\theta,\varphi)
= \frac{1}{T} \int_0^T \sigma_{ij}(t) \, \dot{\epsilon}_{ij}(t) \, dt
= \frac{\omega}{2} \, {\rm Im}\left( \tilde \sigma_{ij}(\omega) \, \tilde \epsilon_{ij}^{\,*}(\omega) \right) ,
\label{AveragedPower}
\end{equation}
where the asterisk denotes complex conjugation.
The tensors $\sigma_{ij}$ and $\epsilon_{ij}$ denote stress and strain in the time domain (without tilde) or frequency domain (with tilde).
Henceforth I work with frequency-domain variables which depend implicitly on $\omega$, and I drop the `tilde' notation.
According to the correspondence principle, linear viscoelasticity is introduced through complex moduli (shear modulus $\mu$ and bulk modulus $K$) in the frequency-domain stress-strain relation:
\begin{equation}
\sigma_{ij} = 2\mu \, \epsilon_{ij} + \left( K-2\mu/3 \right) \epsilon \, \delta_{ij} \, ,
\label{ConstitutiveRel}
\end{equation}
where $\epsilon=\epsilon_{rr}+\epsilon_{\theta\theta}+\epsilon_{\varphi\varphi}$ denotes the trace of the 3D strain tensor and $\delta_{ij}$ is the Kronecker delta.
If Eq.~(\ref{ConstitutiveRel}) holds, the dissipation rate becomes
\begin{equation}
P(r,\theta,\varphi) =  \omega \, {\rm Im}(\mu) \left( \epsilon_{ij}^{} \, \epsilon_{ij}^{\,*} - \frac{1}{3} \left| \epsilon \right|^2 \right) + \frac{\omega}{2} \, {\rm Im}(K) \left| \epsilon \right|^2 \, .
\label{PowerGeneralFormula}
\end{equation}
The term proportional to ${\rm Im}(\mu)$ represents dissipation resulting from shear friction.
The term proportional to ${\rm Im}(K)$ represents bulk dissipation, which is poorly constrained although seismic data suggest that it is much smaller than shear dissipation, at least at seismic frequencies \citep{durek1995,resovsky2005}.
Thus, one usually assumes the condition of `no bulk dissipation': ${\rm Im}(K)=0$ or equivalently $K=K_e$.
This assumption has been criticized on theoretical grounds \citep{morozov2015} and might be invalid in the presence of melt \citep{takei2009}.
Recently, \citet{ricard2014} argued that seismic attenuation could be due to the laminated structure of the mantle, in which case intrinsic dissipation would not be constrained at all.
The occurrence of bulk dissipation is thus an open question.
Here, I cannot impose that $K=K_e$ because it would require that Poisson's ratio $\nu$ varies in tandem with the shear modulus $\mu$ (e.g.\ Appendix~C of \citet{beuthe2014}) according to the $K$-$\mu$-$\nu$ relation:
\begin{equation}
K = \frac{2}{3} \, \frac{1+\nu}{1-2\nu} \, \mu \, .
\label{Knurel}
\end{equation}
This would contradict the assumption of uniform $\nu$ made in the theory of non-uniform thin shells (see Section~\ref{NonUniformThinShellTheory}).
I impose instead the condition of `no Poisson dissipation', i.e.\ that Poisson's ratio remains equal to its elastic value which is real and uniform: $\nu=\nu_{\rm e}$.
This constraint together with Eq.~(\ref{Knurel}) implies that
\begin{equation}
{\rm Im} (K) = \frac{2}{3} \, \frac{1+\nu}{1-2\nu} \, {\rm Im} (\mu) \hspace{5mm} \mbox{($\nu$ is real)} \, .
\label{ImK}
\end{equation}
If there is no Poisson dissipation, the dissipation rate thus reads
\begin{equation}
P(r,\theta,\varphi) =  \omega \, {\rm Im}(\mu) \left( \epsilon_{ij} \, \epsilon_{ij}^{\,*} +   \frac{\nu}{1-2\nu} \, \left| {\rm Tr}\, \epsilon_{ij} \right|^2 \right) .
\label{PowerNoPoissonDissip}
\end{equation}
If there is no bulk dissipation, the factor $\nu/(1-2\nu)$ should be replaced by $-1/3$.
In the incompressible limit ($K\rightarrow\infty$, $\nu\rightarrow1/2$), bulk dissipation always vanishes so that the conditions of `no bulk dissipation' and `no Poisson dissipation' are simultaneously satisfied.
In that case, ${\rm Tr}\,\epsilon_{ij}=0$ and the dissipation rate reduces to the first term of Eq.~(\ref{PowerNoPoissonDissip}).

Before going further, I need to specify the rheology.
The simplest linear viscoelastic model is the one of Maxwell (see Table~\ref{TableVisco}).
In this model, ${\rm Im}(\mu)$ is maximum at the forcing angular frequency $\omega$ if the viscosity of the material is equal to the critical viscosity $\eta_{\rm crit}=\mu_{\rm e}/\omega$, where $\mu_e$ is the elastic shear modulus.
For Enceladus's icy shell, the critical viscosity is equal to $6.6\times10^{13}\rm\,Pa.s$.
But a conductive shell is not at all homogeneous: the viscosity, which controls ${\rm Im}(\mu)$, varies by orders of magnitude between the cold surface and the bottom of the shell at the melting point (the strain varies much less because it is controlled by the upper and colder part of the shell).
Thus, the dissipation rate is maximum in a thin layer where viscosity is closest to the critical viscosity. 
By the same logic, the total power dissipated in the shell is maximum if the viscosity at the bottom of the shell is lower than the critical viscosity.
Since I am interested to maximize tidal dissipation within the shell, I generally assume that the bottom viscosity is equal to $10^{13}\rm\,Pa.s$, which the lower bound for the viscosity of ice in the low stress regime \citep{tobie2003,barr2009}.
Nonetheless, I will also consider higher viscosities for benchmarking.
Although Andrade rheology (e.g.\ \citet{castillo2011}) is more realistic than Maxwell and more dissipative at high viscosity, it makes little difference if the bottom viscosity is lower than the critical viscosity (see example in Section~\ref{TotalPowerCoreShell}).
There is no problem, however, to switch in the thin shell approach from Maxwell to Andrade rheology.

\begin{figure}
   \centering
    \includegraphics[width=15cm]{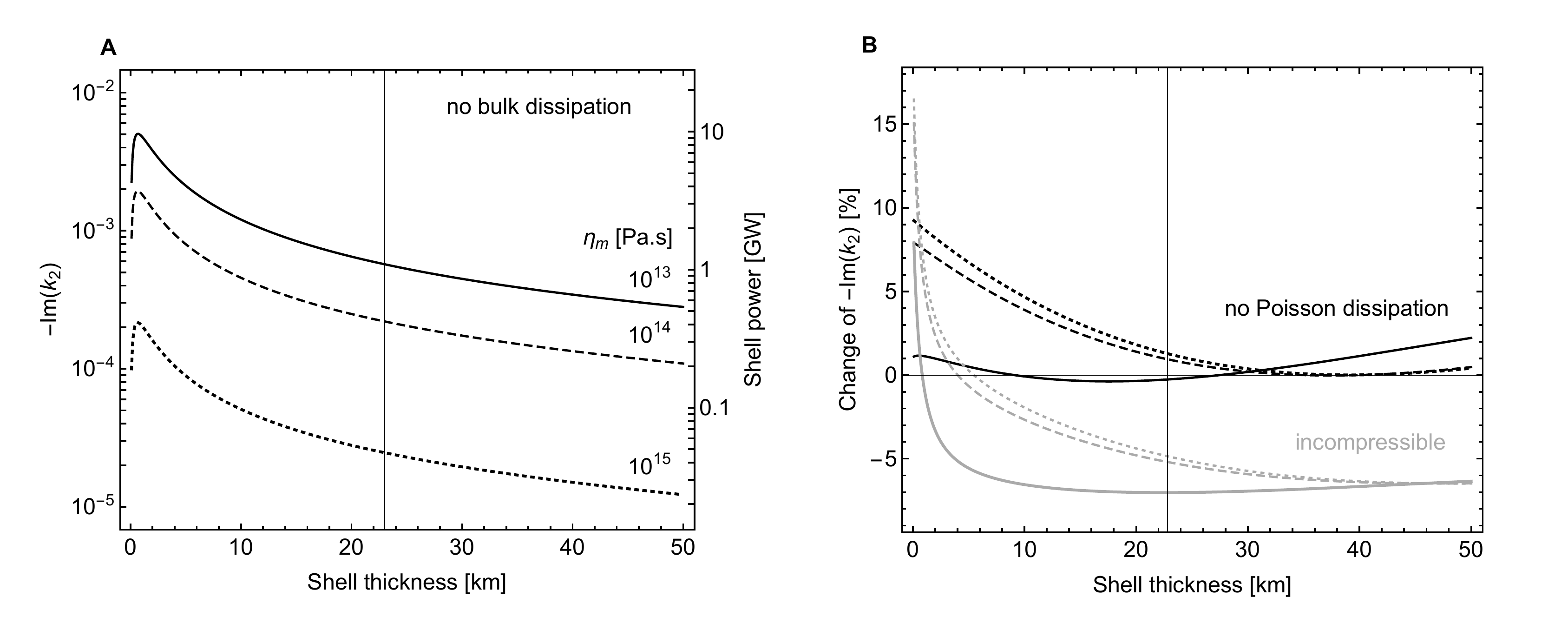}
   \caption[Impact of dissipation condition]
   {Impact of the dissipation condition on the total power dissipated in a thick shell.
   The total power is parameterized by the imaginary part of $k_2$ (see Eq.~(\ref{EdotTotalMacro})).
   (A) $-{\rm Im}(k_2)$ if there is no bulk dissipation ($\nu_e=0.33$) for three values of the bottom viscosity $\eta_{\rm m}$; the right-hand scale gives the shell power including the contribution of the forced libration.
   (B) Relative change of $-{\rm Im}(k_2)$ with respect to Panel~A  if there is no Poisson dissipation (black curves) or if the shell is incompressible (gray curves).
   Solid/dashed/dotted curves correspond to the bottom viscosities specified in Panel~A.
   The vertical line corresponds to $d=23\rm\,km$.
   See Section~\ref{ShellDissipationRate}.
}
   \label{Figk2Thick}
\end{figure}

Fig.~\ref{Figk2Thick} shows how the dissipation condition affects the total power dissipated in a thick shell (`thick' meaning no thin shell approximation) which is laterally uniform and conductive (computational details are given in Section~\ref{Motivation}).
Instead of the power itself, the figure shows the imaginary part of the gravitational Love number $k_2$, which has the advantage that the dependence on the tidal potential (eccentricity and libration) is factored out (Eq.~(\ref{EdotTotalMacro})).
If the shell thickness $d$ is in the range $20-50\rm\,km$, assuming `no Poisson dissipation' changes the total power by less than 2\% whereas assuming an incompressible shell changes it by -5 to -7\% (the precise number depends on the bottom viscosity).
For thinner shells, the effect of the `no Poisson dissipation' condition depends a lot on the bottom viscosity: it is below 1\% if $\eta_{\rm m}=10^{13}\rm\,Pa.s$, but may reach 9\% in the membrane limit if $\eta_{\rm m}=10^{15}\rm\,Pa.s$.
In comparison, the incompressible assumption changes the total power by 9 to 19\% in the membrane limit.
In conclusion, assuming an incompressible shell, as is done in the propagator matrix approach, generally leads to a larger error than the condition of  `no Poisson dissipation'.

\subsection{Dissipation inside the shell}
\label{DissipationShell}

\subsubsection{Shell dissipation rate}
\label{ThinShellDissipationRate}

Apart from the assumption of no Poisson dissipation, the dissipation rate given by Eq.~(\ref{PowerNoPoissonDissip}) is completely general and valid anywhere in the body.
I will now restrict it to the shell and work in the thin shell limit.
The plane stress approximation underlying thin shell theory implies that (Eq.~(C.1) of Paper~I)
\begin{equation}
\epsilon_{\zeta\zeta} = - \frac{\nu}{1-\nu} \left( \epsilon_{\theta\theta} + \epsilon_{\varphi\varphi} \right) .
\label{RadialStrain}
\end{equation}
Substituting this constraint into Eq.~(\ref{PowerNoPoissonDissip}), I write the \textit{shell dissipation rate} in absence of Poisson dissipation as
\begin{equation}
P_{shell}(\zeta,\theta,\varphi) =  \omega \, {\rm Im}(\mu) \left( {\cal E}_2 + \frac{\nu}{1-\nu} \, {\cal E}_{\rm tr} \right) ,
\label{PowerThinShellGeneral}
\end{equation}
where $({\cal E}_2,{\cal E}_{\rm tr})$ are 2D strain invariants:
\begin{eqnarray}
{\cal E}_2 &=&
\left| \epsilon_{\theta\theta} \right|^2 + \left| \epsilon_{\varphi\varphi} \right|^2 + 2 \left| \epsilon_{\theta\varphi} \right|^2 \, ,
\label{E2} \\ 
{\cal E}_{\rm tr} &=&
\left| \epsilon_{\theta\theta} + \epsilon_{\varphi\varphi} \right|^2 \, .
\label{Etr}
\end{eqnarray}
If there is no bulk dissipation, the factor $\nu/(1-\nu)$ in Eq.~(\ref{PowerThinShellGeneral}) should be replaced by the factor $(4{\rm Re}(\nu)-1-|\nu|^2)/(3|1-\nu|^2)$; both factors tend to 1 in the incompressible limit.

If the tidal thin shell equations are solved for $(F,w)$, the shell dissipation rate can be written as a bilinear form in these variables and their complex conjugates (see Appendix~\ref{DissipationRate}):
\begin{equation}
P_{shell}(\zeta,\theta,\varphi) =  \omega \, {\rm Im}(\mu) \left( {\cal E}^{FF} + {\cal E}^{Fw} + {\cal E}^{ww} \right) .
\label{PowerThinShell}
\end{equation}
If there is no Poisson dissipation, the terms in the RHS are given by
\begin{eqnarray}
{\cal E}^{FF} &=& |\alpha|^2 \, (1+\nu) \left( \left| \Delta' F \right|^2 - (1+\nu) \, {\cal A}(F \,; F^*) \right) ,
\nonumber \\
{\cal E}^{Fw} &=& - \frac{1+\nu}{R} \Big( \alpha \, z^* \, {\cal A}(F \,; w^*) + \alpha^* z \, {\cal A}(F^* \,; w)  \Big) \, ,
\nonumber \\
{\cal E}^{ww} &=& \frac{|z|^2}{R^2} \, \frac{1}{1-\nu} \left( \left| \Delta' w \right|^2 - (1-\nu) \, {\cal A}(w \,; w^*) \right) .
\label{PowerThinShellComp}
\end{eqnarray}
The shell dissipation rate depends on depth through ${\rm Im}(\mu)$ and the depth parameter $z$ (Table \ref{TableVisco}).
It does not depend on the choice of reference surface, since $(w,\alpha{}F,z/R)$ are invariant under a change of $R$ (see Appendix~G of Paper~I and Eq.~(J.3) of Paper~I).

\subsubsection{Shell dissipation flux}

The \textit{shell dissipation flux} is the energy flux, due to dissipation within the shell, through the reference surface of the shell (chosen here to be the outer surface of the body).
If the heat transfer is radial, the shell dissipation flux is equal to the dissipation rate integrated over the shell thickness,
\begin{equation}
{\cal F}_{shell}(\theta,\varphi) = \int_d P_{shell}(\zeta,\theta,\varphi) \left( 1 + \zeta/R \right)^2 d\zeta \, ,
\label{Fshell}
\end{equation}
where the factor $(1+\zeta/R)^2$ is associated with the integration measure in spherical coordinates.
The shell dissipation flux can be expressed in terms of the variables $(F,w)$ and of the parameters $(\nu,\alpha,D,\chi)$ (see Appendix~\ref{AppendixSurfaceFlux}):
\begin{equation}
{\cal F}_{shell}(\theta,\varphi) = {\cal F}_{mem} + {\cal F}_{mix} + {\cal F}_{bend} \, .
\label{SurfaceFluxThinShell}
\end{equation}
If there is no Poisson dissipation, the terms in the RHS are given by
\begin{eqnarray}
{\cal F}_{mem} &=& \!\!\!\! - \frac{\omega}{2} \, {\rm Im}(\alpha) \left( \left| \Delta' F \right|^2 - (1+\nu) \, {\cal A}(F \,; F^*) \right) ,
\nonumber \\
{\cal F}_{mix} &=& \frac{\omega}{2} \, \frac{ {\rm Im}(\chi)}{R} \, \Big( {\cal A}(F \,; w^*) + {\cal A}(F^* \,; w) \Big) \, ,
\nonumber \\
{\cal F}_{bend} &=& \frac{\omega}{2} \, \frac{{\rm Im}(D) }{R^4} \left( \left| \Delta' w \right|^2 - (1-\nu) \, {\cal A}(w \,; w^*) \right) .
\label{SurfaceFluxThinShellComp}
\end{eqnarray}
The subscripts \textit{mem, mix, bend} stand for \textit{membrane}, \textit{mixed} and \textit{bending} contributions.
This is analogous to the decomposition of the elastic energy in extensional-shearing, mixed, and bending-twisting terms \citep{novozhilov1964,axelrad1987}.
Membrane and bending contributions are always positive whereas the mixed contribution can be negative.

The shell dissipation flux has several nice properties:
\begin{itemize}
\item
It depends on rheology through depth-integrated shell parameters: ${\rm Im}(\alpha)$, ${\rm Im}(\chi)$, ${\rm Im}(D)$.
\item
It depends on scalar quantities and scalar differential operators, making it easy to compute it with the pseudospectral transform method.
\item
It does not depend on the degree-1 spherical harmonic components of $w$ (invariance under rigid displacements) and of $F$ (degree-1 gauge freedom), which belong to the null space of $\Delta'$ and ${\cal A}$.
\item
It is inversely proportional to the area of the reference surface of the shell (as it should be), because $w$ is invariant under a change of $R$ while other quantities scale as $F\sim{}R^{-2}$, $\alpha\sim{}D\sim{}R^2$, and $\chi\sim{}R$ (see Appendix~G of Paper~I).
\item
It satisfies the \textit{static-geometric duality} (Eq.~(17) of Paper~I), a transformation exchanging the LHS of the governing equations:
$(w,D,\nu,\chi)\leftrightarrow(R^2F,-\alpha,-\nu,\chi)$.
\end{itemize}

\subsubsection{Shell power}
\label{ThinShellPower}

The total power dissipated in the shell, or \textit{shell power}, is obtained by integrating the shell dissipation flux over the reference surface with surface element $dS=R^2 \sin\theta \, d\theta \, d\varphi$:
\begin{eqnarray}
\dot{E}_{shell}
&=&  \int {\cal F}_{shell}(\theta,\varphi) \, dS
\nonumber \\
&=&  \int \left( {\cal F}_{mem} + {\cal F}_{mix} + {\cal F}_{bend} \right) dS
\nonumber \\
&\equiv& \dot{E}_{mem} + \dot{E}_{mix} + \dot{E}_{bend} \, .
\label{EdotShellMicro}
\end{eqnarray}
In general, this integral must be evaluated numerically, but it can be done analytically if the shell is laterally uniform (see Appendix~\ref{PatternsUniformThinShell}).
Eq.~(\ref{EdotShellMicro}) embodies the \textit{micro approach} to tidal dissipation, in which the total power dissipated in the shell is evaluated by integrating the microscopic dissipation rate over the volume of the shell.

An alternative approach to tidal dissipation (\textit{macro approach}) consists in computing the total power dissipated in the body from the work done by the tidal potential (Zschau-Platzman formula, see Eq.~(7) of \citet{platzman1984}):
\begin{eqnarray}
\dot E_{tot} = \frac{\omega}{2} \, \frac{1}{4\pi{}GR} \sum_n \left(2n+1\right) \int_S {\rm Im} \left( U_n^T \, {U_n'}^* \right) dS \, ,
\label{Edot1}
\end{eqnarray}
where $U_n'$ is the secondary potential due to the deformation of the body:
\begin{equation}
U_n' =\Gamma_n-U_n^T \, ,
\label{Unprime}
\end{equation}
in which $\Gamma_n$ is the total perturbing potential (Eq.~(\ref{qn})).
If the primary tidal potential is of degree two, substituting Eq.~(\ref{Gamman}) into Eq.~(\ref{Edot1}) yields the total power in terms of $(U_2^T,w_2)$:
\begin{equation}
\dot E_{tot} = - \frac{5}{2} \, \frac{\omega R}{G} \, {\rm Im}(\upsilon_2) \, \langle |U_2^T|^2 \rangle + 2 \pi R^2 \omega \rho \, {\rm Im} \langle \, U_2^T \, \upsilon_2^* \, w_2^* \, \rangle \, ,
\label{Edot2}
\end{equation}
where the bracket notation $\langle x \rangle$ denotes the angular average of $x$
(or degree-0 spherical harmonic coefficient of $x$).
Once the tidal thin shell equations have been solved for $(F,w)$, Eq.~(\ref{Edot2}) immediately yields the total dissipated power.

If the core is elastic (${\rm Im}(\upsilon_2)=0$) and the ocean is inviscid, the total dissipated power is equal to the shell power.
Setting ${\rm Im}(\upsilon_2)=0$ into Eq.~(\ref{Edot2}) and using Eqs.~(\ref{qn})-(\ref{Gamman}), I can write
\begin{equation}
\dot E_{tot} \, = \, \dot E_{shell} = \frac{\omega}{2} \, {\rm Im} \int_S \, q_2 \, w_2^* \, dS \, ,
\label{Edot3}
\end{equation}
which can be interpreted as follows: $\dot E_{shell}$ is equal to the dissipative part of the power developed by the bottom load acting on the shell (see Appendix~E of \citet{beuthe2014}).
Dissipation only occurs at degree~2 because $(q_n,w_n)$ are in phase if $\upsilon_n$ is real and $U_n^T=0$ (see Eqs.~(\ref{qn})-(\ref{Gamman})).
Eq.~(\ref{Edot3}) provides a non-trivial check on the integrated dissipation rate given by Eq.~(\ref{EdotShellMicro}).

If the core is viscoelastic, it is tempting -- but not correct -- to interpret the two terms of the RHS of Eq.~(\ref{Edot2}) as core and shell contributions.
In Section~\ref{ThinShellPowerPartition}, I will explain how to split $\dot E_{tot}$ between core and shell using the effective tidal potential.

\subsection{Dissipation inside the core}
\label{DissipationCore}

\subsubsection{Effective tidal potential for the core}
\label{EffectiveTidalPotential}

Another source of dissipation arises from the tidal deformations of the viscoelastic core.
If the whole body has a spherically symmetric structure, one first solves the (radial) viscoelastic-gravitational equations from the center to the surface of the body before evaluating the dissipation rate at each point.
Here, the laterally non-uniform shell breaks spherical symmetry.
Nevertheless, radial viscoelastic-gravitational equations are still valid in the core (and ocean) as long as the interior structure is spherically symmetric under the shell.
The effect of the viscoelastic shell can be represented as a pressure load
\begin{equation}
U_n^P=\xi_n q_n/\rho \, ,
\label{UnP}
\end{equation}
which acts on the associated fluid-crust body (defined as the body equivalent to the original one except that the shell has no rigidity; see Section 3.3\ of Paper~I).
Besides this pressure load, tides deform the fluid-crust body (Eq.~(26) of Paper~I).

The radial viscoelastic-gravitational equations should now be solved in a 2-layer body (core plus surface ocean) submitted to tidal and pressure loads.
Beware that `2-layer' does not imply here uniform layers: density and rheology can be depth-dependent.
In the standard approach \citep{takeuchi1972}, the displacements, stresses and gravitational perturbations due to a forcing of degree~$n$ are represented in a solid layer by 6 radial response functions $y_{i n}(r)$ ($i=1...6$).
Since the tidal and pressure load solutions correspond to different boundary conditions, the full solution reads
\begin{equation}
\Upsilon_i(r,\theta,\varphi) = \sum_n \Big( \, y_{i n}^{\circ T}(r) \, U_n^T(\theta,\varphi) + y_{i n}^{\circ P}(r) \, U_n^P(\theta,\varphi) \, \Big) \, ,
\label{Upsilon1}
\end{equation}
where the tidal potential is evaluated at radius $R$.
The variables $y_{i n}^{\circ J}$ ({\small \textit{J=T,P}}) are propagated from the center (where they depend on 3 unknown constants) to the surface of the solid core by solving 6 viscoelastic-gravitational differential equations.
At the core-ocean boundary, a first boundary condition is given by the condition of zero shear stress ($y_{4 n}^{\circ J}=0$).
In the limit of static deformations \citep{saito1974}, a second homogeneous boundary condition is provided by the fluid constraint (e.g.\ \citet{beuthe2015}), which relates $y_{1 n}^{\circ J}$ (radial displacement), $y_{2 n}^{\circ J}$ (radial stress) and $y_{5 n}^{\circ J}$ (gravitational potential perturbation).
As there is no third boundary condition at the core-ocean boundary, one must go on and solve the differential equations in the fluid layer.
Under the assumption of static deformations, the gravitational potential decouples from displacements: it becomes sufficient to propagate the variables $y_{5 n}^{\circ J}$ and $y_{7 n}^{\circ J}=y_{6 n}^{\circ J}+(4\pi{}G/g)y_{2 n}^{\circ J}$ \citep{saito1974}.
At the outer fluid surface, there is only one (inhomogeneous) boundary condition given by
\begin{eqnarray}
y_{7 n}^{\circ T}(R) &=& \frac{2n+1}{R} \hspace{7mm} \mbox{(tidal load)} \, ,
\\
y_{7 n}^{\circ P}(R) &=& - \frac{2n+1}{R} \hspace{5mm} \mbox{(pressure load)} \, ,
\end{eqnarray}
which result from the usual boundary conditions for tidal forcing and pressure loading (Eqs.~(C.5) and (E.3) of \citet{beuthe2016a}).
Thus, the pressure and tidal load solutions are related within the core by $y_{i n}^{\circ P}=-y_{i n}^{\circ T}$ ($i=1...6$).

Defining the \textit{effective tidal potential} for the core by
\begin{equation}
U_n^\circ = U_n^T - U_n^P \, ,
\label{Un0a}
\end{equation}
I can write the full solution in the core (Eq.~(\ref{Upsilon1})) as the tidal solution forced by $U_n^\circ$:
\begin{equation}
\Upsilon_i(r,\theta,\varphi) = \sum_n y_{i n}^{\circ T}(r) \, U_n^\circ \, .
\label{Upsilon2}
\end{equation}
The similar property $y_{i n}^{\circ P}=-y_{i n}^{\circ T}$ ($i=5,7$) holding within the fluid leads to relations between pressure and tidal Love numbers: $k_n^{\circ P} = - h_n^\circ$ and $h_n^{\circ P} = -1/\xi_n - h_n^\circ$, which were already noted as Eqs.~(29)-(30) of Paper~I (the superscript $T$ is omitted for tidal Love numbers).

For evaluation purposes, it is practical to write the effective tidal potential in terms of $U_n^T$ and the flexure solution $w_n$:
\begin{equation}
U_n^\circ = \frac{1}{1+\xi_n h_n^\circ} \left( U_n^T + g \, \xi_n  w_n \right) .
\label{Un0b}
\end{equation}
This formula gives us a preliminary estimate of how much lateral variations of the shell structure influence the core dissipation pattern. 
If the shell is laterally uniform, the radial displacement is related to the tidal potential by $w_n=h_nU_n^T/g$ (see Section~\ref{Motivation}), so that the second term in the brackets of Eq.~(\ref{Un0b}) is small with respect to the first: $\xi_n \, h_n \, U_n^T\ll{}U_n^T$ ($\xi_2h_2\sim10^{-2}$, see Section~\ref{Motivation}).
Lateral variations in the shell result in small deviations from this prediction.
The maximum deviation occurs at the south pole, where the radial displacement at the south pole changes from $-0.5\rm\,m$ to $-0.7\rm\,m$ (see Fig.~8 of Paper~I). 
Thus, lateral variations in the shell change the effective tidal potential by less than 1\% and cause negligible deviations in the core dissipation pattern with respect to the case of a laterally uniform shell.

\subsubsection{Core-shell partition of total power}
\label{ThinShellPowerPartition}

In Section~\ref{ThinShellPower}, I used the macro approach in order to compute the total power dissipated in the body, and showed that it is equal to the shell power if the core is elastic.
If the core is viscoelastic, the total power can be split into core and shell contributions with the help of the effective tidal potential.
Note first that the total perturbing potential can be written as
\begin{equation}
\Gamma_n = g  w_n + U_n^P/\xi_n \, ,
\label{Gamman2}
\end{equation}
which results from Eqs.~(\ref{qn}) and (\ref{UnP}).
Besides, $\Gamma_n$ is related to the effective tidal potential for the core by
\begin{equation}
\Gamma_n =  \left( 1 + k_n^\circ \right) U_n^\circ \, ,
\label{Un0c}
\end{equation}
which results from Eqs.~(\ref{Gamman}) and (\ref{Un0b}).

Using Eqs.~(\ref{Un0a}), (\ref{Gamman2}), and (\ref{Un0c}), I write the integrand of Eq.~(\ref{Edot1}) as
\begin{eqnarray}
{\rm Im} \left( U_n^T \, {U_n'}^* \right)
&=& {\rm Im} \left( U_n^\circ \, \Gamma_n^* \right) + {\rm Im} \left( U_n^P \, \Gamma_n^* \right)
\nonumber \\
&=& - \, {\rm Im} (k_n^\circ) | U_n^\circ |^2 + \frac{4\pi G R}{2n+1} \, {\rm Im} ( q_n \, w_n^* ) \, .
\end{eqnarray}
The total power dissipated in the body thus reads
\begin{equation}
\dot E_{tot} = \dot E_{core} + \dot E_{shell} \, ,
\end{equation}
where
\begin{eqnarray}
\dot E_{core} &=&  - \frac{\omega R}{2G} \sum_n \left(2n+1\right) {\rm Im} (k_n^\circ) \, \langle | U_n^\circ |^2 \rangle \, ,
\label{EdotCoreMacro} \\
\dot E_{shell} &=& \frac{\omega}{2} \sum_n {\rm Im} \int_S \, q_n \, w_n^* \, dS \, .
\label{EdotShellMacro}
\end{eqnarray}
Similarly to Eq.~(\ref{Edot3}), $\dot E_{shell}$ can be identified as the dissipative part of the power developed by the bottom load acting on the massless shell.
It is thus equal to the shell power, the difference with Eq.~(\ref{Edot3}) being that dissipation now occurs at all harmonic degrees.
If the shell is elastic, $\dot E_{shell}$ vanishes because the tidal thin shell equations impose that $(q_n,w_n)$ are in phase.

If $\dot E_{shell}$ is the shell power in Eq.~(\ref{EdotShellMacro}), $\dot E_{core}$ must be equal by subtraction to the power dissipated in the core.
This claim is confirmed in all generality by integrating the dissipation rate over the volume below the shell, using energy conservation (Eq.~(179) of \citet{takeuchi1972}) and the surface boundary conditions for tidal and pressure loading solutions.
As a caveat, beware that the above core-shell partition relies on the assumption of no differential rotation between shell and core (see Section~\ref{TidalLoad}).

For a laterally varying shell, $\dot E_{core}$ and $\dot E_{shell}$ can be expressed in terms of $(w_n,U_n^T)$ by substituting Eq.~(\ref{qn})-(\ref{Gamman}) and Eq.~(\ref{Un0b}) into Eqs.~(\ref{EdotCoreMacro})-(\ref{EdotShellMacro}).
Once the tidal thin shell equations have been solved, it is straightforward to evaluate the core and shell contributions to the total dissipated power.
Comparing the numerical values yielded by the micro and macro  formulas (Eqs.~(\ref{EdotShellMicro}) and (\ref{EdotShellMacro})) for the shell power provides a self-consistency check for numerical codes.

\subsubsection{Core dissipation rate}
\label{MicroApproachCore}

Spatial patterns of core dissipation can only be computed in the micro approach.
If the forcing is of degree~2 (or more generally of a given degree~$n$), the dissipation rate in a spherically symmetric layer can be factorized into radial and angular parts which depend on the internal structure and on the square of the forcing potential, respectively (see \citet{beuthe2013}).
The non-uniformity of the shell, however, introduces harmonic degrees other than degree~2 in the forcing potential (Eq.~(\ref{Un0a})), which interfere when computing the squared forcing potential.
Thus, it is not sufficient to consider the patterns corresponding a tidal forcing of a given harmonic degree.
Nonetheless, it is advantageous to express the dissipation rate in terms of spherical differential operators acting on scalar radial functions, instead of stresses and strains, because derivatives of scalar fields can be computed efficiently with the pseudospectral transform method.

Following Section~2.2 of \citet{beuthe2013}, I write the dissipation rate within the incompressible core as
\begin{equation}
P_{core} = \frac{\omega}{r^2} \, \mbox{Im}(\mu) \left(E_A + E_B + E_C \right) ,
\label{PowerStrainInv}
\end{equation}
where
\begin{eqnarray}
E_A &=& 6 \left| \Upsilon_1 + \frac{1}{2} \, \Delta  \Upsilon_3 \, \right|^2 ,
\nonumber \\
E_B &=& \frac{1}{2} \, {\cal D}_2 \left( \frac{r \Upsilon_4}{\mu} \, ; \frac{r \Upsilon_4^*}{\mu^*} \right) ,
\nonumber \\
E_C &=& {\cal D}_4 \left( \Upsilon_3 \, ; \Upsilon_3^* \right) - \frac{1}{2} \left| \, \Delta \Upsilon_3 \, \right|^2 .
\label{InvFact}
\end{eqnarray}
$E_A$, $E_B$, and $E_C$ correspond respectively to $E_{radial}+E_{tan2}-E_{dilat}/3$, $E_{shear}$, and $E_{tan1}$ in \citet{beuthe2013}.
The differential operators ${\cal D}_2$ and ${\cal D}_4$ are defined in Appendix~\ref{SphericalOperators}.
The functions $\Upsilon_i$ are defined by Eqs.~(\ref{Upsilon2})-(\ref{Un0b}): $\Upsilon_1$ is the radial displacement, $\Upsilon_3$ is the potential for tangential displacement, and $\Upsilon_4$ is the potential for radial-tangential stress (or strain).
If the core and the ocean are both homogeneous, the functions $y_{in}^T$ appearing in $\Upsilon_i$ are given by Eq.~(\ref{yifun}).

As the shell is laterally non-uniform, the functions $\Upsilon_i$ are superpositions of spherical harmonics of different degrees.
In that case, the various terms appearing in the RHS of Eq.~(\ref{InvFact}) can be evaluated with the transform method mentioned in Section \ref{FlexureEquations}.
Derivatives only appear through the spherical Laplacian $\Delta$ (or the related operator $\Delta'$), as can be seen for the operators ${\cal D}_2$ and ${\cal D}_4$ with identities (b), (c), and (e) of Appendix~\ref{SphericalOperators}.

Under the assumption of radial heat transfer, the heat flux at the core-ocean boundary is equal to the dissipation rate integrated over the core radius,
\begin{equation}
{\cal F}_{core}(\theta,\varphi) = \int_0^{R_c} P_{core}(r,\theta,\varphi) \, \Big( \frac{r}{R_c} \Big)^2 dr \, .
\label{Fcore}
\end{equation}
The total power dissipated in the core is obtained by integrating the flux over the core surface with surface element $dS_c=R_c^2 \sin\theta \, d\theta \, d\varphi$:
\begin{equation}
\dot{E}_{core} =  \int {\cal F}_{core}(\theta,\varphi) \, dS_c \, .
\label{EdotCoreMicro}
\end{equation}
This expression should be equal to the power obtained in the macro approach (Eq.~(\ref{EdotCoreMacro})).
If the core and the ocean are both homogeneous, this integral can be done analytically with the solution of Appendix~\ref{DeformationCore}, yielding Eq.~(\ref{EdotCoreMacro}) in which $k_n^\circ$ is given by Eq.~(\ref{hn0}).
Note that interfering harmonic degrees do not contribute to the angular integral.
The equivalence between Eqs.~(\ref{EdotCoreMacro}) and (\ref{EdotCoreMicro}) is another example of the consistency between micro and macro approaches to tidal dissipation.

\section{Benchmarking against a laterally uniform thick shell}
\label{BenchmarkingU}

In this section, I benchmark the thin shell solution for tidal dissipation against the thick shell solution for a laterally uniform shell.
The primary aim is to quantify the impact of the thin shell approximation on dissipation patterns and on the total power dissipated in the shell and core.
Dissipation patterns will be analyzed with the radial-angular factorization method.
Regarding the total power, it can be expressed in terms of the imaginary part of the Love number $k_2$, making it easy to study the error due to the thin shell approximation.

\subsection{Factorization of dissipation rate}
\label{Motivation}

For this benchmark, Enceladus is modelled as a 3-layer body made of a homogeneous core, a homogeneous ocean, and a conductive icy shell.
The internal structure is spherically symmetric.
Model parameters are given in Table~\ref{TableParam}.
The rheology of the shell is modelled as in Section~4.2.3 of Paper~I: the rheology is described with the Maxwell model; the viscosity depends on temperature through an Arrhenius relation (Table~\ref{TableVisco}); the temperature profile is the solution of the Cartesian 1D heat equation without internal source (Eq.~(56) of Paper~I); the surface temperature is uniform and set to $59\rm\,K$.

Before computing dissipation, one should solve for tidal deformations.
In the thin shell approach, the tidal thin shell equations can be solved analytically in the spherical harmonic basis (see Table~\ref{TableUniform}).
The exact solution for a thick shell is obtained by integrating the elastic-gravitational equations for the spherically symmetric problem in the static limit \citep{takeuchi1972,saito1974}, as done for the core in Section~\ref{EffectiveTidalPotential}.
The solution is a set of six radial functions $y_i$, three of which are needed here: $y_1$ (radial displacement), $y_3$ (potential for tangential displacement), and $y_4$ (potential for radial-tangential stress or strain).

\begin{table}[ht]\centering
\ra{1.3}
\small
\caption[Tidal deformation of a laterally uniform thin shell]
{\small
Degree-$n$ tidal deformation of a laterally uniform thin shell (see Section~4 of Paper~I, except for $U_n^\circ$ which is given by Eq.~(\ref{Un0c})).
}
\begin{tabular}{@{}llll@{}}
 \hline
Symbol & Name &  Solution${}^{a,b}$ & Example${}^{c}$  \\
\hline
\multicolumn{4}{l}{Basic variables} \\
$w_n$ &radial displacement &  $h_n \, U_n^T/g$ & \\
$F_n$ & stress function &  $(\chi/\alpha) / ( \delta_n'-1-\nu ) \, (w_n/R)$ & \\
$\Gamma_n$ & total perturbing potential & $( k_n + 1 ) \, U_n^T$ & \\
$q_n$ & tidal load & $\rho g \, \Lambda_n \, w_n$ & \\
$U_n^\circ$ & effective tidal potential & $((k_n+1)/(k_n^\circ+1)) \, U_n^T$ & \\ 
\hline
\multicolumn{4}{l}{Spring Constants (SC)} \\
$\Lambda^M_n$ & membrane SC & $\delta'_n / (\delta'_n-1-\nu ) \, ( \rho g R^2 \alpha_{\rm inv})^{-1}$ & $21.267 + 0.730 \, i$ \\
$\Lambda^{B\,}_n$ & bending SC & $\delta'_n ( \delta'_n-1+\nu ) \, D_{\rm inv}/(\rho{}gR^4)$ & $0.378 + 0.038 \, i$ \\
$\Lambda^{corr}_n$ & next-to-leading SC & $( \chi/\psi -1 ) \Lambda_n^M + ( \chi \psi -1 ) \Lambda_n^B$ & $0.020 + 0.003 \, i$ \\
$\Lambda_n$ & thin shell SC & $\Lambda_n^M + \Lambda_n^B\, + \Lambda_n^{corr}$ & $21.665 + 0.771 \, i$ \\
\hline
\multicolumn{4}{l}{Tidal Love numbers (TLN)} \\
$h_{n}^\circ$ & fluid-crust radial TLN & Eqs.~(\ref{hn0})-(\ref{hn0R}) & $1.594$\\
$k_{n}^\circ$ & fluid-crust gravitational TLN & $h_{n}^\circ-1$ & $0.594$\\
$h_n$ & radial TLN & $h_{n}^\circ/(1+ (  1 + \xi_n \, h_{n}^\circ ) \Lambda_n )$ & $0.0448- 0.0015 \, i$ \\
$k_n$ &  gravitational TLN & $\left( 1 + \Lambda_n \right) h_n -1$ & $0.0167- 0.0006 \, i$ \\
\hline
\multicolumn{4}{l}{\scriptsize ${}^{a}$ $(\alpha,\alpha_{\rm inv},D_{\rm inv},\chi,\psi)$ are defined in Table~\ref{TableVisco}; numerical values are given in Table~3 of Paper~I.}
\vspace{-1mm}\\
\multicolumn{4}{l}{\scriptsize ${}^{b}$ $\xi_n$ is the degree-$n$ density ratio, see Eq.~(\ref{xin}); $\delta_n'=-(n-1)(n+2)$, see Appendix~\ref{SphericalOperators}.}
\vspace{-1mm}\\
\multicolumn{4}{l}{\scriptsize ${}^{c}$ If $n=2$, $d=23\rm\,km$, $\eta_{\rm m}=10^{13}\rm\,Pa.s$; and the core is elastic (other parameters given in Table~\ref{TableParam}).} \\
\end{tabular}
\label{TableUniform}
\end{table}%

\newpage

If the internal structure is spherically symmetric, the dissipation rate can be factorized \citep{beuthe2013}:
\begin{equation}
P(r,\theta,\varphi) =  \frac{\omega^5 R^4}{2r^2} \, {\rm Im}(\mu) \, \Big( \left( f_A + f_K \right) \Psi_A + f_B \, \Psi_B + f_C \, \Psi_C \Big) \, ,
\label{PowerSpherical}
\end{equation}
where $f_J$ are the \textit{radial weight functions} while $\Psi_J$ are angular functions representing the \textit{basic spatial patterns}.
The ratios $(f_A + f_K, f_B , f_C)/f_T$ (with $f_T=\sum_J f_J$) measure the contributions of the patterns to the angular average of the dissipation rate.
Table~\ref{TableRadialAngular} gives the weight functions in terms of the viscoelastic-gravitational solutions $y_i$,
and specifies the angular functions for degree-$2$ eccentricity tides combined with the forced libration (see Eq.~(\ref{TidalPot}) and Appendix~D of \citet{beuthe2013}).

In a similar fashion, the shell surface flux can be expressed as a weighted sum of the angular functions $\Psi_J$:
\begin{equation}
{\cal F}_{shell}(\theta,\varphi) = \left( {\cal F}_A + {\cal F}_K \right) \Psi_A + {\cal F}_B \, \Psi_B  + {\cal F}_C \, \Psi_C \, ,
\label{SurfaceFluxABC}
\end{equation}
where
\begin{equation}
{\cal F}_J = \frac{\omega^5 R^2}{2}  \int_d {\rm Im}(\mu) \, f_J \, dr \, .
\label{FluxWeight}
\end{equation}
The ratios $({\cal F}_A + {\cal F}_K , {\cal F}_B , {\cal F}_C)/{\cal F}_T$ (with ${\cal F}_T=\sum_J {\cal F}_J$) measure the contributions of the patterns to the average flux (or to the total power).

\begin{table}[htp]
\ra{1.3}
\small
\caption[Radial weights and angular functions factorizing the dissipation rate]
{Radial weights and angular functions factorizing the dissipation rate in a laterally uniform body forced by degree-2 tides (Eq.~(\ref{PowerSpherical})).
The harmonic basis is given for eccentricity tides combined with the forced libration: $\gamma_1=1+\gamma_0/2e$ (Eq.~(\ref{TidalPot})).
The symbols $y_i$ are the viscoelastic-gravitational solutions of \citet{takeuchi1972}.
If the body is incompressible, $r\partial_ry_1=-2y_1+6y_3$ and $f_K=0$.
$f_T$ is the weight for the angular average of the dissipation rate.
$P_{nm}$ denote the unnormalized associated Legendre functions of degree $n$ and order $m$ with argument $\cos\theta$.}
\begin{center}
\begin{tabular}{lll}
\hline
$J$ & Radial weight $f_J$ & Angular function $\Psi_J$\\
$A$ & $(4/3) \left| r \partial_r y_1 - y_1 + 3 y_3 \right|^2$
& $\Psi_0 + \Psi_2 + \Psi_4$
\\
$B$ & $6 \left| r y_4/\mu \right|^2$
& $\Psi_0 + (1/2) \Psi_2 - (2/3) \Psi_4$
\\
$C$ & $24 \left| y_3 \right|^2$
& $\Psi_0 - \Psi_2 + (1/6)\Psi_4$
\\
$K$ & $\frac{{\rm Im}(K)}{{\rm Im}(\mu)} \left| r \partial_r y_1 + 2 y_1 - 6 y_3 \right|^2$
& $\Psi_0 + \Psi_2 + \Psi_4$
\\
$T$ & $f_A + f_B + f_C + f_K$
& $\Psi_0$
\\
\hline
$n$ & \multicolumn{2}{l}{Harmonic basis $\Psi_n$} \\
0 & \multicolumn{2}{l}{ $\frac{3}{5} \left( 3+4\gamma_1^2 \right) e^2$} \\
2 & \multicolumn{2}{l}{ $-\frac{3}{7}  \left( 3+8\gamma_1^2 \right) e^2 P_{20} + \frac{9}{14} e^2 P_{22} \cos2\varphi$} \\
4 & \multicolumn{2}{l}{ $\frac{9}{140} \left( 27+16\gamma_1^2 \right) e^2 P_{40} - \frac{27}{140} \,e^2 P_{42} \cos2\varphi + \frac{3}{1120} \left( 9-16\gamma_1^2 \right) e^2 P_{44} \cos4\varphi$} \\
\hline
\end{tabular}
\end{center}
\label{TableRadialAngular}
\end{table}%

\begin{figure}
   \centering
      \includegraphics[width=15cm]{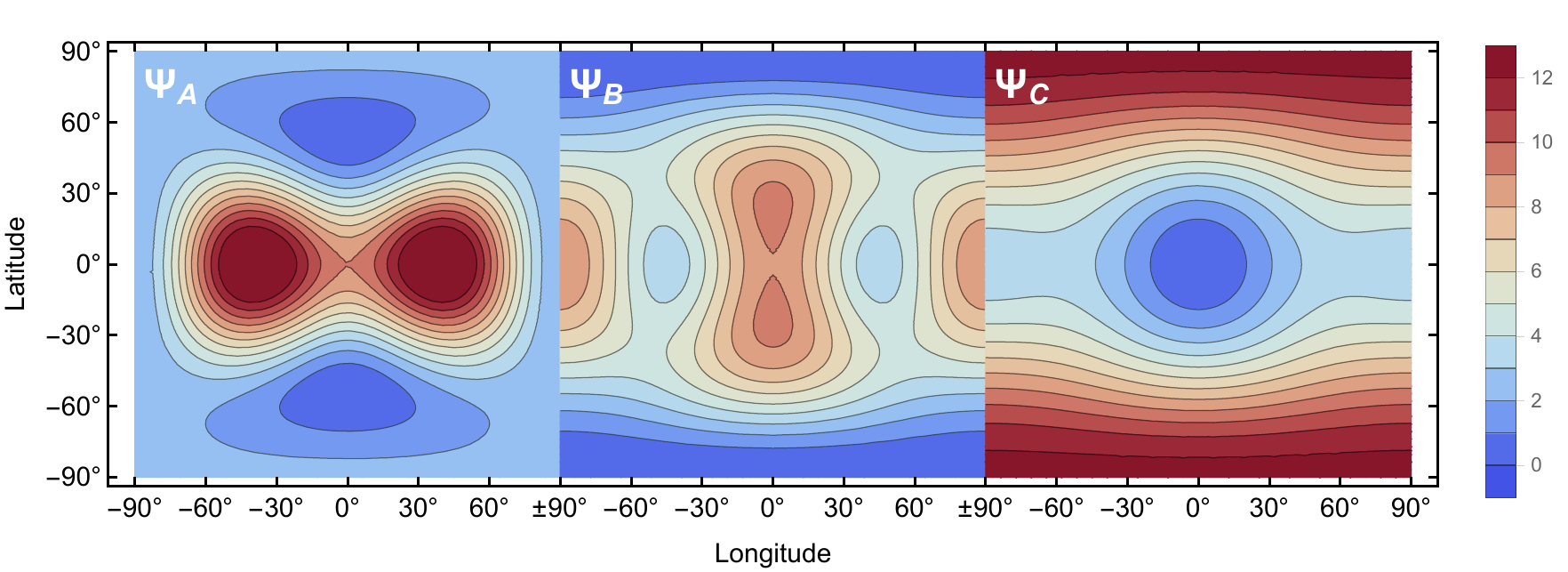}
   \caption[Basic patterns of tidal heating in a laterally uniform shell]
   {Basic patterns of tidal heating in a laterally uniform shell, due to degree-2 eccentricity tides and $0.12^\circ$ forced libration.
   The tidal axis goes through $0^\circ$ longitude.
   Each pattern repeats from $90^\circ$ to $-90^\circ$.
   The amplitude has been divided by $e^2$.
    See Table~\ref{TableRadialAngular} for analytical expressions.}
   \label{FigPatterns}
\end{figure}

Fig.~\ref{FigPatterns} shows the basic dissipation patterns $(\Psi_A,\Psi_B,\Psi_C)$ for degree-2 eccentricity tides combined with forced libration ($\gamma_0=0.12^\circ$ or $\gamma_1=1.223$).
These patterns do not differ much from the patterns without forced libration (see Fig.~1 of \citet{beuthe2013}; beware that this figure shows patterns with zero mean and unit standard deviation).
$\Psi_A$ includes significant contributions from both degrees~2 and 4, has maxima along the equator and minima at middle latitudes.
$\Psi_B$ does not contribute in the thin shell approximation (but it contributes to core dissipation).
$\Psi_C$ is dominated by the degree-2 component of the squared tidal potential, is maximum at the poles, and zero along the tidal axis.

\subsection{Patterns in a laterally uniform thin shell}
\label{DissipationRateUniform}
 
If the shell is laterally uniform, the thin shell dissipation rate must be factorizable in radial weights and angular functions.
Thin shell assumptions (i.e.\ purely tangential stress) imply that only two patterns contribute: $\Psi_A$ and $\Psi_C$.
In Appendix~\ref{PatternsUniformThinShell}, I substitute the analytical solutions for $(F,w)$ given in Table~\ref{TableUniform} into the general expression of the thin shell dissipation rate (Eqs.~(\ref{PowerThinShell})-(\ref{PowerThinShellComp})).
The resulting formulas for the radial weights $f_J$ and ${\cal F}_J$ are given by Eqs.~(\ref{WeightsThinShell}) and (\ref{SurfaceFluxWeightsThinShell}).

\begin{figure}
   \centering
      \includegraphics[width=5.5cm]{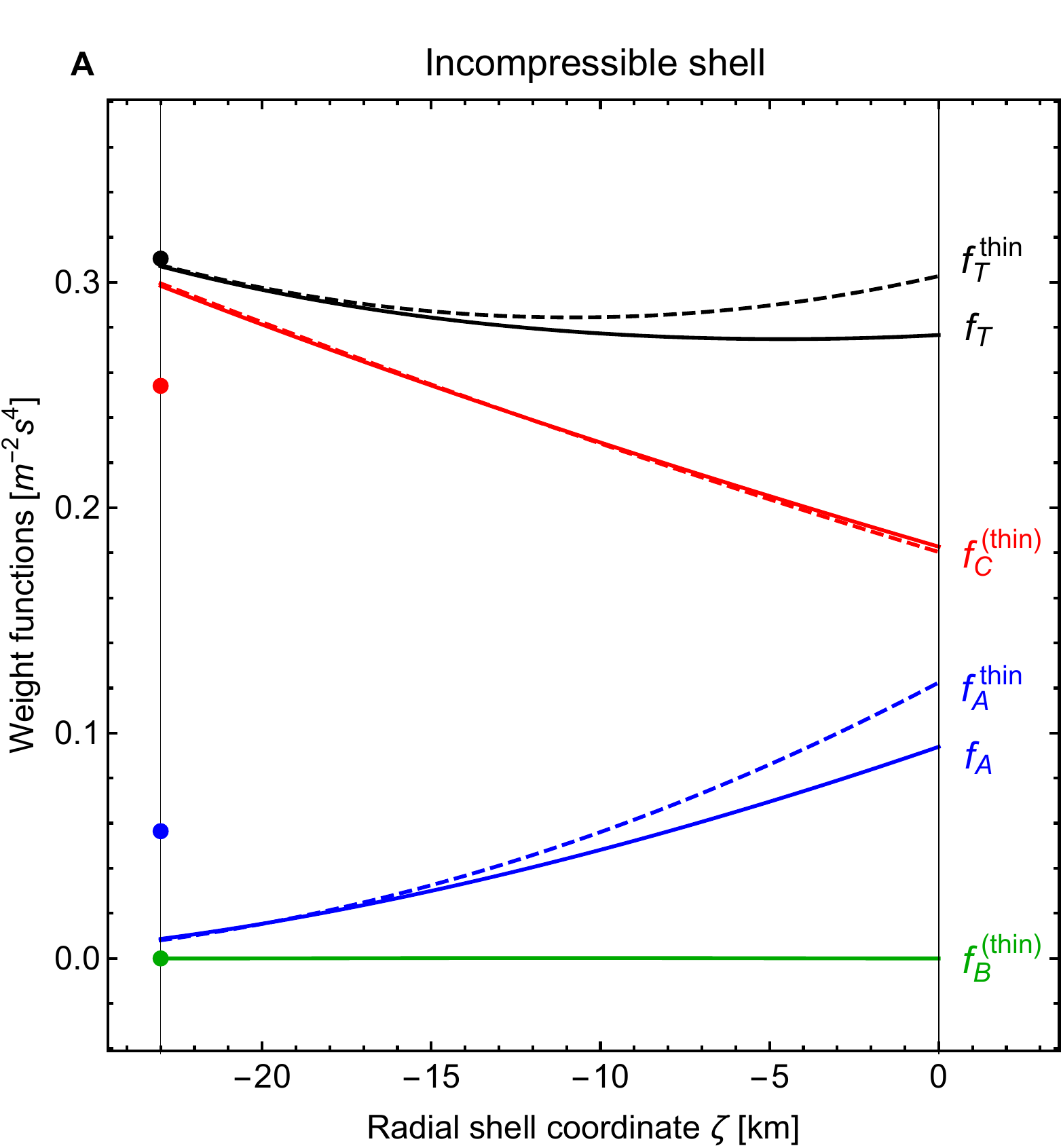}
   \hspace{5mm}
   \includegraphics[width=5.5cm]{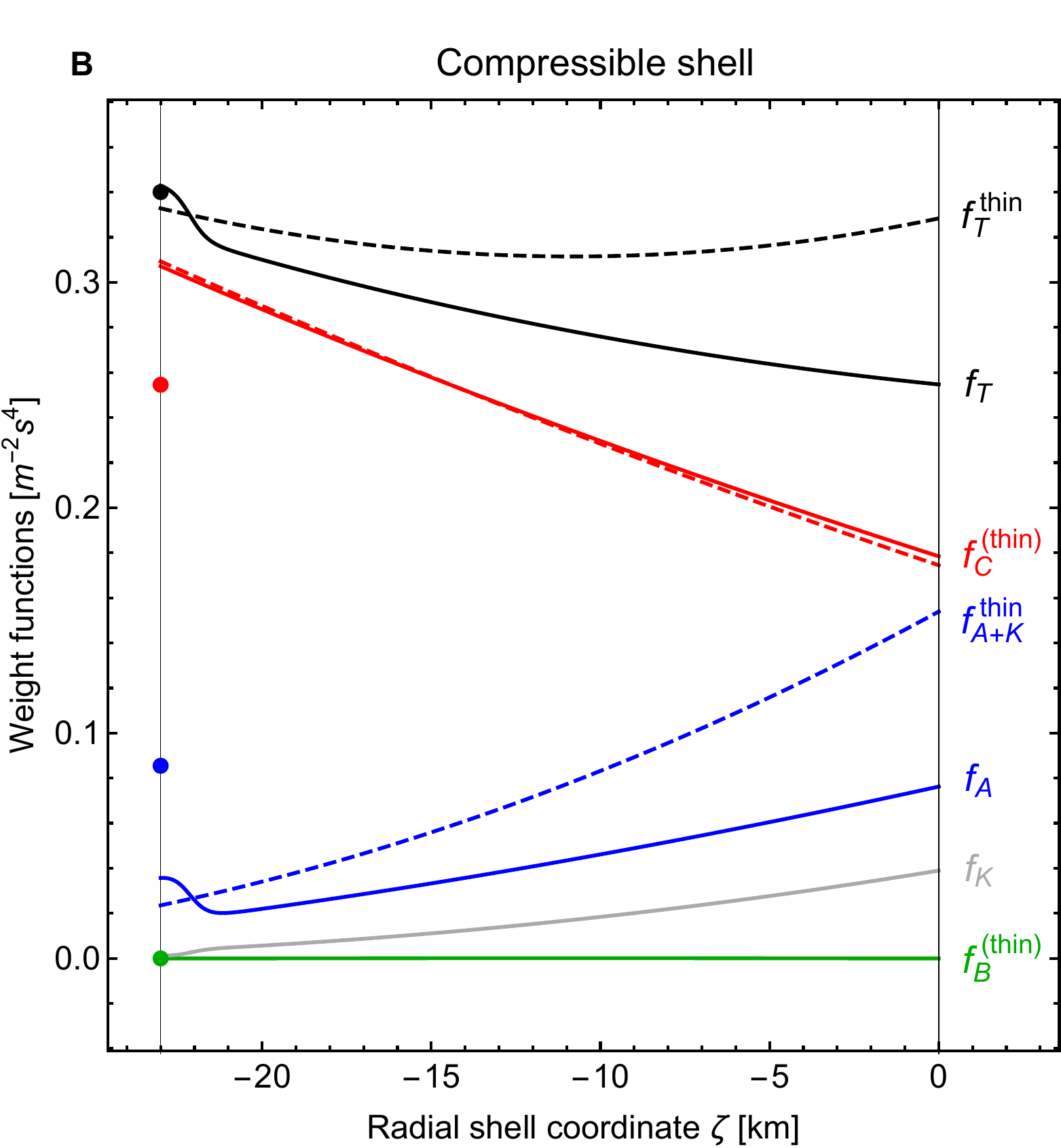}
   \caption[Dissipation rate in a laterally uniform conductive shell]
   {Dissipation rate in a laterally uniform conductive shell ($d=23\rm\,km$, $\eta_{\rm m}=10^{13}\rm\,Pa.s$): radial weight functions for (A) an incompressible shell, and (B) a compressible shell.
   Solid curves show the exact results for a thick shell without bulk dissipation.
   Dashed curves show the thin shell results without Poisson dissipation (Eq.~(\ref{WeightsThinShell})).
   Big dots indicate the values of the radial weights in the membrane limit.
   See Section~\ref{DissipationRateUniform}.
}
   \label{FigWeights}
\end{figure}

Regarding the dissipation rate, Fig.~\ref{FigWeights} shows that the thin shell approximation is very accurate for the dominant weight function $f_C$, with or without compressibility.
It is less accurate for the coefficient of Pattern~A: the difference is in part due to the error on the weight function $f_A$, and in part to bulk dissipation (if the shell is compressible).
However, the error is very small near the bottom of the shell where dissipation is highest.

Regarding the surface flux pattern, Table~\ref{TableRatios} gives the contribution of the dominant Pattern~C to the average surface flux for different types of solutions (thick shell/thin shell/membrane), and for different dissipation conditions.
The thin shell approach predicts much better the dissipation pattern than the membrane approximation: the error is less than 1\% if $\eta_{\rm m}=10^{13}\rm\,Pa.s$ (or less than 3\% if $\eta_{\rm m}=10^{14}\rm\,Pa.s$).
For thinner shells, the contribution of Pattern~C decreases, reaching a lower bound of $75\%$ to $82\%$ in the membrane limit (more details are given in Appendix~\ref{PatternsUniformThinShell}).
Fig.~\ref{FigPatternThinShell} shows the surface flux pattern predicted in the membrane approximation and with the thin shell approach (the latter is indistinguishable from the thick shell solution if $\eta_{\rm m}=10^{13}\rm\,Pa.s$).
All in all, the surface flux pattern remains similar as shell thickness and bottom viscosity vary, but the dissipation contrast between poles and tidal axis is much stronger if the shell is rather thick.
The min/max dissipation contrast in a thin shell is much higher than in a solid body (or in the core, see Section~\ref{CoreDissipationU}), for which the surface flux is about twice as large at the poles than along the tidal axis.

\begin{table}[h]\centering
\ra{1.3}
\small
\caption[Pattern of surface shell flux]{
Surface shell flux of a laterally uniform shell: percentage contribution of Pattern~C to the average surface flux. Pattern~A contributes the remainder.
The shell is $23\rm\,km$ thick and the bottom viscosity is either $10^{13}$ or $10^{14}\rm\,Pa.s$ (results separated by $|$).
See Section~\ref{DissipationRateUniform}.
}
\begin{tabular}{@{}llll@{}}
\hline
Dissipation constraint & Thick shell & Thin shell & Membrane$^a$ \\
\hline
No Poisson dissipation & 92.2 $|$ 93.1 & 91.7 $|$ 91.8  &  75.0 $|$ 75.0 \\
No bulk dissipation  & 92.0 $|$ 94.0 & - &  76.0 $|$ 80.9 \\
Incompressible  & 96.3 $|$ 96.5 & 96.5 $|$ 96.6 & 81.8 $|$ 81.8 \\
\hline
\multicolumn{4}{l}{\scriptsize ${}^a$ Eq.~(\ref{FCratio}) with $\nu=0.33$, $\bar\kappa=0.42\,|\,0.06$, and $\nu=0.5$, respectively.}
\end{tabular}
\label{TableRatios}
\end{table}%

\begin{figure}
\center
      \includegraphics[width=6cm]{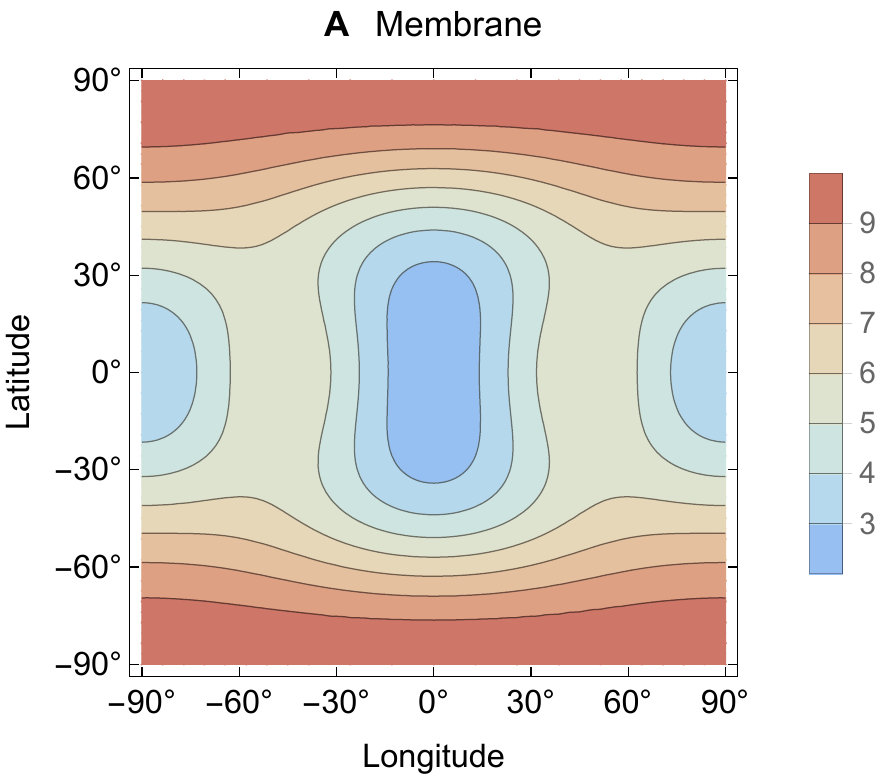}
      \hspace{1mm}
     \includegraphics[width=6cm]{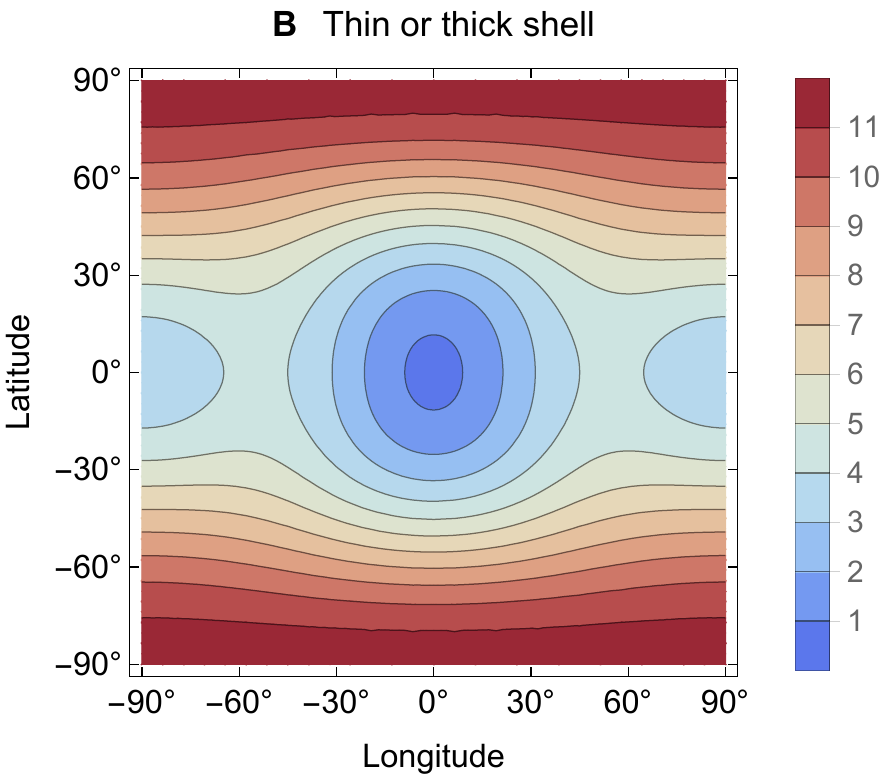}
   \caption[Surface flux pattern in a laterally uniform conductive shell]
   {Surface flux pattern in a laterally uniform conductive shell due to eccentricity tides plus $0.12^\circ$ forced libration:
   (A) compressible membrane (no Poisson dissipation);
   (B) compressible thin shell (the pattern given by the exact thick shell solution is indistinguishable in this case).
   The patterns represent $(1-\bar {\cal F}_C)\Psi_A+\bar {\cal F}_C\Psi_C$, where $\bar {\cal F}_C$ is the percentage contribution of Pattern~C given in Table~\ref{TableRatios}.
   The amplitude has been divided by $e^2$.
   The shell is $23\rm\,km$ thick and the bottom viscosity is $10^{13}\rm\,Pa.s$.
   See Section~\ref{DissipationRateUniform}.
}
   \label{FigPatternThinShell}
\end{figure}

\subsection{Total power in core and shell}
\label{TotalPowerCoreShell}

In the macro approach, the total dissipated power is given by a surface integral, the integrand of which is the product of the primary and secondary tidal potentials (Eq.~(\ref{Edot1})).
Substituting $\Gamma_n=(k_n+1)U_n^T$ (Table~\ref{TableUniform}) in the Zschau-Platzman formula, I get back the well-known formula for the total power dissipated in a body with a spherically symmetric structure:
\begin{equation}
\dot E_{tot} = - \frac{2n+1}{2} \, \frac{\omega R}{G} \, {\rm Im}( k_n ) \, \langle | U_n^T |^2 \rangle \, .
\label{EdotTotalMacro}
\end{equation}
The effect of the internal structure is hidden in the imaginary part of the gravitational Love number, while the external forcing appears as the averaged squared tidal potential \citep{zschau1978,platzman1984}.
For degree-2 eccentricity tides combined with the forced libration, the latter is given by (see Table~\ref{TableRadialAngular})
\begin{eqnarray}
\langle | U_2^T |^2 \rangle
&=& (\omega{}R)^4 \, \Psi_0
\nonumber \\
&=& (\omega{}R)^4 \, e^2 \, \frac{3}{5} \left( 3+4\left(1+\frac{\gamma_0}{2e}\right)^2 \right) .
\label{U2sq}
\end{eqnarray}
The amplification due to the forced libration is the same as for a homogeneous body \citep{wisdom2004}, because of the (arbitrary) assumption that the core and shell librate with the same angle.

\begin{figure}
   \centering
      \includegraphics[width=5.5cm]{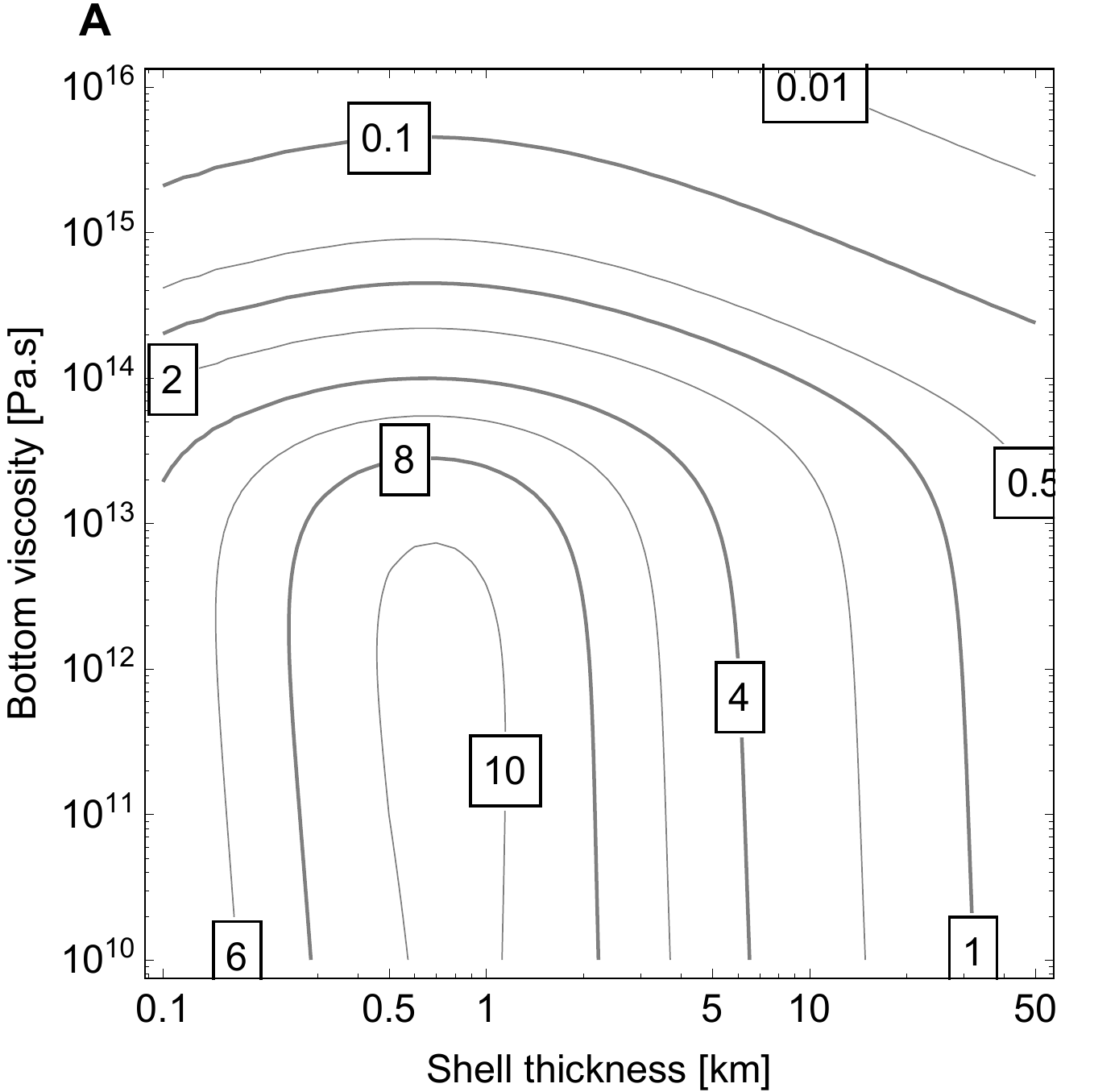}
   \hspace{5mm}
   \includegraphics[width=5.5cm]{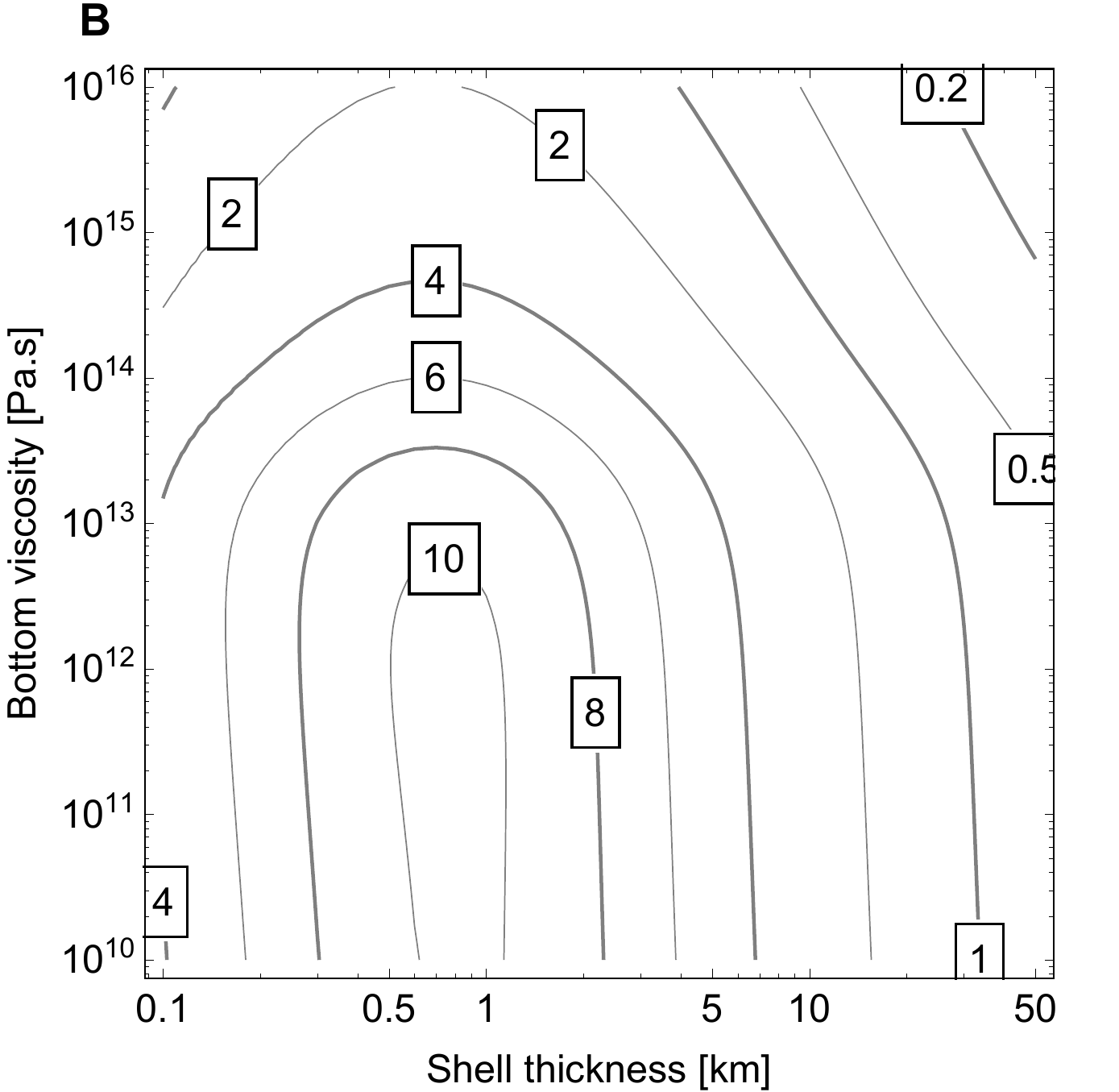}
   \caption[Core dissipation under a laterally uniform shell]{
   Total power (in GW) dissipated in a laterally uniform conductive thin shell as a function of shell thickness and bottom viscosity.
   (A) Maxwell rheology.
   (B) Andrade rheology.
   The tidal potential includes eccentricity tides and the $0.12^\circ$ forced libration. The core is elastic.
   Dissipation increases as the bottom viscosity decreases down to $10^{13}\rm\,Pa.s$, but not much below that threshold if the shell is thicker than a few km.
   See Section~\ref{TotalPowerCoreShell}.}
   \label{FigShellPower}
\end{figure}

If the core is elastic, the total power $\dot E_{tot}$ is equal to the shell power $\dot E_{shell}$.
Eq.~(\ref{U2sq}) shows that the forced libration ($\gamma_0=0.12^\circ$) increases the total power dissipated in the shell by 28\%.
Fig.~\ref{FigShellPower} shows the total power dissipated in a laterally uniform shell as a function of the shell thickness and the viscosity at the bottom of the conductive shell.
Rheology is either Maxwell or Andrade, the latter with parameters equal to $\alpha_A=0.25$ and $\beta_A\approx\mu_{\rm e}^{\alpha_A-1}\eta^{-\alpha_A}$ \citep{castillo2011}.
If the shell is thicker than a few km, dissipation increases as the bottom viscosity decreases from $10^{16}$ to $10^{13}\rm\,Pas.s$.
Lowering the bottom viscosity below that value does not increase dissipation much further.
Andrade rheology leads to more dissipation than Maxwell if the bottom viscosity is larger than $10^{13}\rm\,Pa.s$, but makes little difference below that threshold.

\begin{figure}
   \centering
       \includegraphics[width=6cm]{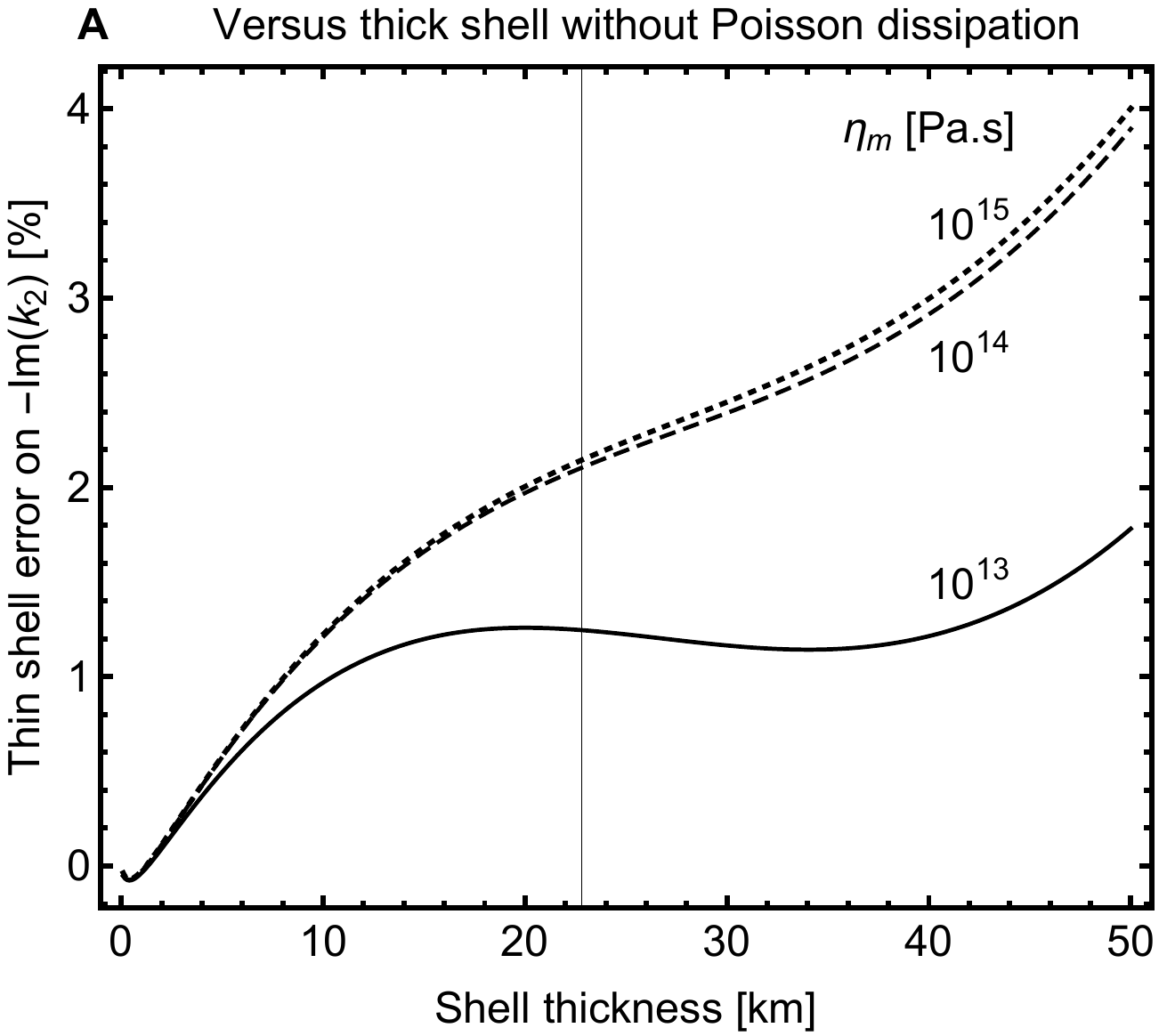}
       \hspace{5mm}
      \includegraphics[width=6cm]{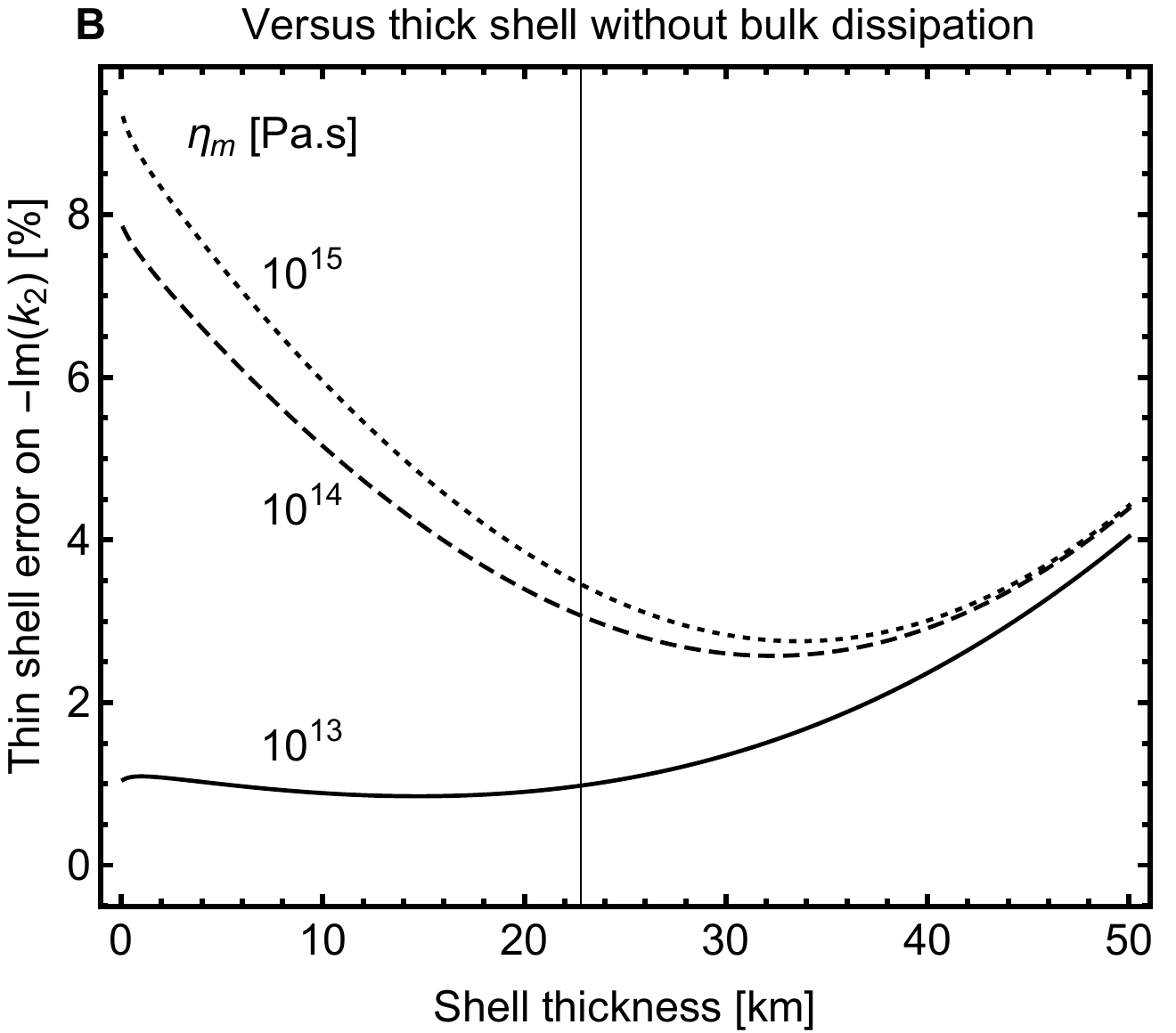}
   \caption[Total power dissipated in a laterally uniform shell]
   {Error on the total power dissipated in a laterally uniform thin shell.
   In Panel~A, the point of comparison is a thick shell with bulk dissipation; in Panel~B, it is a thick shell without bulk dissipation (see Fig.~\ref{Figk2Thick}). 
   The shell is conductive with Maxwell rheology: results are shown for three values of the bottom viscosity $\eta_{\rm m}$.
   The vertical line corresponds to $d=23\rm\,km$.
   The curve for $\eta_{\rm m}=10^{15}\rm\,Pa.s$ provides an approximate upper bound: the error does not get bigger than this.
   See Section~\ref{TotalPowerCoreShell}.}
   \label{Figk2Thin}
\end{figure}

In Fig.~\ref{Figk2Thin}, I compare the thin shell approximation of ${\rm Im}(k_2)$ with the exact thick shell results for different values of the shell thickness and bottom viscosity.
Fig.~\ref{Figk2Thin}A confirms that the thin shell result is a well-behaved approximation of the exact result: the error with respect to the thick shell power with no Poisson dissipation is below 4\% if $d<50\rm\,km$ (or below 2\% if $d<20\rm\,km$) and tends to zero with decreasing shell thickness.
Fig.~\ref{Figk2Thin}B shows that the error with respect to the thick shell power with no bulk dissipation is below 4\% if $20<d<50\rm\,km$.
For thinner shells, the error increases again if the bottom viscosity is higher than $10^{13}\rm\,Pa.s$, but the error is approximately bounded by the curve for $\eta_{\rm m}=10^{15}\rm\,Pa.s$.
In the membrane limit ($d\rightarrow0\rm\,km$) the error ranges from 0 to 9\%, depending on the bottom viscosity (the membrane error is exactly $(1-2\bar\kappa)/(11+2\bar\kappa)$ where $\bar\kappa$ is the effective bulk dissipation varying between 0 and 0.5; see Eq.~(\ref{FCratio})).

Suppose now that the core is viscoelastic.
The partition of the total power (Eqs.~(\ref{EdotCoreMacro})-(\ref{EdotShellMacro})) becomes, after substituting the expressions for $(w_n,q_n,U_n^\circ)$ from Table~\ref{TableUniform},
\begin{eqnarray}
\dot E_{core} &=& - \frac{2n+1}{2} \, \frac{\omega R}{G} \, \left| \frac{k_n+1}{k_n^\circ+1} \right|^2 {\rm Im} (k_n^\circ) \, \langle | U_n^T |^2 \rangle \, ,
\label{EdotCoreMacroU} \\
\dot{E}_{shell} &=& \frac{2n+1}{2} \, \frac{\omega R}{G} \, \xi_n \, |h_n|^2 \, {\rm Im}(\Lambda_n) \, \langle | U_n^T |^2 \rangle \, .
\label{EdotShellMacroU}
\end{eqnarray}
In the membrane limit ($\Lambda_n\rightarrow\Lambda_n^M$), these formulas are identical to Eqs.~(98)-(102) of \citet{beuthe2014}.
All quantities in the RHS can be numerically evaluated.
In particular, the tidal Love numbers $(k_n^\circ, k_n, h_n)$ and the thin shell spring constant $\Lambda_n$ can be computed with the analytical formulas of Table~\ref{TableUniform}, which remain valid if the core is viscoelastic (see Section~\ref{CoreDissipationU}).

The macro formula for $\dot{E}_{shell}$ (Eq.~(\ref{EdotShellMacroU})) is identical to the volume-integrated dissipation rate (Eq.~(\ref{EshellLambda})): this is again an example of the micro-macro equivalence in tidal dissipation \citep{beuthe2013,beuthe2014}.
The substitutions $\Lambda_n\rightarrow\Lambda_n^M$ and $\Lambda_n\rightarrow\Lambda_n^B$ in Eq.~(\ref{EdotShellMacroU}) yield to first order the sum of the membrane and mixed contributions, and the bending contribution, respectively (Eq.~(\ref{EshellLambda}) again).
The bending contribution is about 5\% of the total power for a laterally uniform shell with $d=23\rm\,km$ and $\eta_{\rm m}=10^{13}\rm\,Pa.s$ (it increases if the shell is laterally non-uniform, see Section~\ref{Results}).
Finally, energy conservation ($\dot E_{core}+\dot{E}_{shell}=\dot E_{tot}$) is guaranteed by the identity (\ref{IdentityMicroMacro}).

\subsection{Dissipation in the core}
\label{CoreDissipationU}

In this section, I will first show that the thin shell approximation causes a very small error on core dissipation, before studying the conditions required for high core dissipation and examining the core dissipation pattern.

First, what is the error on core dissipation due to the thin shell approach? 
The formula for core dissipation (Eq.~(\ref{EdotCoreMacroU})) is actually valid beyond the thin shell approximation if the shell density is homogeneous and there is no density contrast at the shell-ocean boundary (the ocean density can increase with depth).
In that case, the full solution in the core can be obtained from the fluid-crust solution in the core by \textit{gravity scaling} (see Appendix~F of \citet{beuthe2014}):
\begin{equation}
y_{i n}^{T}(r) = \frac{k_n+1}{k_n^\circ+1} \, y_{i n}^{\circ T}(r) \, .
\label{solyinU}
\end{equation}
As dissipation depends on the product of stress and strain, this procedure accounts for the factor $|(k_n+1)/(k_n^\circ+1)|^2$ in Eq.~(\ref{EdotCoreMacroU}). 
For thin shells, gravity scaling is equivalent to the effective tidal potential trick, as is seen by substituting $U_n^\circ$ from Table~\ref{TableUniform} into the core solution given by Eq.~(\ref{Upsilon2}):
\begin{equation}
\Upsilon_i(r,\theta,\varphi)
= \sum_n y_{i n}^{\circ T}(r) \, U_n^\circ
= \sum_n y_{i n}^{T}(r) \, U_n^T \, .
\end{equation}
Using the property that Eq.~(\ref{EdotCoreMacroU}) is applicable to both thin and thick shells (with the restrictions on the shell density mentioned above), I can estimate the thin shell error on $\dot E_{core}$ from the thin shell error on $|k_2+1|^2$ which is less than 0.1\% ($k_2\sim10^{-2}$ and the thin shell error on $k_2$ is a few percent).
Core dissipation thus mainly depends on internal structure through the factor ${\rm Im} (k_n^\circ)/|k_n^\circ+1|^2$.

Tidal heating reaches several tens of GW if the unconsolidated core is modelled as a very soft viscoelastic material \citep{choblet2017} (\citet{roberts2015} studied before the enhancement of tidal heating in a fluffy core, but without global ocean and with the shear modulus of the core larger than the shear modulus of ice).
The complex shear modulus of the homogeneous core is parameterized in terms of the elastic shear modulus $\mu_{c \rm e}$ and a nondimensional parameter $\delta$ (zero if the core is elastic, otherwise positive):
\begin{eqnarray}
\mu_c &=& \frac{\mu_{c \rm e}}{1-i\delta}
\nonumber \\
&=&  |\mu_{c}| \, \frac{1+i \delta}{\sqrt{1+\delta^2}} \, .
\label{CoreRheology}
\end{eqnarray}
For Maxwell rheology, $\delta$ is related to the core viscosity $\eta_c$ by $\delta=\mu_{c\rm e}/(\omega\eta_c)$, but the above expression is generally valid for any linear rheological model.
\citet{choblet2017} parameterize core rheology with the effective shear modulus $\mu_{\rm eff}=|\mu_{c}|$ and the dissipation function $Q_\mu^{-1}=\delta/\sqrt{1+\delta^2}$.
The latter ranges from 0.2 to 0.8, corresponding to $\delta$ values between 0.5 and 2.

The total power dissipated in the core is maximum if ${\rm Im} (k_n^\circ)/|k_n^\circ+1|^2$ is maximum (see above).
At constant $\delta$, this occurs for a homogeneous core if
\begin{equation}
|\mu_{c}| \cong \frac{A_2}{5} \, \frac{R^4}{R_c^4} \, \rho_b g R \, \cong \, 6.6\times10^6\rm\,Pa.s \, ,
\end{equation}
where $A_2\cong0.24$ (see Appendix~\ref{DeformationCore}).
Fig.~\ref{FigCorePower}A shows the total power dissipated in the core as a function of $|\mu_{c}|$ and $\delta$.
The core is homogeneous and incompressible, the ocean is homogeneous and the shell is conductive.
The assumption of no differential rotation (Section~\ref{TidalLoad}) results in overestimating core dissipation by about 25\%, but it does not matter since the core rheology is unknown.
In Section~\ref{ThermalEquilibrium}, I set $\delta=1$ and adjust $|\mu_c|$ so that the core power is equal to the difference between the conductive power and the shell power.

\begin{figure}
   \centering
      \includegraphics[width=6cm]{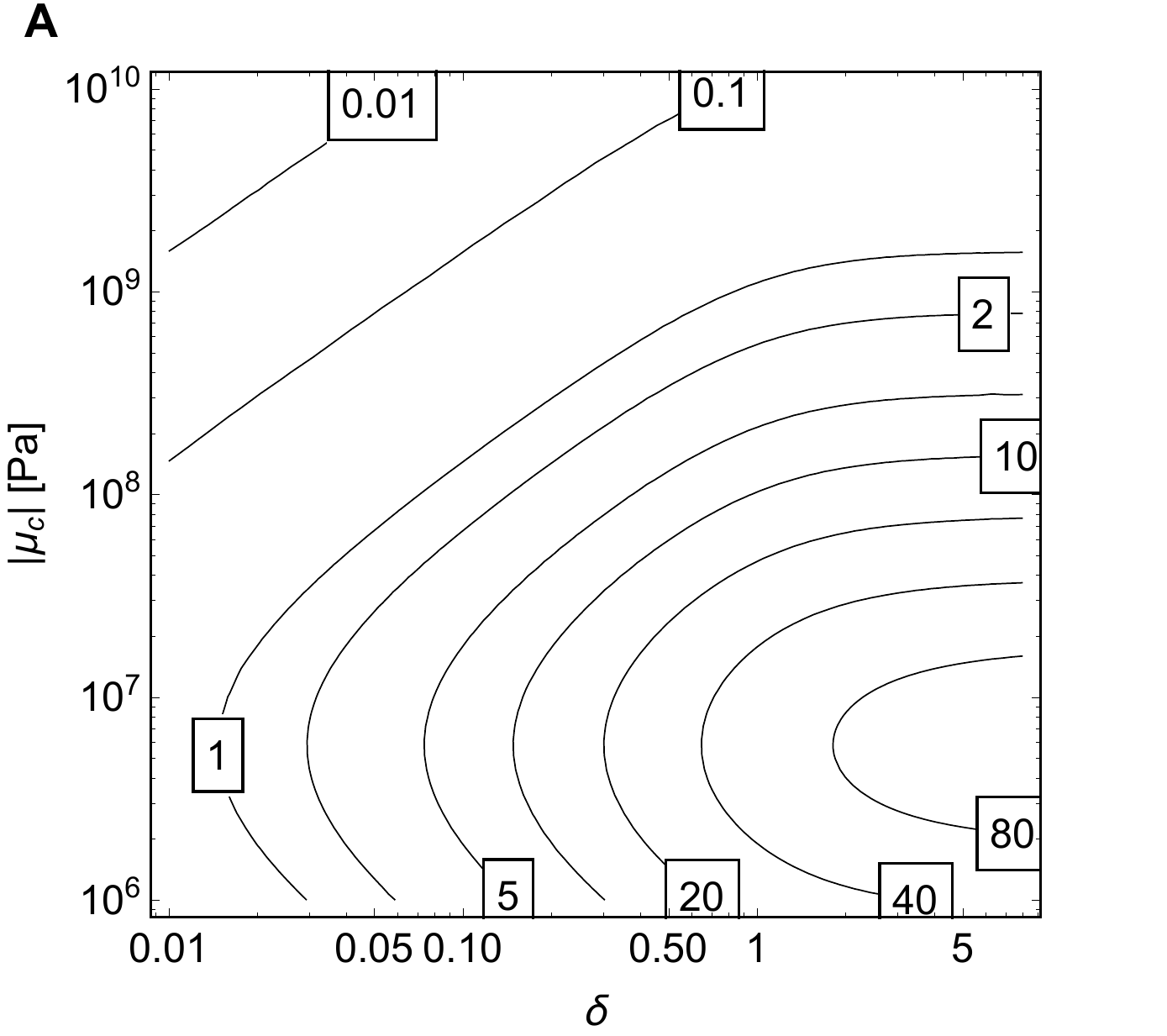}
      \includegraphics[width=6cm]{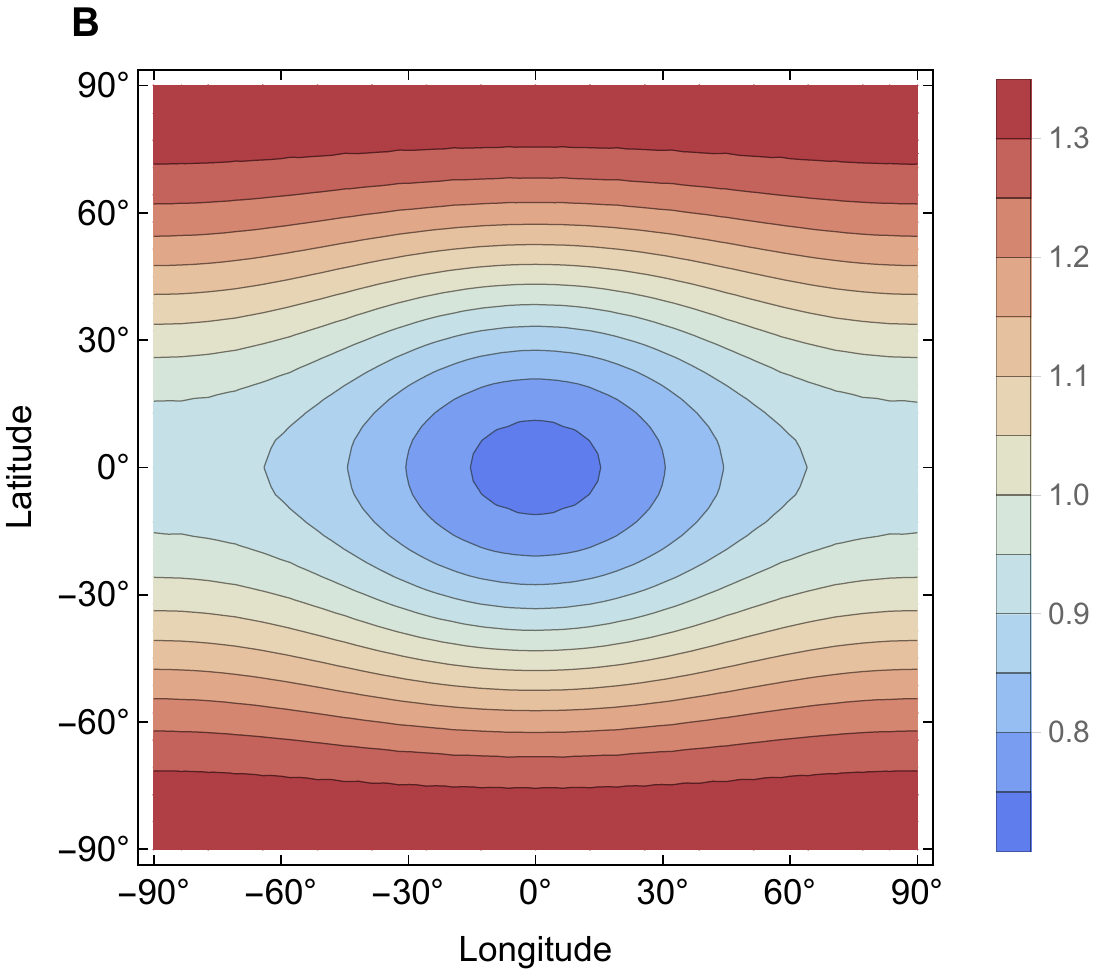}
   \caption[Core dissipation under a laterally uniform shell]
   {Core dissipation under a laterally uniform conductive shell ($d=23\rm\,km$): (A) total power dissipated in the core (in GW) as a function of the rheological parameters ($|\mu_c|$, $\delta$);
   (B) heat flux pattern at the core surface, normalized by the average flux.
   The core is submitted to the same forced libration as the shell.
   See Section~\ref{CoreDissipationU}.}
   \label{FigCorePower}
\end{figure}

Although the spatial pattern of dissipation could be obtained from the core dissipation rate under a non-uniform shell (Eq.~(\ref{PowerStrainInv})), it is simpler to use the radial-angular factorization method.
After substituting Eq.~(\ref{solyinU}) and Eq.~(\ref{yifun}) into the formulas of Table~\ref{TableRadialAngular}, I can write the radial weights within the core as
\begin{equation}
f_J =  \left| \frac{k_2+1}{k_2^\circ+1} \, \frac{h_2^{\circ c}}{g} \right|^2 \bar f_J \hspace{10mm} (J=A,B,C) \, ,
\label{CoreWeights}
\end{equation}
where $h_2^{\circ c}$ is the fluid-crust radial Love number at the core-ocean boundary (Eq.~(\ref{hn0core})).
The reduced radial weights $\bar f_J$ are functions of the reduced radius $\hat r=r/R_c$:
\begin{equation}
\left( \bar f_A, \bar f_B, \bar f_C \right) =  \frac{3}{25} \, \hat r^2 \left(  \left( 8 - 9 \hat r^2 \right)^2, \, 2 \left( 8 - 8 \hat  r^2 \right)^2, \, 2 \left( 8 - 5\hat  r^2 \right)^2 \right) .
\label{WeightsHomogeneous}
\end{equation}
Apart from a global scaling factor, the radial weights are identical to those for a homogeneous body of radius $R_C$ (Eq.~(55) of \citet{beuthe2013}).
Therefore, the dissipation pattern in a homogeneous core is identical to the one within a homogeneous body.
If heat is transported radially, the flux patterns $(A,B,C)$ at the core surface are weighted by
\begin{equation}
\int_0^1 \left( \bar f_A , \bar f_B , \bar f_C \right) d\hat r = \left(0.13, 0.31,0.56 \right) \frac{19}{5} \, .
\label{IntegratedWeights}
\end{equation}
Patterns A, B, and C thus contribute respectively 13, 31, and 56 \% of the average flux.

The core dissipation flux (Fig.~\ref{FigCorePower}B) has nearly no degree-4 harmonic component.
The inclusion of the forced libration does not change much the pattern: the flux enhancement due to libration varies between 23\% (along the leading-trailing axis) and 30\% (at the poles and along the tidal axis). 
Similarly to shell dissipation, core dissipation is higher at the poles than along the tidal axis.
The maximum dissipation contrast, however, is only about a factor of two, whereas it can be larger by an order of magnitude for a conductive shell.
Thus, core dissipation hardly explains the spatial variations of the observed surface flux.

\section{Thermal equilibrium in a conductive shell}
\label{ThermalEquilibrium}

In this section, I study the conditions under which Enceladus's non-uniform shell is in thermal equilibrium between tidal heating and conductive cooling.
I assume here that thermal equilibrium implies a shell in a steady state, but this is not necessarily true because shell thickness variations are progressively destroyed by viscous relaxation at the shell-ocean boundary.
This mechanism must be dynamically compensated by ocean freezing or ice melting at the shell-ocean boundary \citep{cadek2019}.

\subsection{Coupling dissipation to heat transfer}
\label{CouplingDissipation}

\citet{ojakangas1989a} were the first to compute lateral variations of shell thickness by balancing heat production (due to tidal dissipation within the shell and heat flow from the core) against conductive heat transfer.
Their model, however, does not take into account the lateral variations of shell thickness and rheology when computing tidal dissipation.
Conversely, one should in principle include the effect of tidal dissipation when computing the local temperature profile which determines the rheology of ice.
Therefore, tidal dissipation and heat transfer should be solved as a coupled system.

The conductive equilibrium solution is found by iteration.
The viscoelastic shell parameters $(\alpha,D,\chi)$ are initially evaluated for the no-dissipation temperature profile (Eq.~(\ref{Tprofile})).
One iteration consists of the following three steps:
\begin{enumerate}
\item
solving the tidal thin shell equations for the stress function $F$ and the radial displacement $w$ (as in Paper~I).
\item
computing the shell dissipation rate $P_{shell}$ (Eqs.~(\ref{PowerThinShell})-(\ref{PowerThinShellComp})) and the shell dissipation flux ${\cal F}_{shell}$ (Eqs.~(\ref{SurfaceFluxThinShell})-(\ref{SurfaceFluxThinShellComp})).
\item
solving numerically the heat equation with as a source term.
The viscoelastic shell parameters corresponding to this new temperature profile are simultaneously evaluated.
\end{enumerate}
The procedure is reiterated until the value of the dissipated power stabilizes, which normally happens after a few iterations.
The relative increase in shell power between the no-dissipation solution and the iterated solution will be called \textit{rheology feedback}.
More iterations are needed if the rheology feedback is large.

\subsection{Shell structure and core rheology}
\label{ShellStructure}

\begin{figure}
   \centering
      \includegraphics[width=6cm]{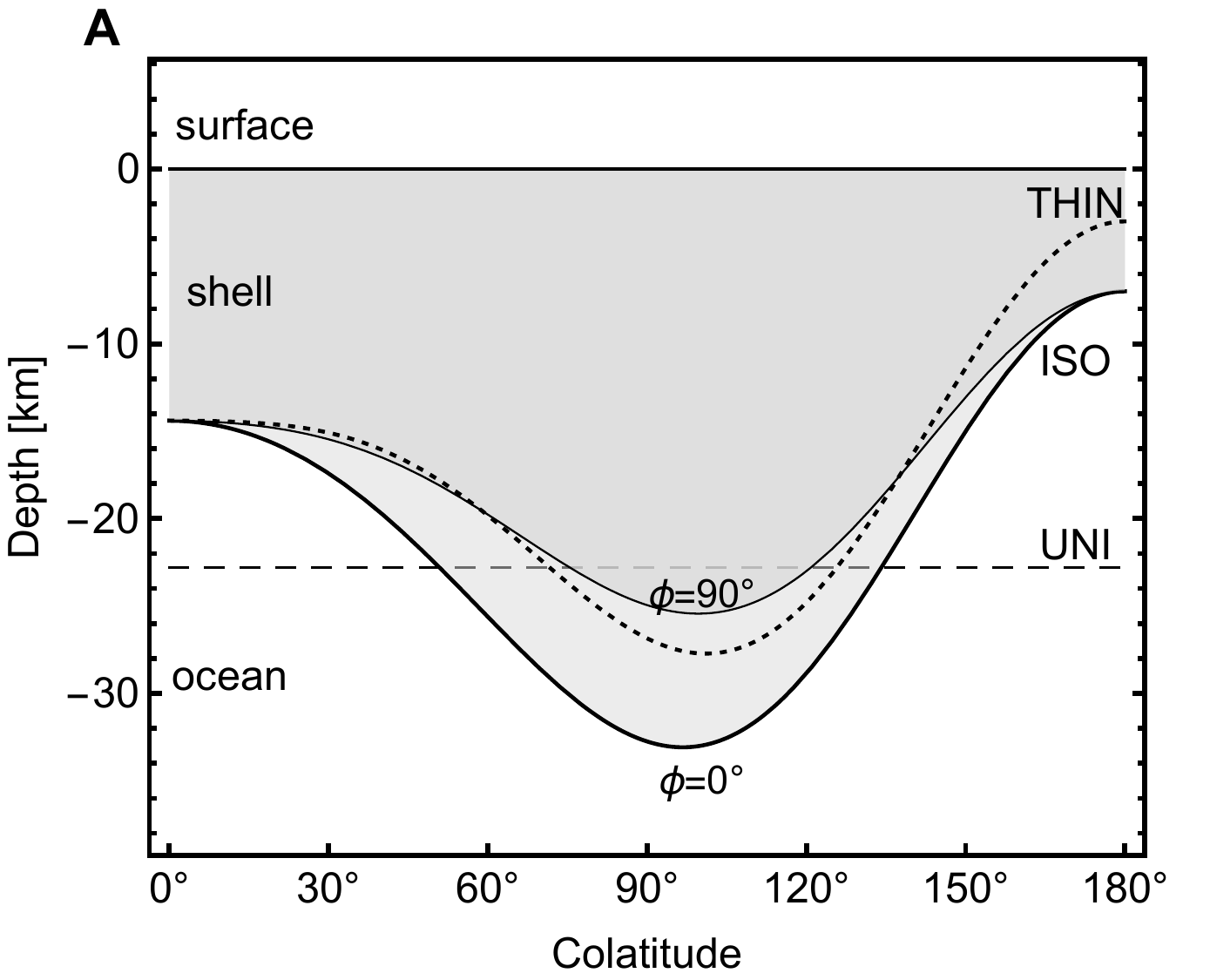}
   \hspace{5mm}
    \includegraphics[width=6cm]{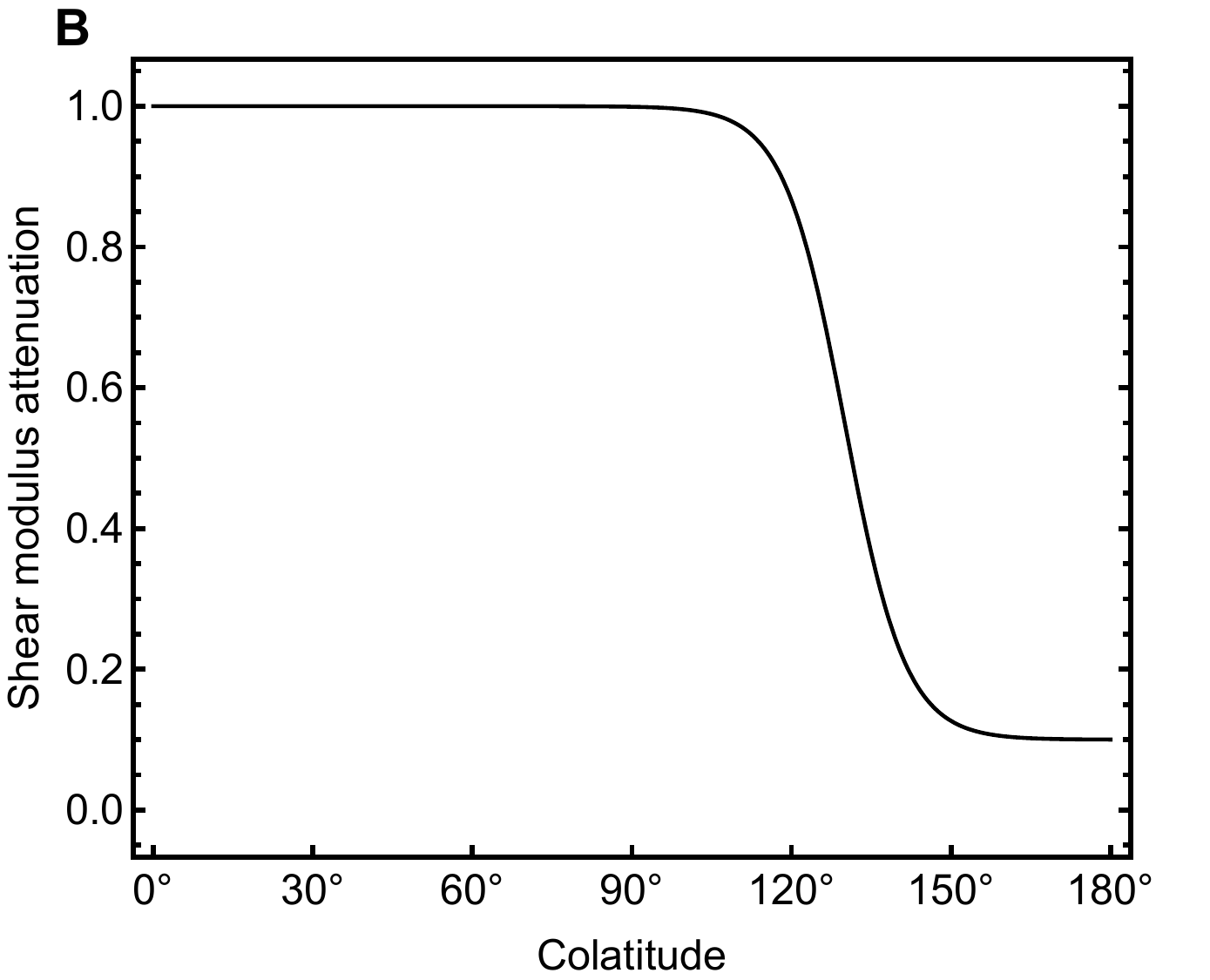}
   \caption[Shell structure]
   {Latitudinal shell structure: (A)
    isostatic profile ISO (thick and thin solid curves), ad hoc profile THIN with thinner crust at south pole (dotted curve), and uniform thickness profile UNI (dashed line);
  (B) reduction of elastic shear modulus in model THIN (Eq.~(\ref{reduction})).
   See Section~\ref{ShellStructure}.
   }
   \label{FigShellStructure}
\end{figure}

On the basis of gravity, topography, and libration data, Enceladus is thought to be made of a large silicate core, surrounded by a deep ocean and a thin icy shell.
Gravity-topography data combined with the hypothesis of isostasy result in the following model of shell thickness variations (see discussion in Section 5.1 of Paper~I):
\begin{equation}
d = d_{00} + d_{20} \, P_{20}(\cos\theta) + d_{22} \, P_{22}(\cos\theta) \cos2\varphi + d_{30} \, P_{30}(\cos\theta) \, ,
\label{ThicknessVarGRL}
\end{equation}
where $(d_{00},d_{20},d_{22},d_{30})=(22.8,-12.1,1.3,3.7)\rm\,km$ with $1\sigma$ errors of $(4,2.4,0.3,0.7)\rm\,km$ (uncertainties are ignored below).
$P_{nm}$ are the unnormalized associated Legendre functions.
Contrary to Paper~I, our models include non-zonal variations of shell thickness.
The resulting shell thickness is $14.4$ and $7\rm\,km$ at the north and south poles, respectively, and varies between $24.95$ and $32.75\rm\,km$ along the equator (the thickest shell is along the tidal axis).
Such models are denoted `ISO' (solid curves in Fig.~\ref{FigShellStructure}A) whereas models with uniform thickness (equal to $22.8\rm\,km$) are denoted `UNI' (dashed curve in Fig.~\ref{FigShellStructure}A).
When studying asymmetric core dissipation, I also consider the model `THIN' in which the shell is very thin at the south pole.
It is parameterized by Eq.~(\ref{ThicknessVarGRL}) in which $(d_{00},d_{20},d_{22},d_{30})=(20.8,-12.1,0,5.7)\rm\,km$ (dotted curve in Fig.~\ref{FigShellStructure}A).
The resulting shell thickness is $14.4$, $26.85$, and $3\rm\,km$ at the north pole, equator, and south pole, respectively.

The icy shell responds to deformations as a linear viscoelastic material with Maxwell rheology.
Elastic and viscoelastic parameters are given in Table~\ref{TableParam}.
The viscosity depends on temperature through an Arrhenius relation (Table~\ref{TableVisco}).
In this paper, I assume that the bottom viscosity is $10^{13}\rm\,Pa.s$ in order to maximize shell dissipation (see Section~\ref{TotalPowerCoreShell}).
This value is at the lower end of the range usually considered for melting ice \citep{tobie2003,barr2009} and could lead to fast viscous ice flow destroying the topography at the bottom of the shell.
Balancing viscous flow against ice-water phase change, \citet{cadek2019} argue that the bottom viscosity should be larger than $3\times10^{14}\rm\,Pa.s$.

Newly published laboratory studies of the anelastic response of ice at tidal frequencies suggest that dissipation could be an order of magnitude higher than predicted by Andrade rheology.
\citep{mccarthy2016}.
This phenomenon is simulated here by increasing the nominal dissipation rate by a factor of 10.
These models are denoted by the letter `H' (for `High') whereas the models with nominal dissipation are denoted by the letter `L' (for `Low').

In the model `THIN', the shell is further weakened at the south pole (because of faulting) by multiplying the elastic shear modulus by
\begin{equation}
\mbox{reduction factor} =
\frac{a+1}{2} + \frac{a-1}{2} \, \tanh \left( \frac{\pi}{180} \, b \left( \theta-\theta_0 \right) \right) ,
\label{reduction}
\end{equation}
where $a=0.1$, $b=5$, and $\theta_0=130^\circ$ (see Fig.~\ref{FigShellStructure}B).

If the core is non-porous and elastic, the shear modulus of the core is set to $\mu_{\rm ce}=40\rm\,GPa$.
If the core is porous and viscoelastic, core rheology is parameterized by Eq.~(\ref{CoreRheology}) with $\delta=1$, and the elastic shear modulus is adjusted so that the core power and shell power sum to the conductive power.
The shear modulus must be about 1000 times smaller than its elastic value for a non-porous silicate core:
$\mu_{\rm ce}=34$,  $68.5$, and $32.4\rm\,MPa$ in models ISO-LC, ISO-HC, and THIN-LC, respectively
(or $|\mu_{\rm c}|=24.0$,  $48.4$, and $22.9\rm\,MPa$, respectively).

\subsection{Conductive model}
\label{ConductiveModel}

\begin{figure}
   \centering
      \includegraphics[width=6cm]{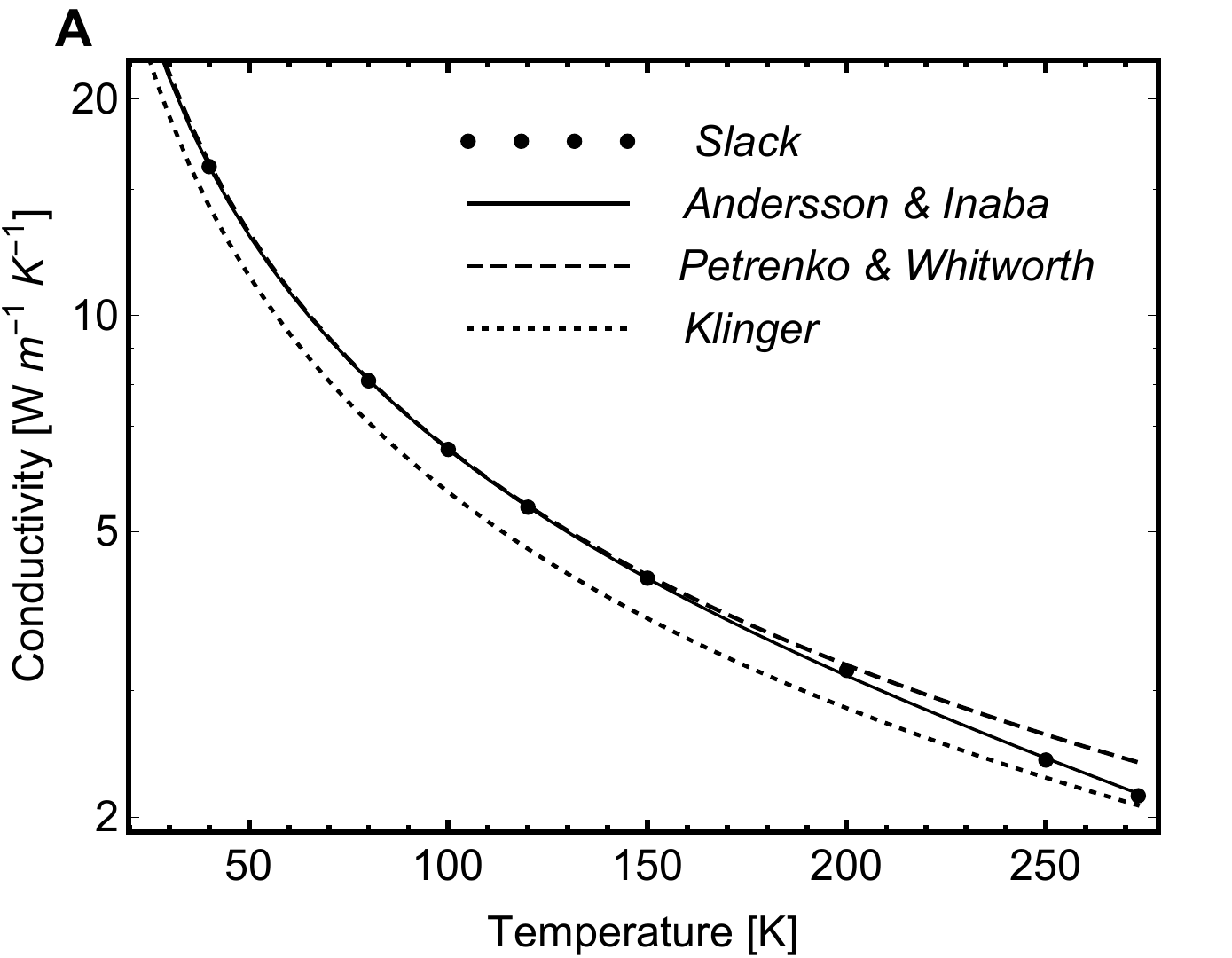}
   \hspace{5mm}
     \includegraphics[width=6cm]{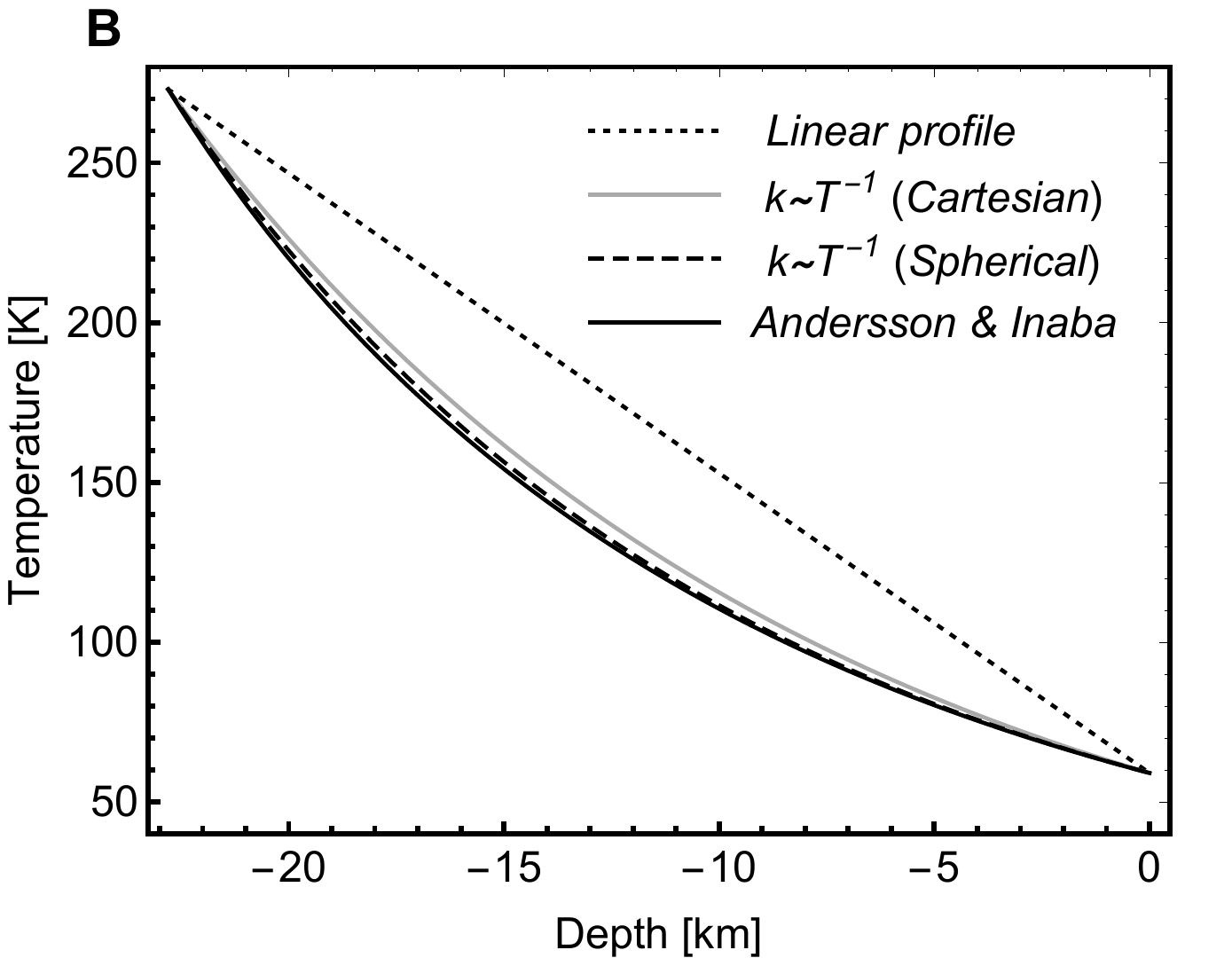}
   \caption[Conductive model]
   {Conductive model: (A) conductivity of ice as a function of temperature.
   The dots represent the `best estimate' of \citet{slack1980}.
   The solid, dashed, and dotted curves show three published fits (see Section~\ref{ConductiveModel}).
   (B) Temperature profile in absence of dissipation. The shell thickness and surface temperature are set to $59\rm\,K$ and $23\rm\,km$, respectively.
   The different profiles are discussed in Section~\ref{ConductiveModel}.
   }
   \label{FigConductivity}
\end{figure}

Given that the shell thickness is smaller than $40\rm\,km$, the shell is most likely in a conductive state \citep{barr2007,mitri2008}.
If the shell is radially in thermal equilibrium and the lateral heat transfer is negligible, the temperature $T(r,\theta,\varphi)$ satisfies the radial heat equation:
\begin{equation}
\frac{1}{r^2} \, \frac{d}{dr} \left( r^2 \, k_{ice} \, \frac{dT}{dr} \right) = - P_{shell}(r,\theta,\varphi) \, .
\label{HeatEq}
\end{equation}
The \textit{conductive flux} denotes the conductive heat flux at the surface of the shell:
\begin{equation}
{\cal F}_{\rm cond} =- k_{ice} \, \frac{dT}{dr} \bigg|_{r=R} \, .
\end{equation}
The conductivity of ice decreases as $1/T$ in the range $(40,175)\rm\,K$ but  falls below this line at high temperature \citep{slack1980,petrenko1999}.
In planetology, the conductivity of ice is often approximated by $k_{ice}=a/T$ with either $a=567{\rm\,W/m}$ \citep{klinger1980} or $a=651{\rm\,W/m}$ \citep{petrenko1999}.
The former relation underestimates the conductivity over the whole range but is only 3\% too low at the melting temperature, whereas the latter fits well the data up to $200\rm\,K$ but is 11\% too high at the melting temperature (see Fig.~\ref{FigConductivity}A).
When solving numerically the heat equation, I adopt the more accurate fit of \citet{andersson2005},
\begin{equation}
k_{ice} = 632/T + 0.38 - 0.00197 \, T \hspace{5mm} \mbox{(SI units)} \, ,
\label{condfit}
\end{equation}
with an estimated error of 5\% in the range $(40,273\rm\,K)$.
Note that the data reviewed by \citet{slack1980} pertain to single crystals of pure ice: the conductivity is certainly modified by polycrystalline anisotropy and salt contamination.
Another factor neglected here is the insulating effect of a $100\rm\,m$-thick snow cover which, if present, lowers the near-surface conductivity by 1 or 2 orders of magnitudes (Fig.~4 of \citet{travis2015}).

Boundary conditions are given by fixing the temperature at the top and bottom of the shell.
The surface  temperature $T_{\rm s}$ is mainly determined by radiative equilibrium with the annual solar insolation, with a small contribution due to internal heating (see Section~\ref{Temperature}).
The temperature at the bottom of the shell (of radius $R_o$) is equal to the melting temperature $T_{\rm m}=273\rm\,K$, because the temperature of the ocean is expected to be only slightly less than the freezing point of pure water \citep{glein2018}.

The heat equation in presence of a source (Eq.~(\ref{HeatEq})) is solved with the BVP solver for boundary value ordinary differential equations \citep{shampine2006}.
The starting guess is the analytical solution obtained by neglecting dissipation and approximating the conductivity with $k_{ice}=a/T$:
\begin{equation}
T(r) = T_{\rm m}^{\,\, \frac{R_o}{r} \, \frac{R-r}{d}} \, T_{\rm s}^{\,\, \frac{R}{r} \, \frac{r-R_o}{d}} \, .
\label{Tprofile}
\end{equation}
This profile predicts slightly lower temperatures than the Cartesian profile
(identical to Eq.~(\ref{Tprofile}) except for the factors $R_o/R$ and $R/r$ in the exponents; see Eq.~(56) of Paper~I).
The conductive flux associated with Eq.~(\ref{Tprofile}) reads
\begin{equation}
{\cal F}_{\rm cond} = \frac{a}{d} \, \frac{R_o}{R} \, \ln \left( \frac{T_{\rm m}}{T_{\rm s}} \right) ,
\label{ConductiveFlux}
\end{equation}
where the factor $R_o/R$ is the correction due to spherical geometry.

Fig.~\ref{FigConductivity}B shows the temperature profile in the shell if there is no dissipation.
The top curve (dotted straight line) is associated with constant conductivity in Cartesian geometry.
The two intermediate curves result from a conductivity inversely proportional to temperature (Eq.~(\ref{Tprofile}), either in Cartesian (solid gray) or in spherical geometry (dashed black).
The lowest curve (solid black) is the solution of the heat equation in spherical geometry for the best-fitting conductivity (Eq.~(\ref{condfit})).
The three curves obtained with a variable conductivity do not differ much between themselves, but decrease much more steeply than the linear profile at the bottom of the shell.
Thus, models with constant conductivity overestimate the thickness of the most dissipative layer and the resulting dissipation.

\subsection{Surface temperature}
\label{Temperature}

The surface temperature depends on solar insolation, albedo, and internal heating sources, all of which vary with latitude and possibly with longitude too.
Enceladus's albedo $A$ is high and varies between $0.74$ and $0.81$ north of $60^\circ\rm\,S$ \citep{howett2010}, while $A=0.80$ fits well the data close to the South Pole \citep{howett2011}.
Without much error, we can assume a uniform albedo of $A=0.81$ as in \citet{spencer2006}.

\begin{figure}
   \centering
       \includegraphics[width=6cm]{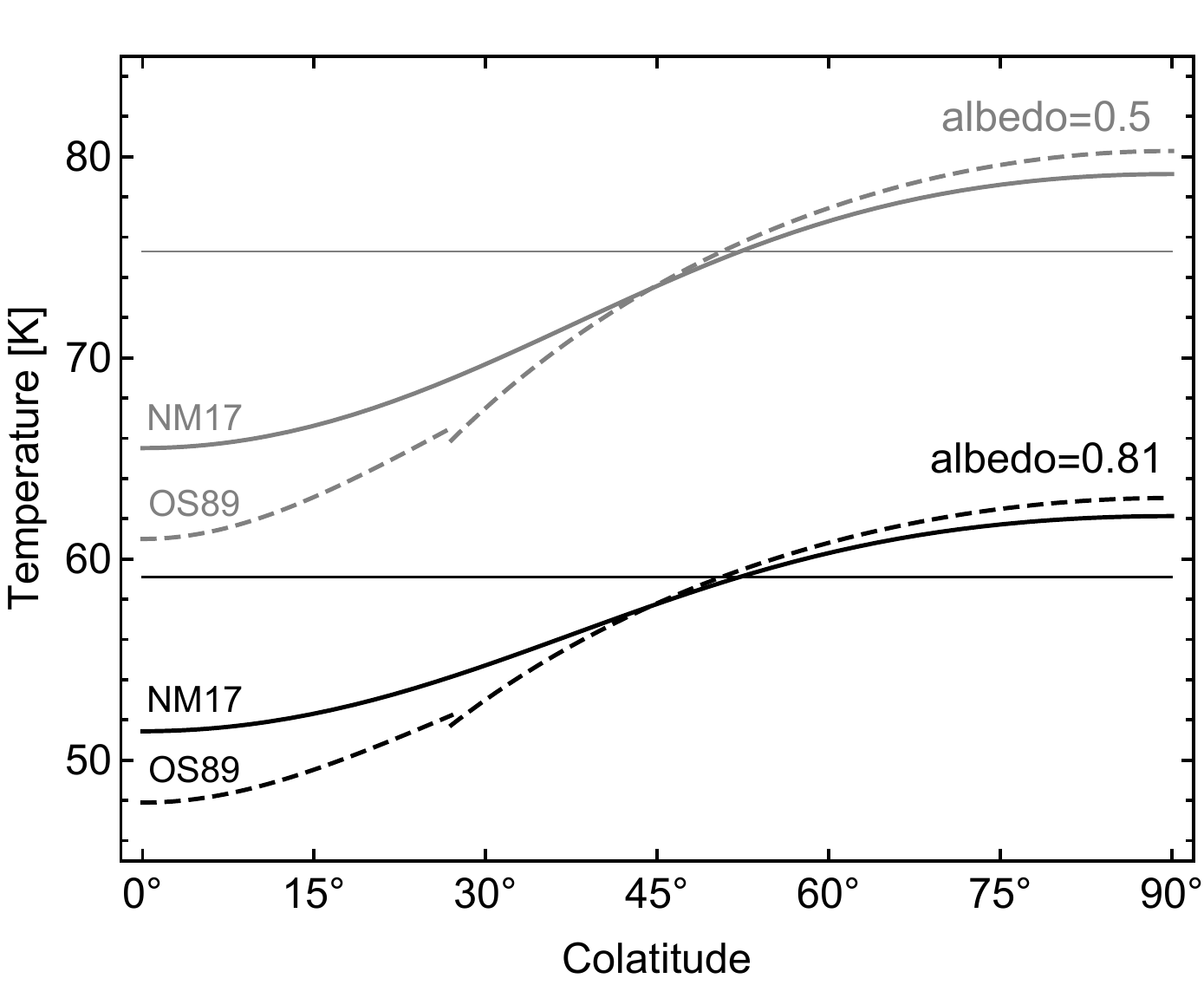}
   \caption[Surface temperature]
   {Near-surface temperature of Enceladus as a function of colatitude, assuming no internal heating.
   Dashed and solid curves show the approximations OS89 \citep{ojakangas1989a} and NM17 \citep{nadeau2017}, respectively.
   Enceladus's high albedo results in lower temperatures (black curves) than would be predicted with Europa's albedo (gray curves).
   Models using only the total solar insolation, as in Paper~I, predict uniform surface temperatures (shown as horizontal lines).
   See Section~\ref{Temperature}.
}
   \label{FigTemp}
\end{figure}

Solar insolation is globally proportional to the solar irradiance at Saturn (${\cal F}_{\rm sat}=14.8\rm\,Wm^{-2}$) and varies locally with latitude.
\citet{roberts2008} took the latter factor into account with the approximate formula of \citet{ojakangas1989a}.
Neglecting internal heating and assuming unit emissivity, they predicted surface temperatures between 61 and $80\rm\,K$.
This range, however, results from setting the average equatorial temperature to $80\rm\,K$, above the subsolar temperature of $76\rm\,K$ \citep{spencer2006}.
Such values are higher than the near-surface temperature used in models of tidal dissipation, i.e.\ the approximately constant temperature below the penetration depth of diurnal and seasonal temperature oscillations (respectively about $1\rm\,cm$ and $1\rm\,m$, see \citet{howett2010,howett2011}).
Using the formula of \citet{ojakangas1989a} with an albedo $A=0.81$ yields instead an equatorial temperature of $63\rm\,K$ (Fig.~\ref{FigTemp}).
A second problem is that \citet{ojakangas1989a} rightly assume that Europa's obliquity with respect to Jupiter's orbital plane is small ($i\sim3^\circ$), whereas Enceladus's obliquity with respect to Saturn's orbital plane is large ($i\sim27^\circ$).
In the latter case, the insolation formula of \citet{nadeau2017} gives a better fit of the mean annual insolation with the additional advantage of being continuous:
\begin{equation}
{\cal F}_{\rm in} = \frac{1}{4} \, {\cal F}_{\rm sat} \, s(\cos\theta,\cos i) \, ,
\end{equation}
where ${\cal F}_{\rm sat}/4$ is the global annual average insolation.
The distribution function $s(\cos\theta,\cos i)$ is approximated by a 6th-order expansion in Legendre polynomials (with unit average on the sphere).
As an aside, note that the approximation of \citet{nadeau2017} overestimates the polar insolation if $i<12^\circ$; the formula of \citet{ojakangas1989a} actually gives a better fit near the poles if $i<6^\circ$ (\textit{A.~Nadeau, private comm.}).

At the poles, the conductive flux becomes comparable to the radiative flux.
For example,  Eq.~(\ref{ConductiveFlux}) yields ${\cal F}_{\rm cond}=150\rm\,mW/m^2$ if $d=7\rm\,km$ at the south pole, i.e.\ 40\% of the radiative flux in equilibrium with solar insolation ($380\rm\,mW/m^2$ at the same location).
For a black body, the equilibrium between solar insolation, reemitted radiation, and internal heating reads
\begin{equation}
\sigma\, T_{\rm s}^4 = (1-A) \, {\cal F}_{\rm in} + {\cal F}_{\rm cond} \, ,
\label{Tequil}
\end{equation}
where $\sigma=5.67\times10^{-8}\rm\,Wm^{-2}K^{-4}$ is the Stefan-Boltzmann constant.
If internal heating is ignored, the mean annual near-surface temperature varies between $51\rm\,K$ (at the poles) and $62\rm\,K$ (at the equator), as shown in Fig.~\ref{FigTemp}.
If the crust is nowhere much thinner than $7\rm\,km$, the conductive flux can be approximated by Eq.~(\ref{ConductiveFlux}) in which $T_{\rm s}$ is the surface temperature before the correction and $a=632\rm\,W/m$.
With this correction, the surface temperature in the Model ISO increases by 2 and 4 degrees at the north and south poles, respectively, while the increase is only about half a degree along the equator, with a very small longitudinal variation.
In a model with a very thin crust at the south pole (e.g.\ $d=2\rm\,km$), the correction is larger and Eq.~(\ref{Tequil}) should be solved self-consistently.

\subsection{Results and discussion}
\label{Results}

\subsubsection{Shell dissipation}
\label{Results1}

\begin{table}[h]
\ra{1.3}
\small
\caption[Thermal equilibrium models]
{Thermal equilibrium models: characteristics, total power (conductive/shell/core), partition of the shell power and feedback effect.
UNI/ISO/THIN specify the shell thickness model; L/H denote the low/high level of shell dissipation; C indicates that core dissipation occurs
(shell and core structures are detailed in Section~\ref{ShellStructure}).}
\begin{tabular}{@{}llllllll@{}}
\hline
Models & UNI-L & ISO-L & UNI-H & ISO-H & ISO-LC & ISO-HC & THIN-LC \\
Shell thickness & uniform & isostatic & uniform & isostatic & isostatic & isostatic & thin SPT \\
Shell dissipation & low & low & high & high & low & high & low \\
Core dissipation & -- & -- & -- & -- & high & high & high \\ 
\hline
Power (GW) & & & & & & & \\
Conductive power  & 31.0 & 34.5 & 31.8 & 35.4 & 34.5  & 35.5 & 40.3 \\
Shell power & 0.96 & 1.16 &  12.9 & 15.3 & 1.22  & 15.9 & 2.99 \\
Core power & 0 & 0 & 0  & 0 & 33.3 & 19.5 & 37.3 \\
\hline
Shell partition (\%) & & & & & & &\\ 
Membrane & 86.7 & 89.8 & 85.1  & 89.2 & 89.8 & 89.2 & 57.4 \\
Mixing & 7.8 & 2.0 & 9.7   &  3.2 & 2.0 & 3.3 & -3.0 \\
Bending & 5.6 & 8.2 & 5.2  & 7.6 & 8.2 & 7.5 & 45.6 \\
\hline
Rheology feedback (\%) 
& 2 & 2 & 37  & 35 & 2 & 37 & 5 \\
\hline
\end{tabular}
\label{TablePower}
\end{table}%

\begin{figure}
   \centering
           \includegraphics[width=6cm]{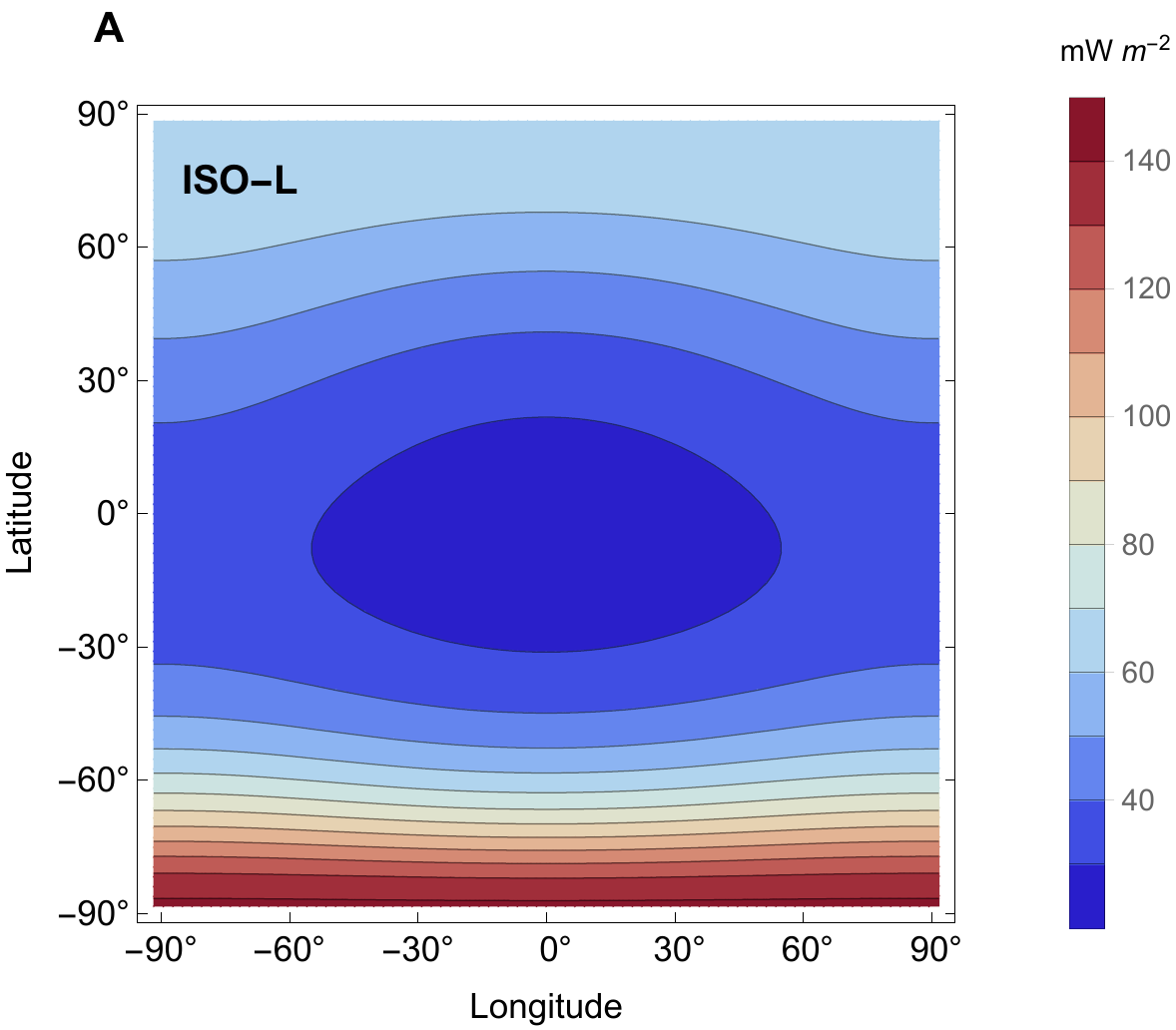}
            \hspace{3mm}
           \includegraphics[width=6cm]{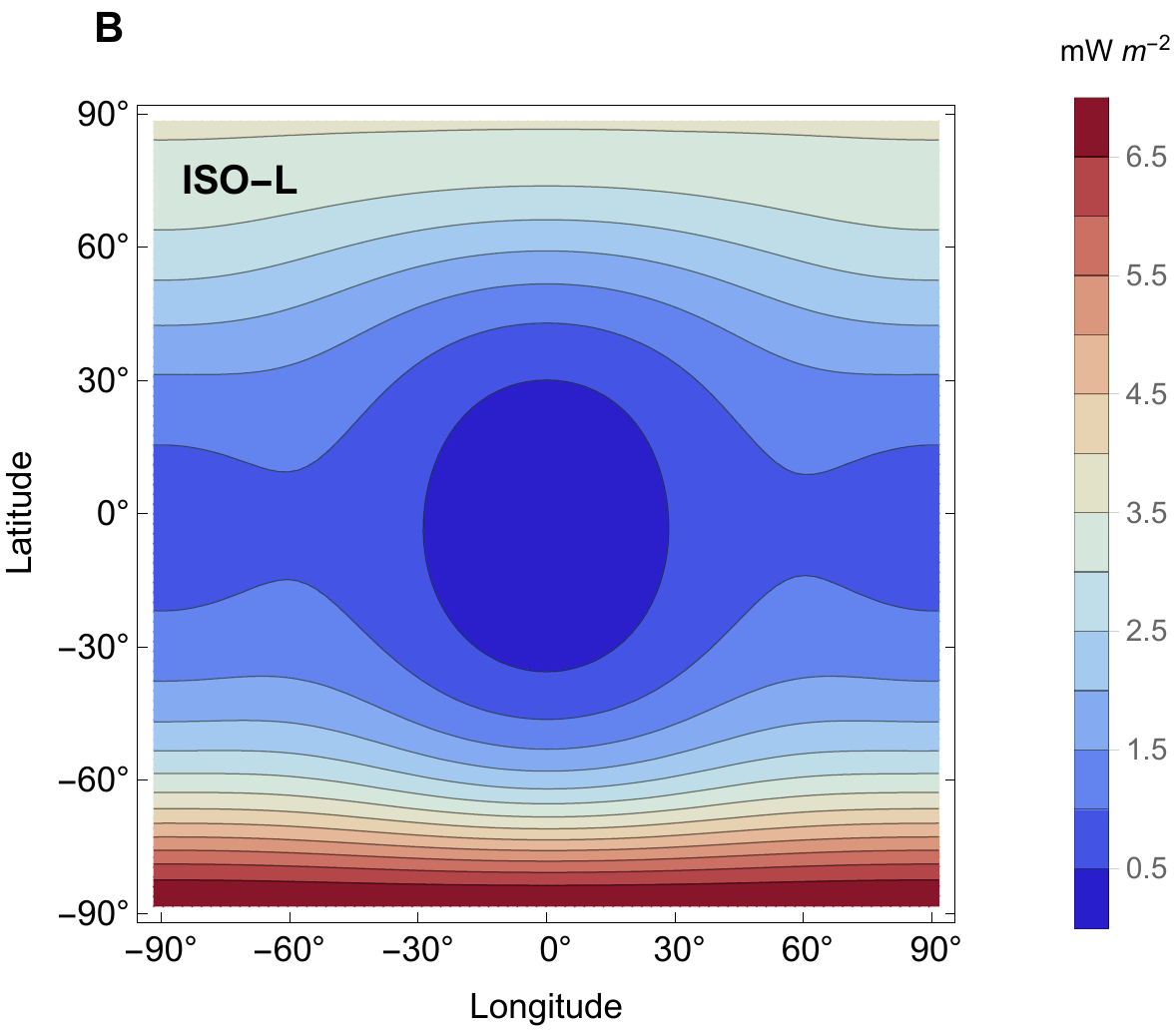}
   \caption[Heat flux pattern in model ISO-L]{
   Heat flux pattern in model ISO-L: (A) conductive flux; (B) shell dissipation flux.
   Patterns repeat between $90^\circ$ and $270^\circ$; longitude $0^\circ$ corresponds to the tidal axis.
      See Section~\ref{Results1}.
      }
   \label{FigFluxPattern}
\end{figure}

In good approximation, the conductive flux is inversely proportional to the shell thickness, with marginal influences of surface temperature and in-shell dissipation (Eq.~(\ref{ConductiveFlux})).
If the shell thickness is isostatic, the conductive flux is highest at the south pole where it reaches $150\rm\,mW/m^2$ (Fig.~\ref{FigFluxPattern}A), with an average over the SPT (below $55^\circ\rm\,S$ latitude) of $85\rm\,mW/m^2$.
The conductive power emitted by the SPT is then $6\rm\,GW$, which is comparable to the SPT power deduced from Cassini infrared data (between 4 and $20\rm\,GW$, see review by \citet{spencer2018}).
It is thus reasonable to treat the conductive power of the isostatic shell as an observational constraint, to be matched by the power generated within Enceladus.

\begin{figure}
   \centering
        \includegraphics[width=6cm]{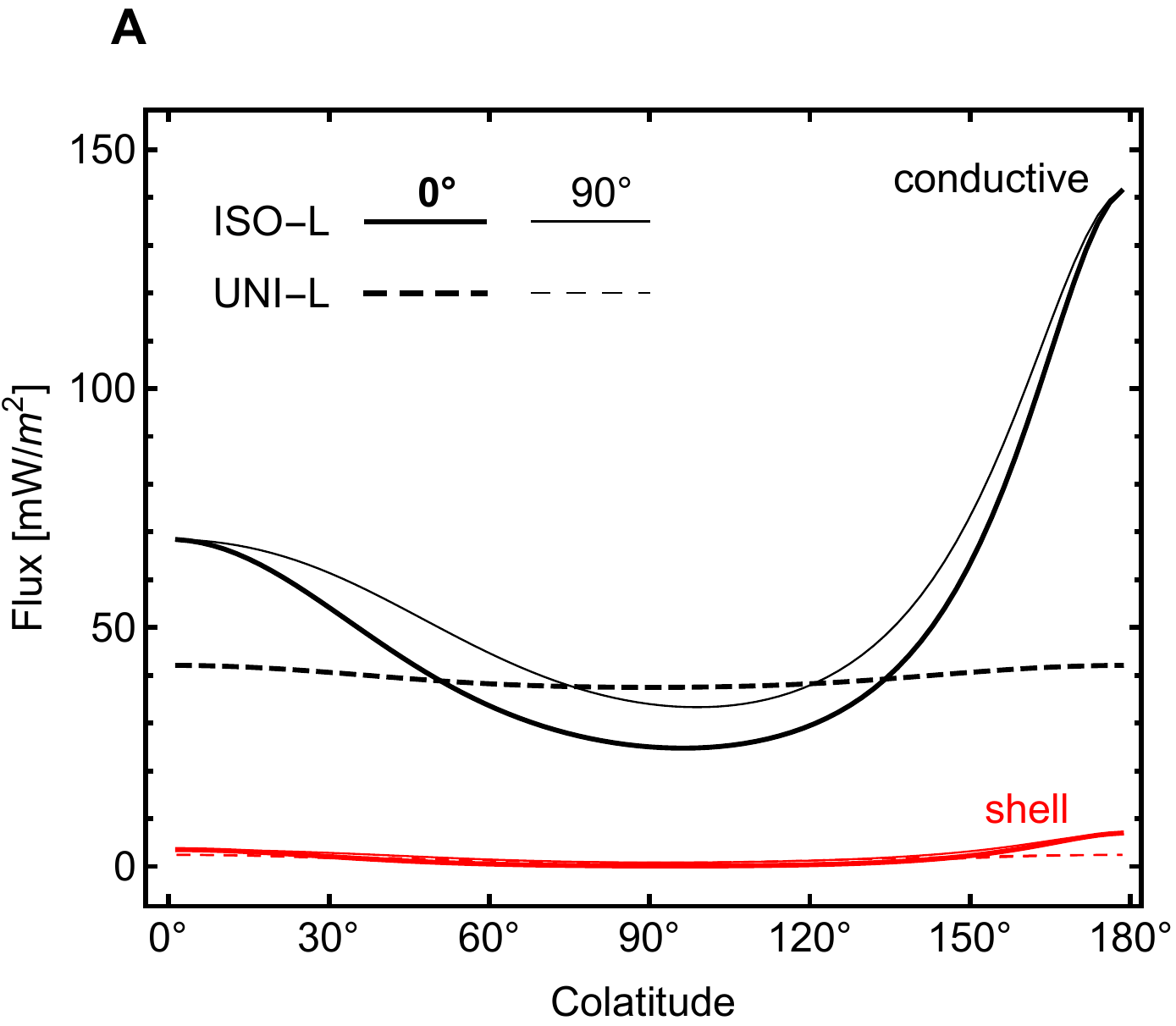}
         \hspace{3mm}
         \includegraphics[width=6cm]{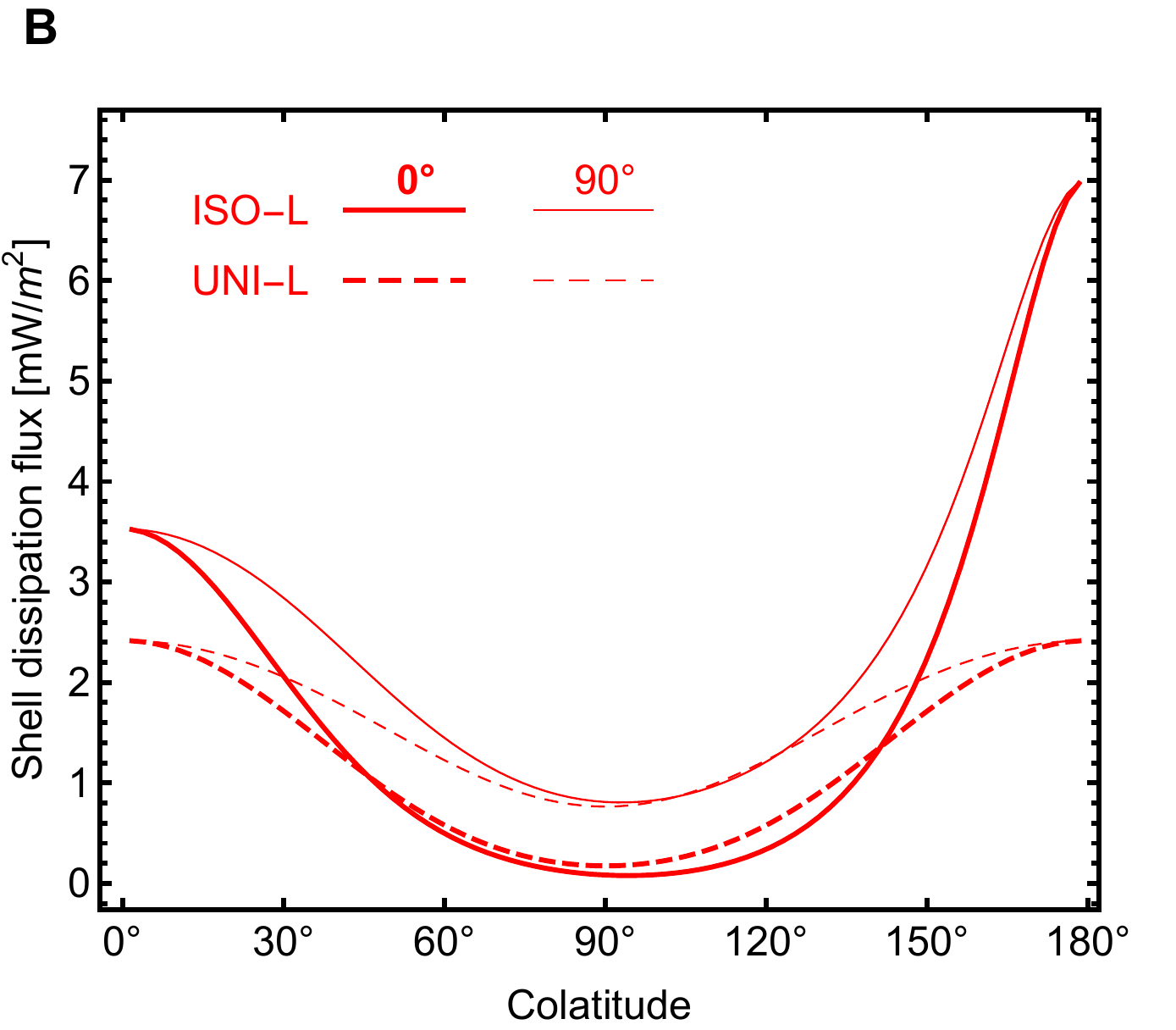}
         \includegraphics[width=6cm]{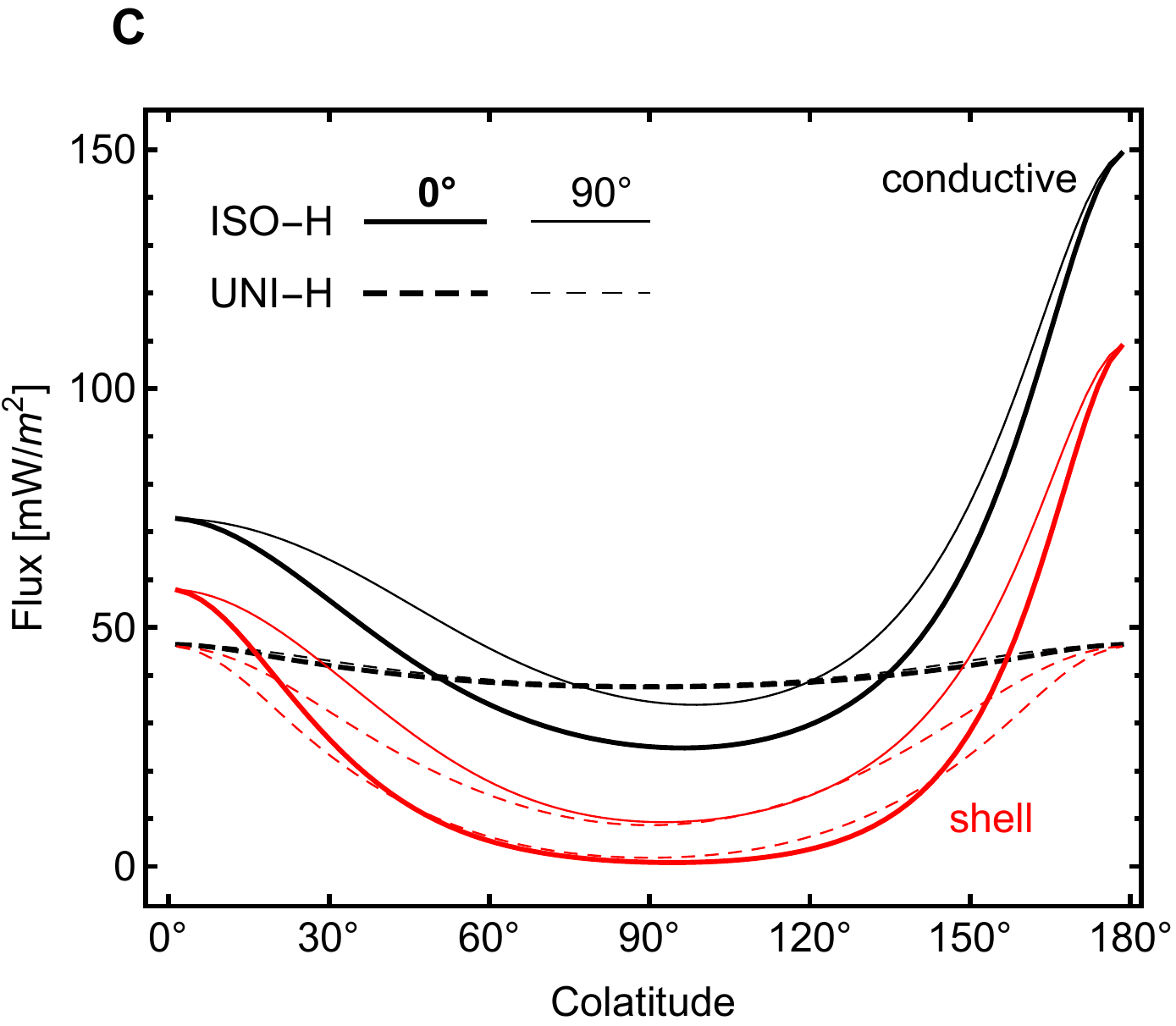}
          \hspace{3mm}
        \includegraphics[width=6cm]{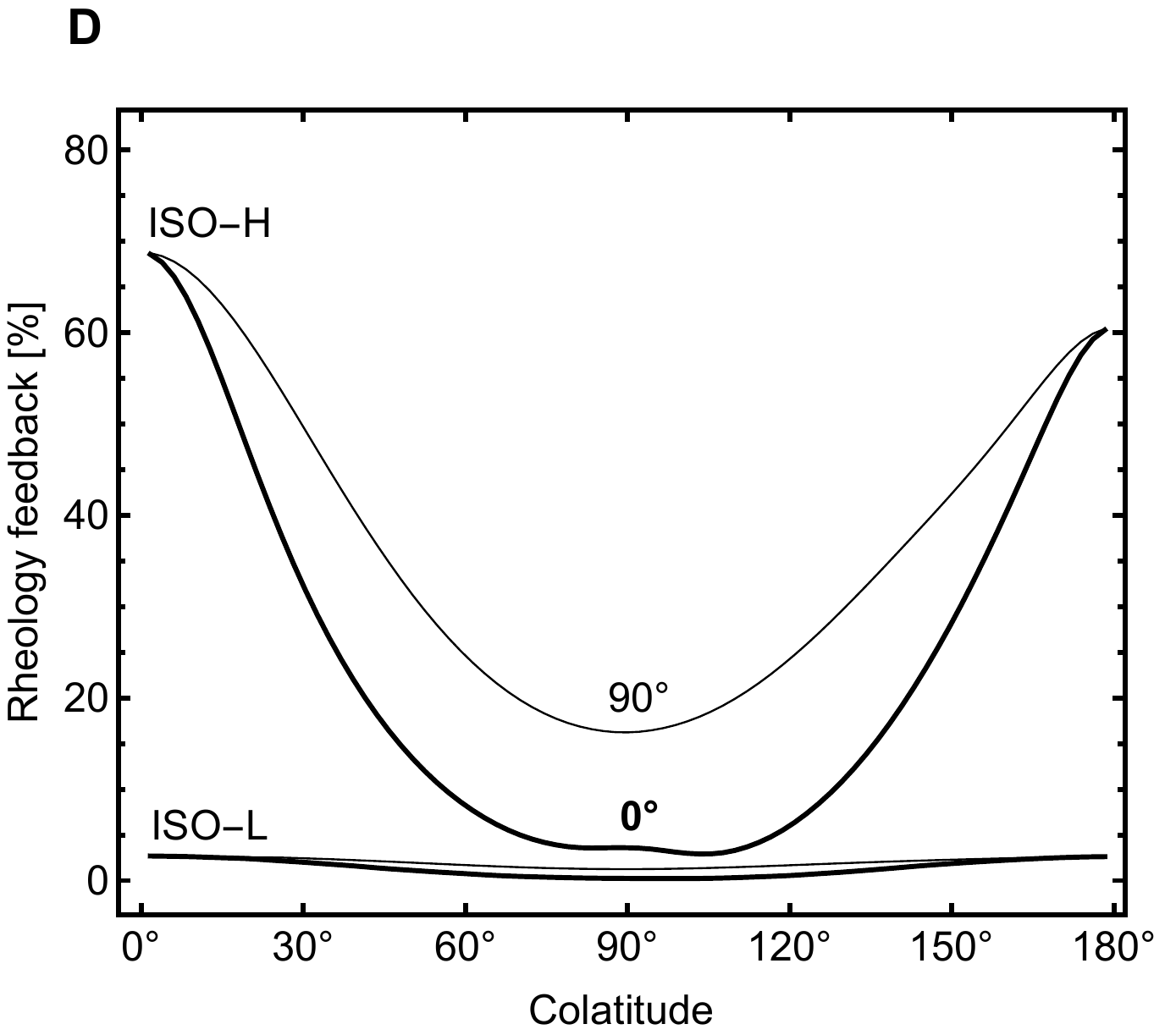}
   \caption[Heat flux and rheology feedback if no core dissipation]{
   Heat flux and rheology feedback if no core dissipation: meridional profiles.
    (A) conductive flux and shell dissipation flux if nominal heating (models UNI-L and ISO-L);
    (B) zoom on the shell dissipation flux in these models;
   (C) conductive flux and shell dissipation flux if high heating (models UNI-H and ISO-H);
    (D) rheology feedback for ISO models.
   Solid (resp.\ dashed) curves correspond to models with isostatic (resp.\ uniform) shell thickness.
   Thick (resp.\ thin) curves correspond to longitude $0^\circ$ (resp.\ $90^\circ$).
   See Section~\ref{Results1}.
   }
   \label{FigFluxProfile}
\end{figure}

For nominal (i.e.\ low) dissipation, the shell dissipation flux is only a few percent of the conductive flux (Fig.~\ref{FigFluxPattern}B), as found by \citet{soucek2019}.
While the non-uniformity of the shell only slightly increases the shell power (by 20\%, see Table~\ref{TablePower}), it has a major effect on the pattern of the shell dissipation flux.
Dissipation within a floating thin shell is typically highest at the poles and lowest along the tidal axis.
This is already true for a laterally uniform shell (Fig.~\ref{FigPatternThinShell}) but these contrasts are further enhanced by isostatic variations of shell thickness (Fig.~\ref{FigFluxPattern}B and Fig.~\ref{FigFluxProfile}).
In that case, the flux is amplified by nearly a factor of 3 at the south pole, where the shell is thinner by a factor of 3 (with respect to the average thickness).

If shell dissipation is ten times higher, the shell power increases more than ten times because of the significant rheology feedback (35\% instead of 2\% on average, see Table~\ref{TablePower}).
For the same reason, a non-zero forced libration has a larger effect on the shell power if shell dissipation is high (42\% increase instead of 29\% in the nominal case).
In the end, the shell power of the model ISO-H makes up 43\% of the conductive power.
Furthermore, rheology feedback is much stronger at the poles than at the equator; in particular, it significantly enhances the flux at the south pole (Fig.~\ref{FigFluxProfile}D).
Thus, high dissipation within the shell can generate the major part of the conductive flux pattern (Fig.~\ref{FigFluxProfile}C).
Nevertheless, it cannot explain the whole conductive flux, whatever the dissipation enhancement, because shell dissipation within the shell remains close to zero along the tidal axis.
Thermal equilibrium requires a non-zero heat flux from below the shell, either from core or ocean dissipation.

\subsubsection{Partition and scaling}
\label{Results2}

\begin{figure}
   \centering
         \includegraphics[width=5.5cm]{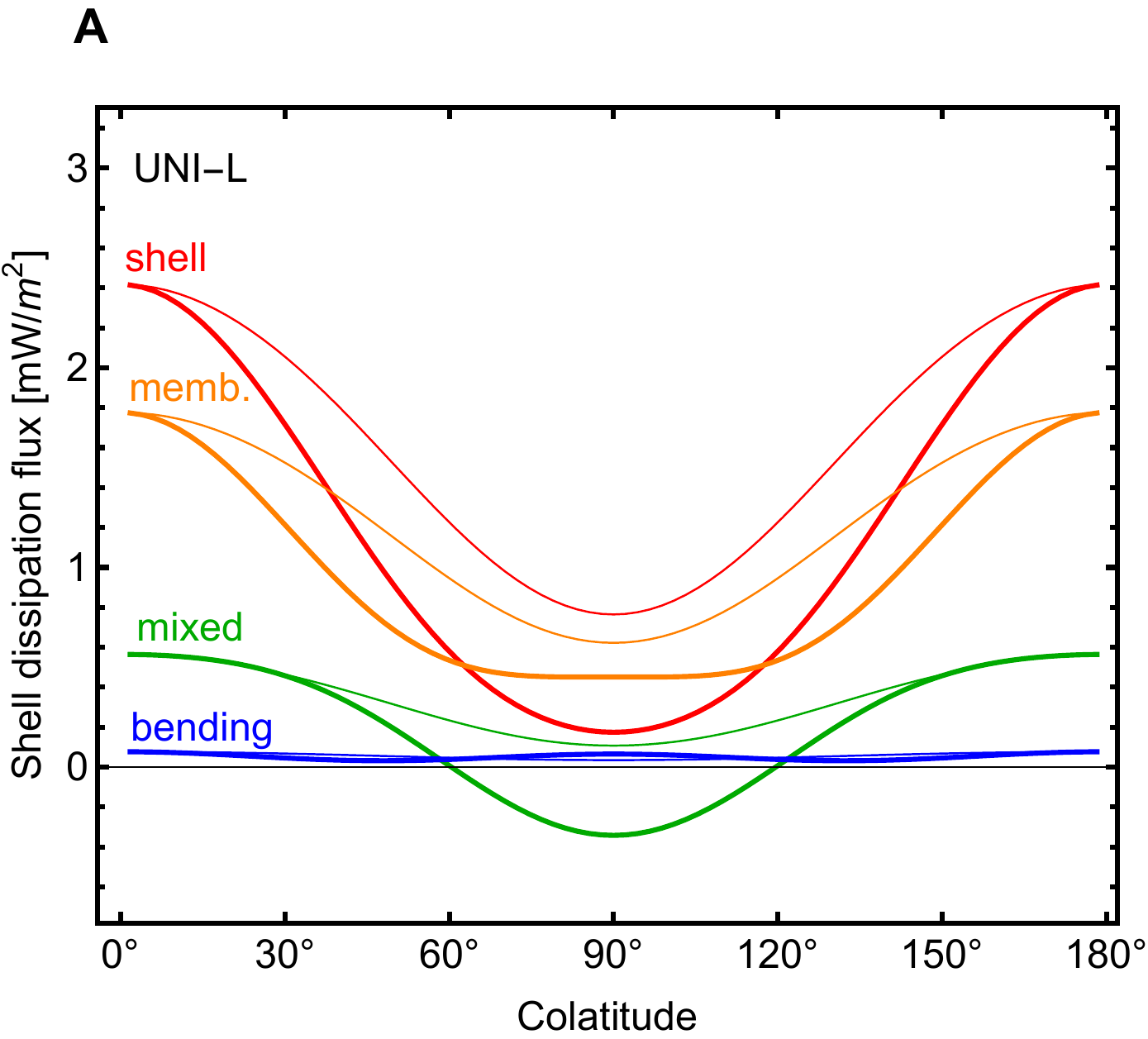}
          \hspace{3mm}
          \includegraphics[width=5.5cm]{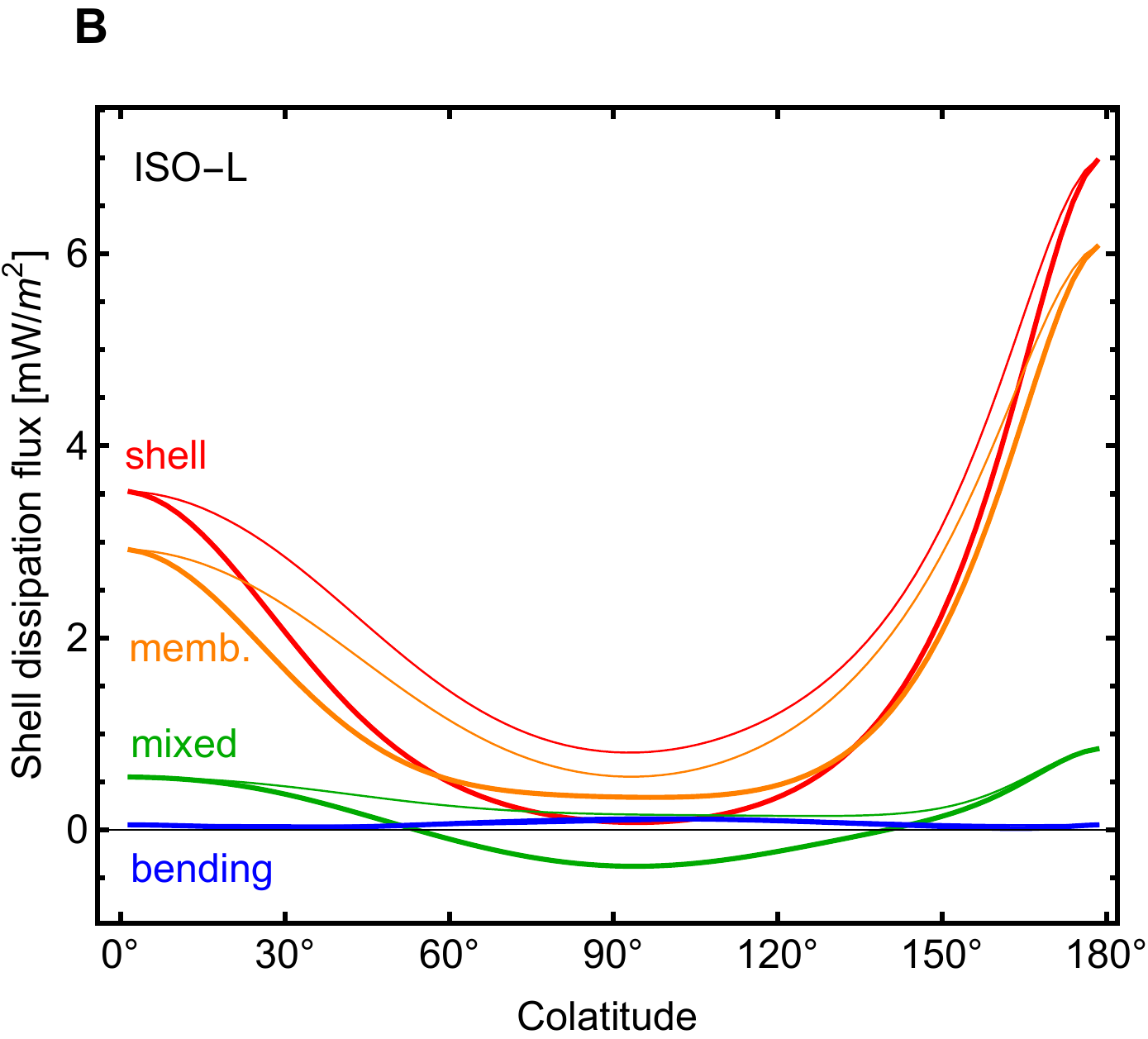}
           \includegraphics[width=5.5cm]{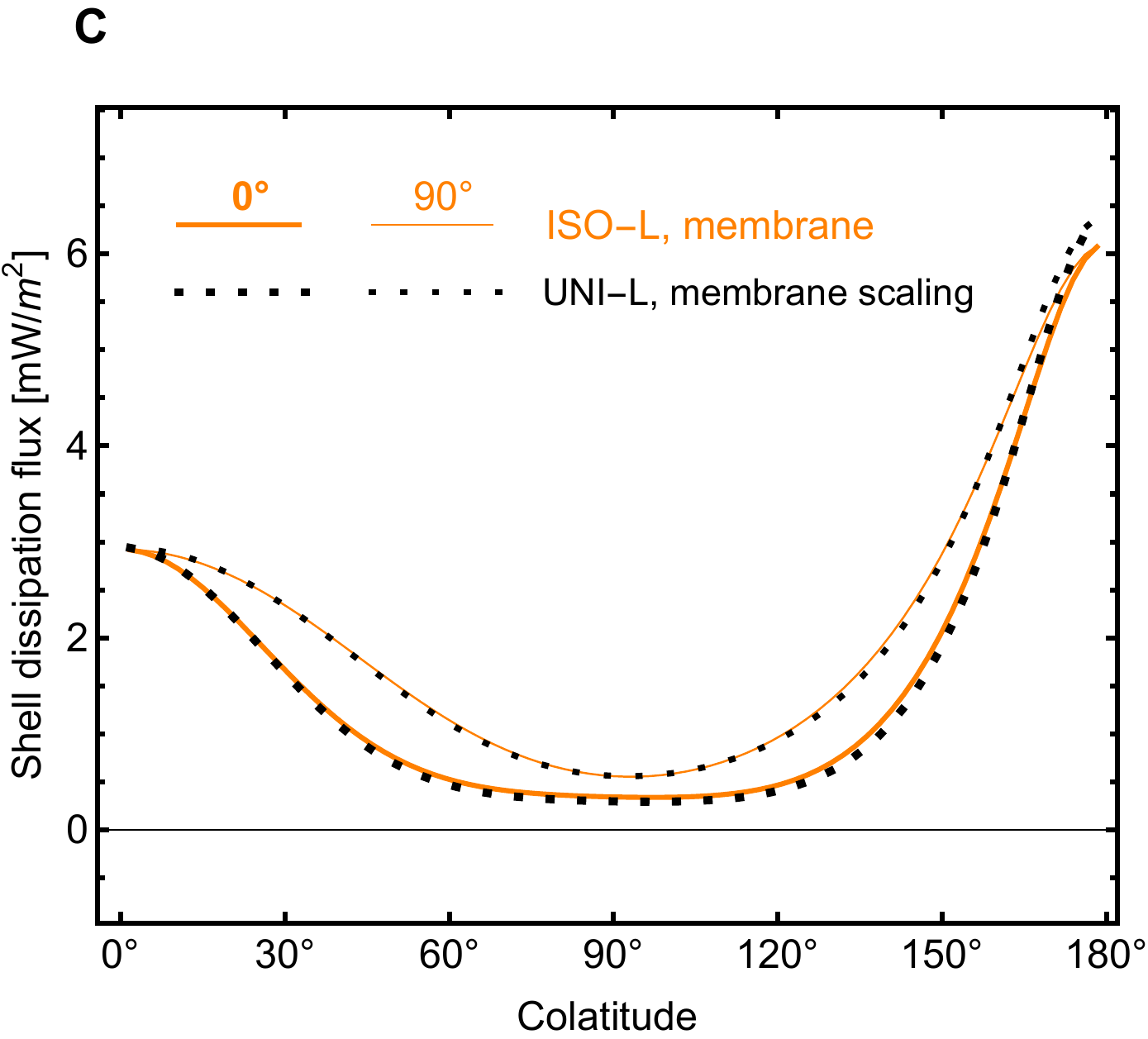}
          \hspace{3mm}
          \includegraphics[width=5.5cm]{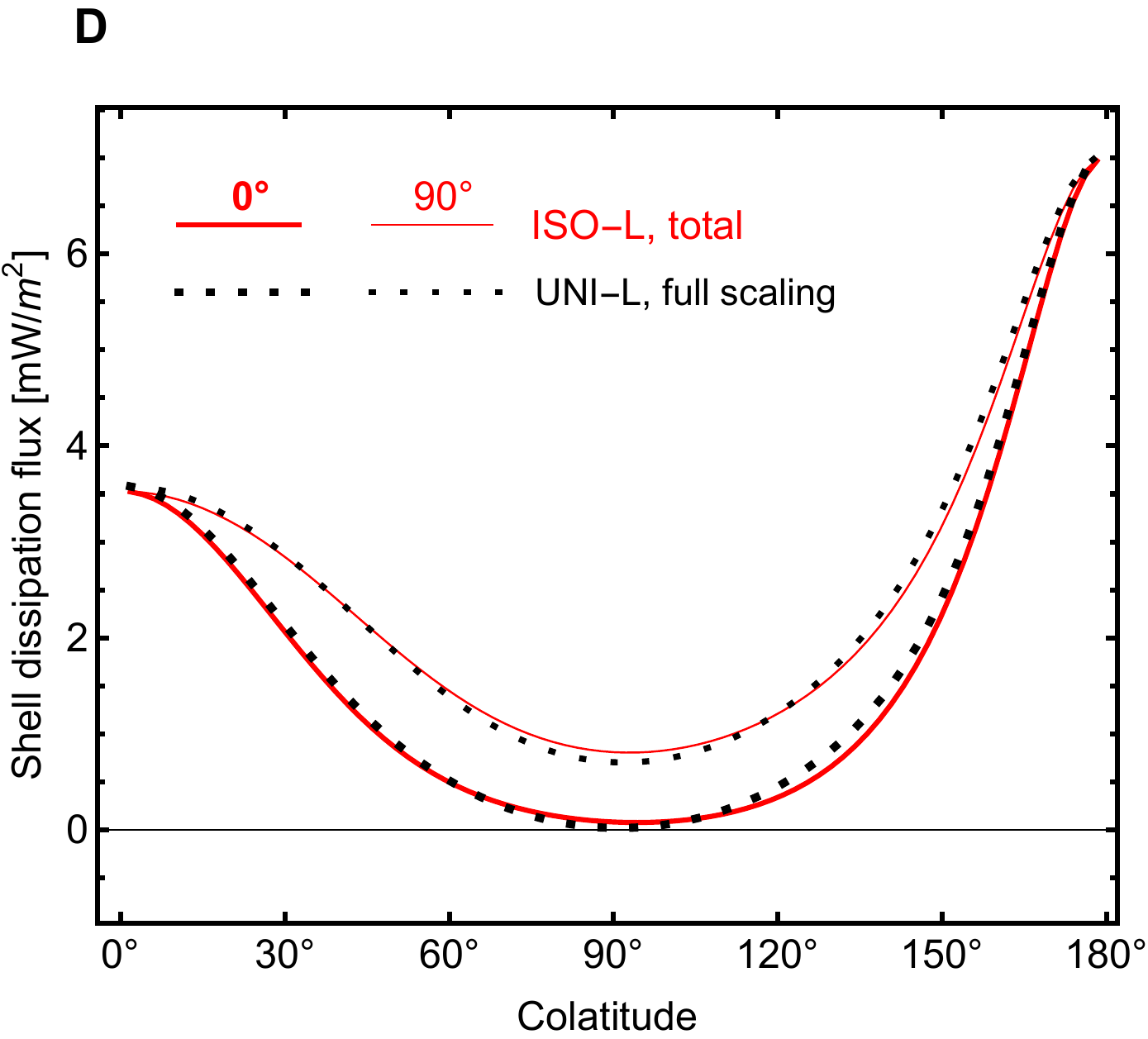}
   \caption[Partition and scaling of the shell dissipation flux]{
   Partition and scaling of the shell dissipation flux: meridional profiles.
   (A) Membrane/mixed/bending contributions in model UNI-L.
   (B) Same for model ISO-L.
   (C) Membrane scaling from UNI-L to ISO-L.
   (D) Full scaling from UNI-L to ISO-L.
   Thick and thin curves correspond to longitudes $0^\circ$ and $90^\circ$, respectively.
      Dotted curves are scaling predictions.
   Full scaling is done with  Eq.~(\ref{scaling}), while only the first term of this equation is retained for membrane scaling.
   See Section~\ref{Results2}.
   }
   \label{FigFluxPartition}
\end{figure}

Before looking at models with core dissipation, it is instructive to examine the partition of shell dissipation into membrane/mixed/bending contributions.
The shell power is clearly dominated by the membrane contribution (between 85 and 90\%, see Table~\ref{TablePower}).
This power partition is slightly misleading because the mixed term switches sign between the poles and the equator (Fig.~\ref{FigFluxPartition}).
At the equator, the mixed contribution largely cancels the (always positive) membrane and bending contributions, whereas it makes up a significant portion of the south polar flux (23\% and 12\% in models UNI-L and ISO-L, respectively).
It is thus important to include mixed and bending contributions, and also to keep both of them: neglecting the latter may result in a negative surface flux at the equator.

\begin{figure}
   \centering
\includegraphics[width=6cm]{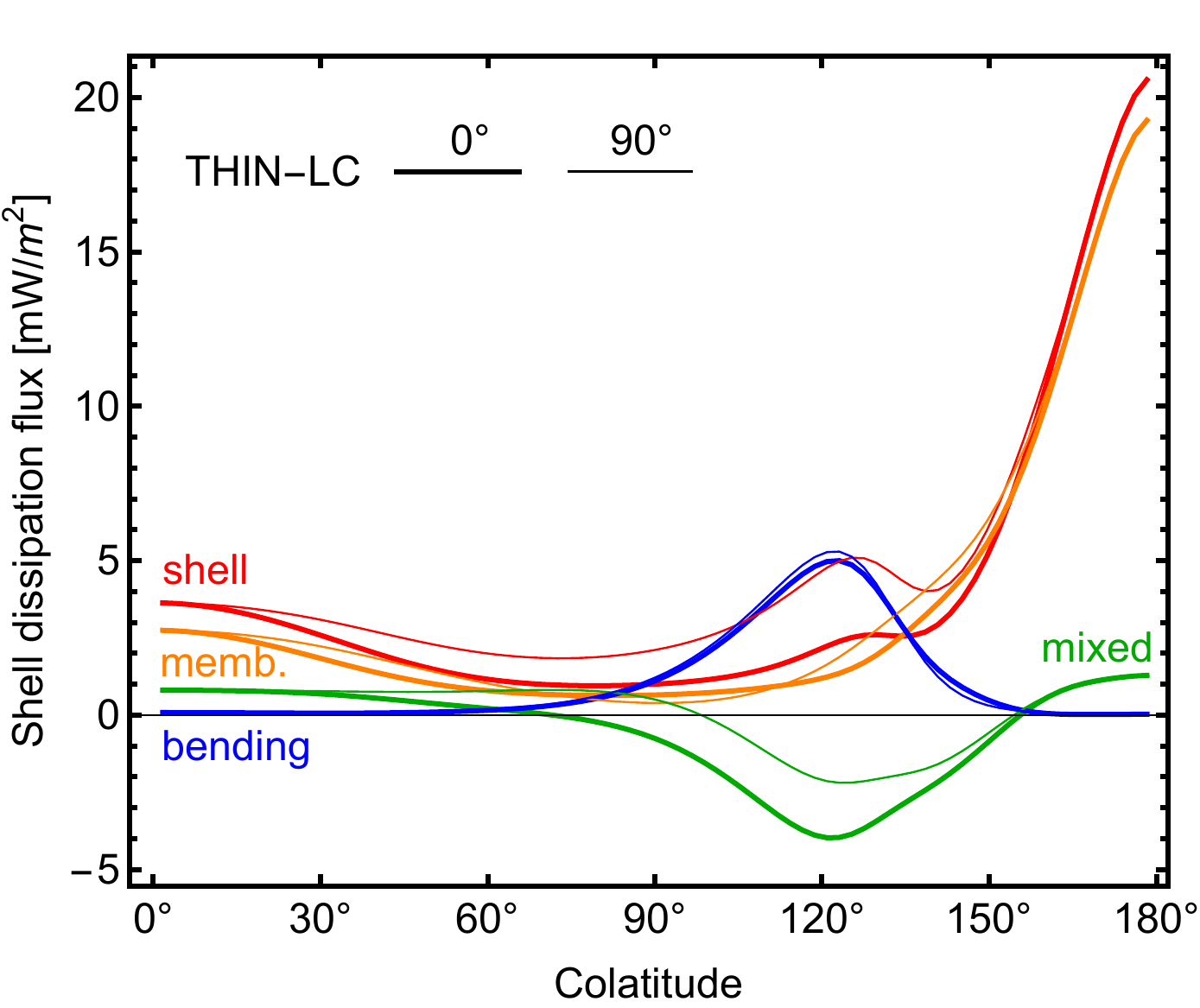}
   \caption[Partition of the shell dissipation flux into membrane/mixed/bending contributions]{
     Partition of the shell dissipation flux into membrane/mixed/bending contributions: meridional profiles for model THIN-LC.
        Thick and thin curves correspond to longitudes $0^\circ$ and $90^\circ$, respectively.
    See Section~\ref{Results2}.}
   \label{FigFluxPartitionThin}
\end{figure}

The membrane term is mainly responsible for amplifying the flux in areas with thinner shell (Fig.~\ref{FigFluxPartition}).
The decomposition of the shell dissipation flux (Eq.~(\ref{SurfaceFluxThinShellComp})) explain this behaviour: membrane, mixed and bending terms term are proportional to ${\rm Im}(\alpha)\sim1/d$, ${\rm Im}(\chi)\sim{}d$, and ${\rm Im}(D)\sim{}d^3$, respectively. The factors of Eq.~(\ref{SurfaceFluxThinShellComp}) that depend on $F$ and $w$ are less sensitive, in the membrane approximation, to the local shell thickness $d$.
On the one hand, the stress function -- in the membrane limit of a hard shell -- does not depend on the shell properties (see Eq.~(80) of Paper~I; a shell is hard if $(\mu_e/\rho{}gR)(d/R)\gg1$, see Section~4.3.2. of Paper~I).
On the other, the shell structure weakly affects the radial displacement of the shell (Fig.~8 of Paper~I).
Thus, the shell dissipation flux in a laterally non-uniform shell can be approximated by the following (full) scaling rule:
\begin{equation}
{\cal F}_{shell}^{\rm \, iso} \approx A^{mem} \, {\cal F}_{mem}^{\rm \, uni} + {\cal F}_{mix}^{\rm \, uni} +{\cal F}_{bend}^{\rm \, uni} \, ,
\label{scaling}
\end{equation}
where superscripts `uni' and `iso' refer to the laterally uniform (UNI) and non-uniform (ISO) models.
The terms in the RHS can be computed with analytical formulas (Eq.~(\ref{SurfaceFluxThinShellCompU})).
The amplification factor is defined by
\begin{equation}
A^{mem} = \frac{ {\rm Im}( \alpha^{\rm iso \,} ) }{ {\rm Im}(\alpha^{\rm uni})}
\sim \frac{ d^{\rm \, uni} }{ d^{\rm \, iso} } \, .
\label{amplifactor}
\end{equation}
Thus, if the rheology is approximately laterally uniform, the dissipation flux is inversely proportional to the shell thickness.
Beware that the scaling rule (\ref{amplifactor}) only holds if lateral variations of shell properties are of long wavelength and if the shell is everywhere hard, as in models ISO-L and ISO-H.
It breaks down, for example, in the model THIN-LC in which bending effects are important and the shell is locally soft at the south pole (Fig.~\ref{FigFluxPartitionThin}).

In Paper~I, I showed that the surface stress (and strain) of Enceladus scales with the factor $|\alpha|\sim1/d$
(see Eq.~(88) of Paper~I).
Why does the shell dissipation flux scale in the same way?
By definition, it is proportional to the time average of stress times strain rate, integrated over the shell thickness.
In a conductive shell, dissipation mostly occurs in a thin layer at the bottom of the shell.
If dissipation is too low to perturb the temperature profile, the thickness of the thin dissipative layer is proportional to the local shell thickness.
The amplitudes of the stress and strain rate at the bottom of the shell both scale with $\alpha$.
Therefore, the shell dissipation flux scales as $\alpha^2d\sim1/d$ for a hard shell in the membrane limit, which gives back Eq.~(\ref{amplifactor}).

\subsubsection{Core dissipation}
\label{Results3}

Since shell dissipation cannot account for the full conductive flux, we now turn to models with high core dissipation.
Two pieces of (indirect) evidence for a highly porous and dissipative core are, first, its low density (between 2300 and $2600\rm\,kg/m^3$, see \citet{mckinnon2015,beuthe2016b}) and, second, the hydrothermal activity inferred from silica nanoparticles and molecular hydrogen observed in the plume \citep{hsu2015,sekine2015,waite2017,glein2018}.
As already seen with the laterally uniform model (Fig.~\ref{FigCorePower}), it is possible to tune the viscoelastic core parameters so that core dissipation provides the missing part of the conductive power; it can be similarly done in the laterally non-uniform model by successive adjustments.
Unfortunately, laboratory experiments on porous materials have only been done at conditions very different (higher frequency, lower pressure) of those to which the core is submitted during tidal loading.
Thus, we don't know whether the effective shear modulus of the core $|\mu_c|$ can really be 1000 times smaller than its elastic value for non-porous silicates.

\begin{figure}
   \centering
          \includegraphics[width=6cm]{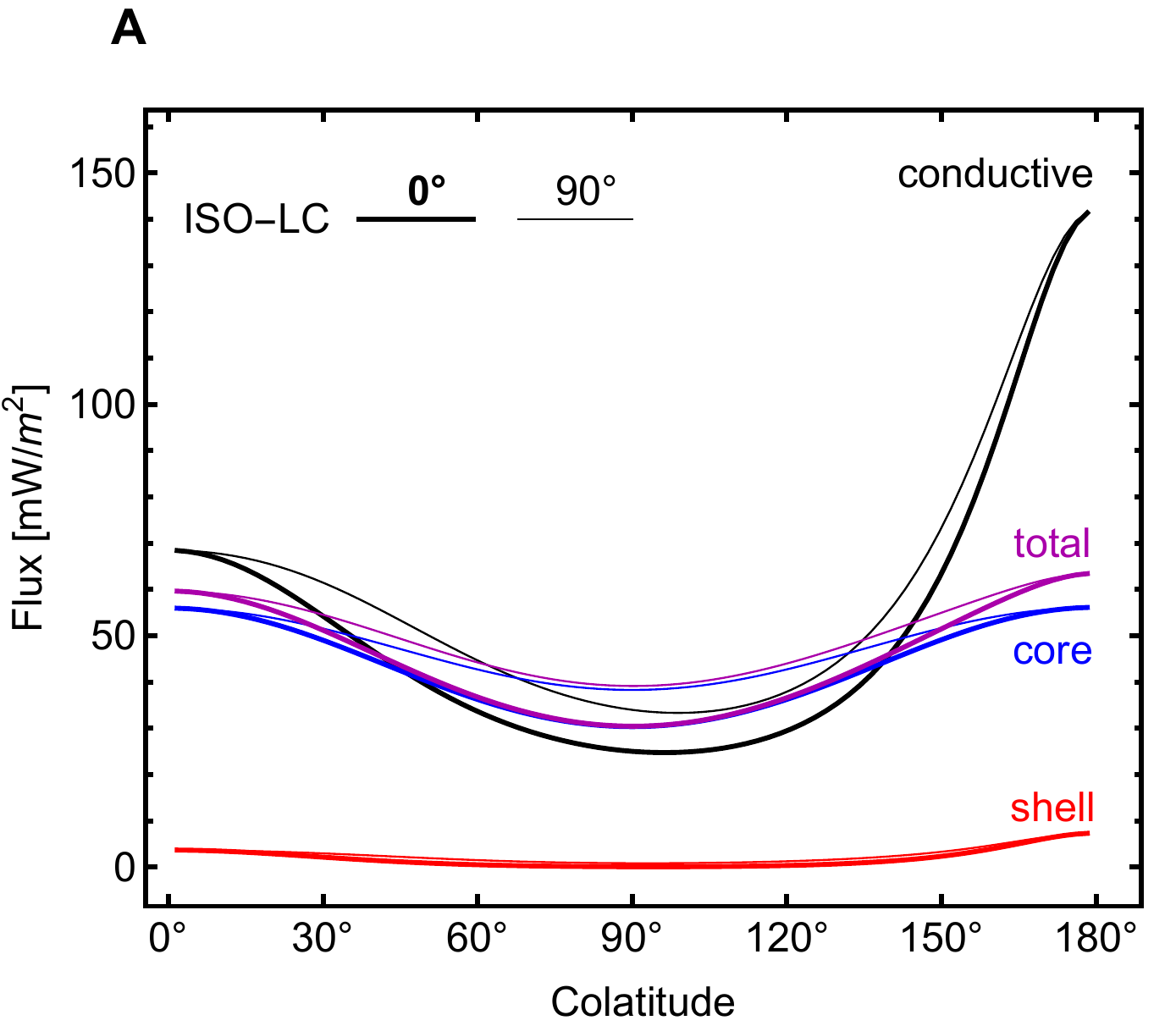}
          \hspace{3mm}
          \includegraphics[width=6cm]{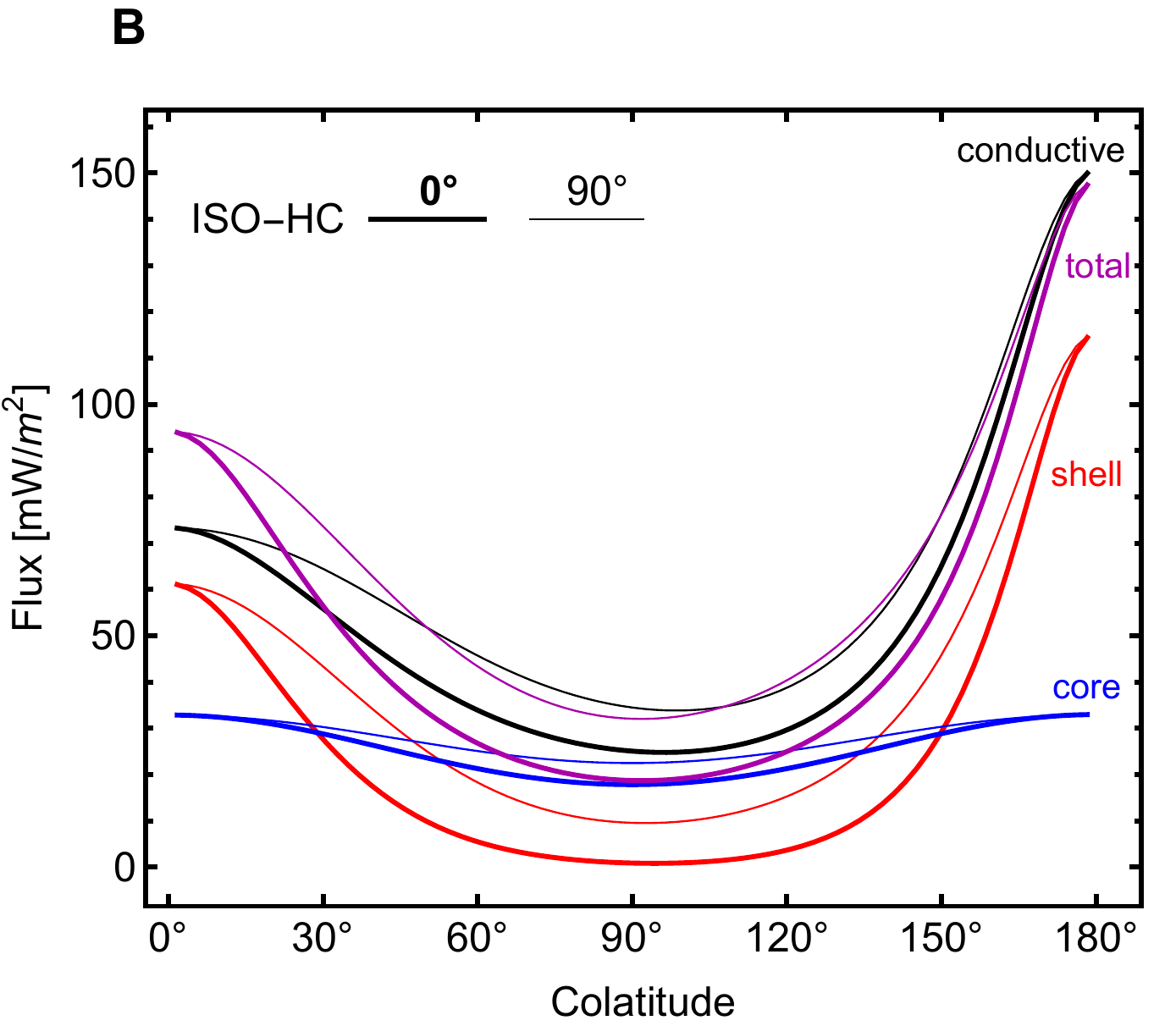}
   \caption[Surface heat flux if high dissipation within the core]{
      Heat flux if high dissipation within the core: meridional profiles.
      (A) Conductive flux, shell dissipation flux, core dissipation flux (at the surface), and the sum of the latter two in model ISO-LC.
      (B) Same for model ISO-HC.
      Thick and thin curves correspond to longitudes $0^\circ$ and $90^\circ$, respectively.
      See Section~\ref{Results3}.
      }
   \label{FigFluxProfileCore}
\end{figure}

\begin{figure}
   \centering
            \includegraphics[width=4.9cm]{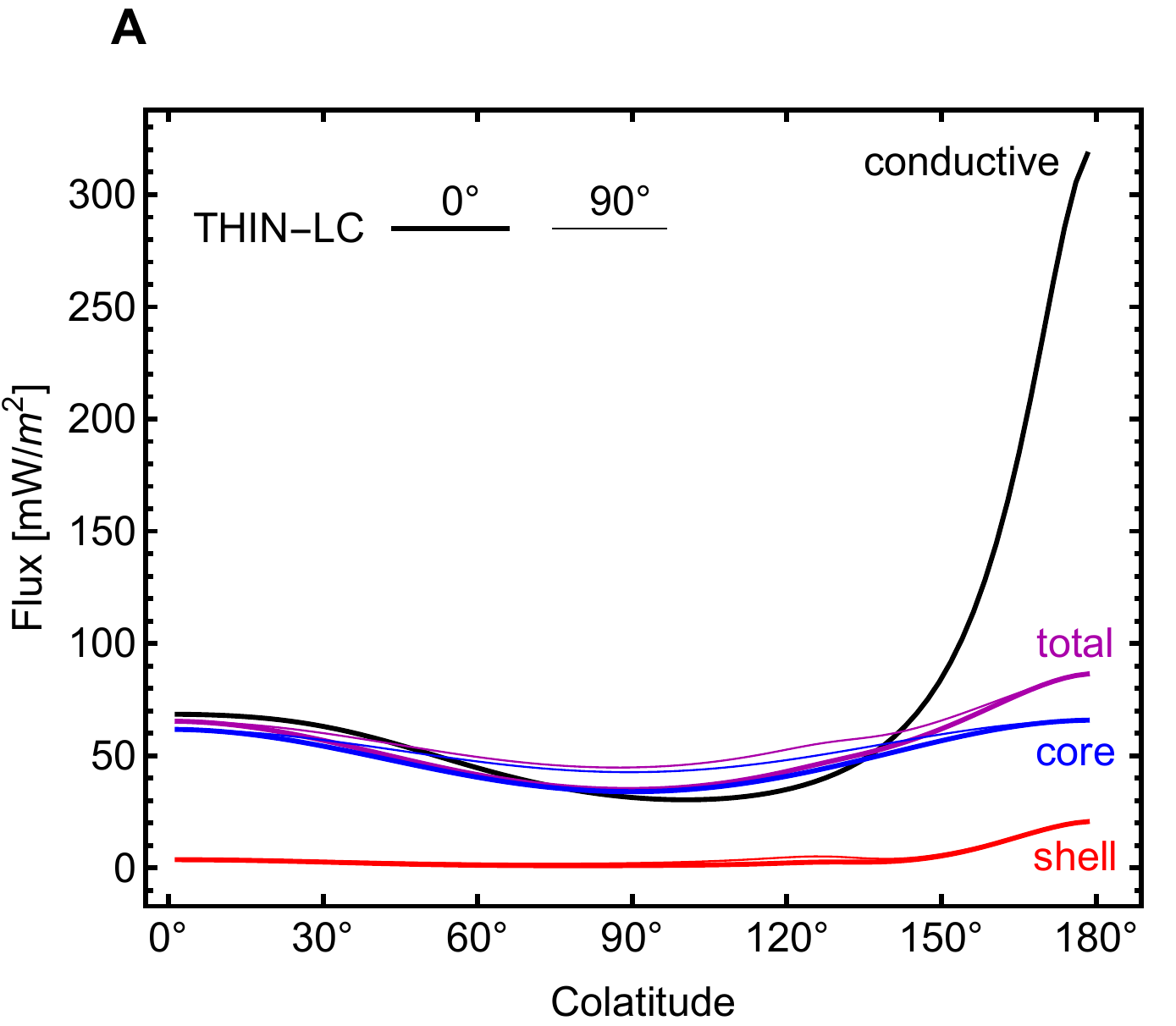}
         \includegraphics[width=4.9cm]{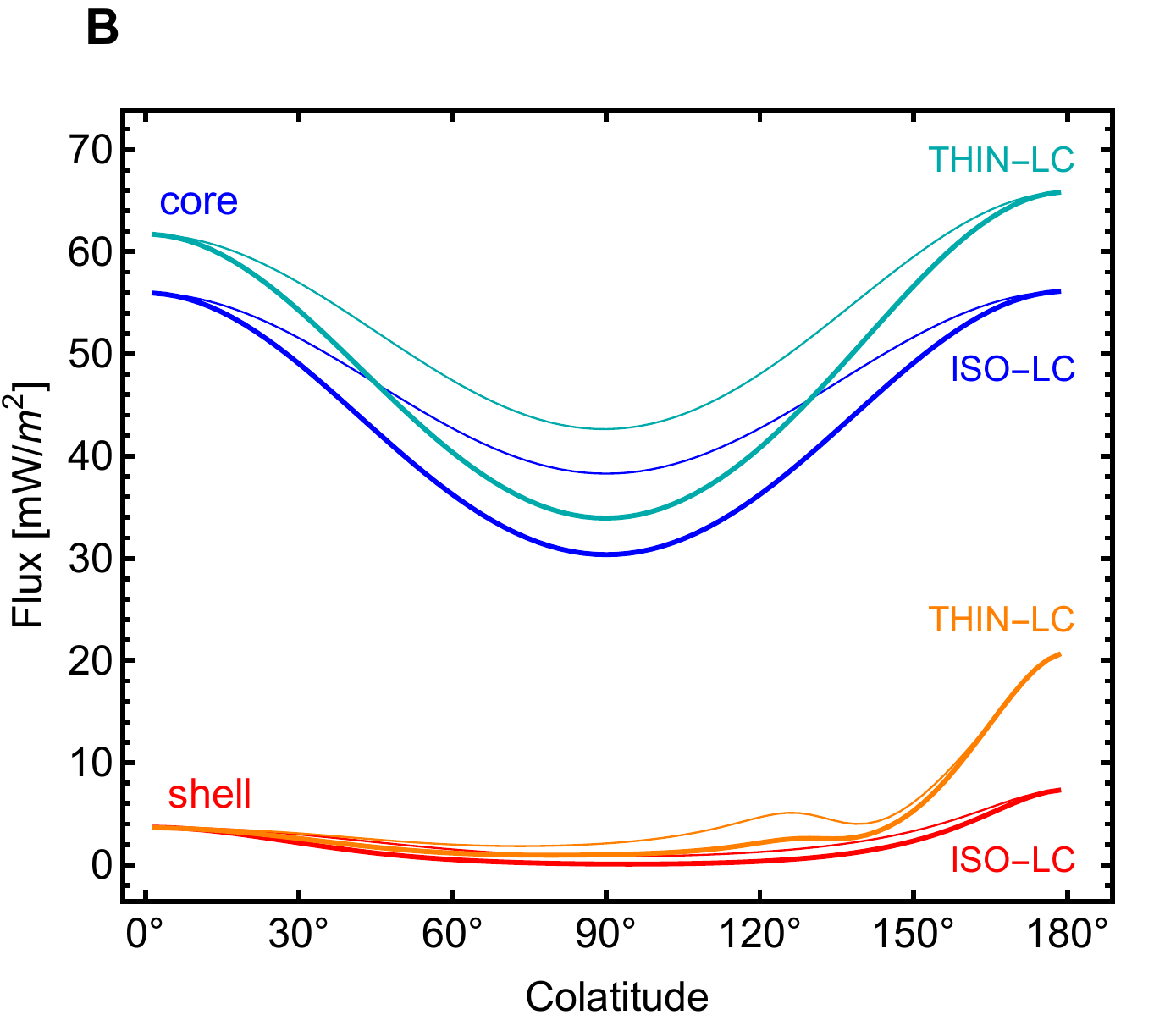}
         \includegraphics[width=4.9cm]{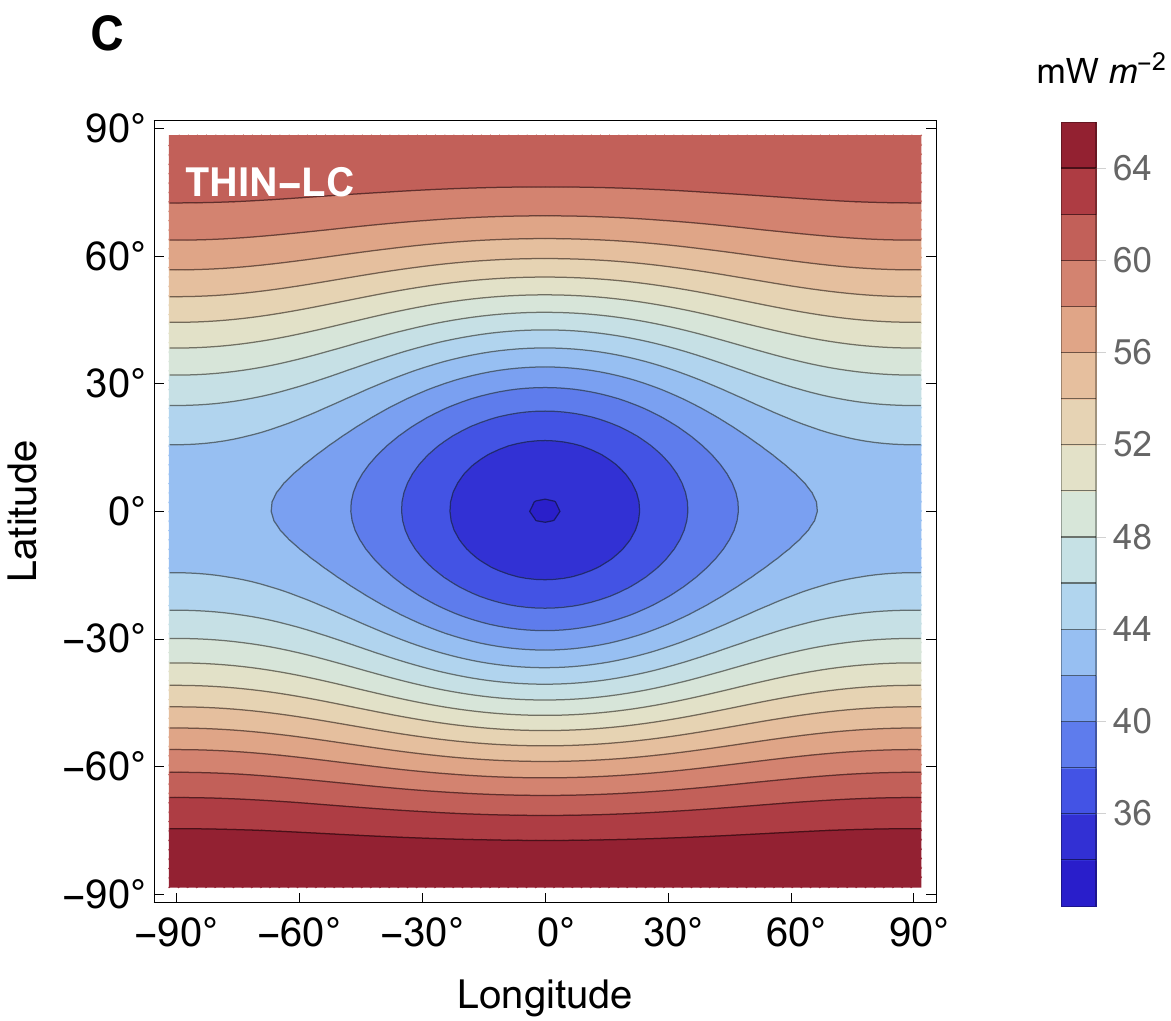}
   \caption[Surface heat flux if weak shell in SPT]{
   Heat flux if weak shell in SPT.
   (A) Meridional profiles of conductive flux, shell dissipation flux, core dissipation flux (at the surface), and the sum of the latter two in model THIN-LC.
   (B) Meridional profiles of core and shell dissipation fluxes in models ISO-LC and THIN-LC.
    (C) 2D pattern of core dissipation flux in model THIN-LC.
    The core dissipation flux is barely north-south asymmetric.}
   \label{FigFluxThin}
\end{figure}

Although the core can potentially provide whatever is needed for the total heat budget, it faces a more difficult job in explaining the north-south asymmetry of the observed heat flux.
The problem is probably made worse by the ocean circulation which partially averages the core heat flux, but I will ignore this complication here.
In models with isostatic thickness variation, the core dissipation flux varies only by a factor of two between the poles and the equator (Fig.~\ref{FigFluxProfileCore} and Fig.~\ref{FigFluxThin}B).
In model ISO-LC, shell dissipation provides only 5\% of the south polar flux while the core contribution reaches 40\%, leaving more than half of the conductive flux unaccounted for.
If shell dissipation is high (model ISO-HC), shell and core dissipation provide respectively 76\% and 22\% of the south polar flux, the total matching well the conductive flux.
Alhough there is a mismatch of about 25\% at the north pole and at the equator, one can probably find a model of lateral variation in thickness and rheology for which the fit is perfect.

\citet{choblet2017} floated the idea that the shell structure induces a north-south asymmetry in heat production within the core.
I will test their proposal with the model THIN-LC, which has a shell structure similar to the one proposed in the Supplementary Information of their paper: the shell is very thin ($3\rm\,km$ thick) and very soft ($\mu_{\rm e}=0.35\rm\,GPa$) in the SPT (Fig.~\ref{FigShellStructure}).
An immediate problem with this model is that the conductive flux more than doubles at the south pole and cannot be matched by internal dissipation.
Leaving this problem aside, we observe that core dissipation differs by only a few percent between the north and south poles (Fig.~\ref{FigFluxThin}), confirming thus the preliminary analysis made after Eq.~(\ref{Un0b}).
Therefore the very asymmetric shell structure does not translate into a large asymmetry in core dissipation.

Although hydrothermal flow within the core focuses the tidally dissipated heat towards the poles \citep{travis2015,choblet2017}, it cannot by itself produce a strong north-south asymmetry unless the core is very inhomogeneous.
An asymmetrical core structure could be due to its conditions of formation.
This impact of this assumption on tidal dissipation and hydrothermal flow remains to be investigated.

\section{Summary}
\label{Summary}

Enceladus's high heat output is attributed to dissipative eccentricity tides, but more by default than by a true understanding of its modus operandi.
So far, geodynamical models have been unable to account for the anomalous heating at the south pole both in magnitude and in localization.
The north-south asymmetry in geophysical activity is particularly telling, because it negates the usual assumption of a spherically symmetric internal structure which entails equal dissipation at the north and south poles.
The only other information we have about an asymmetric interior resides in gravity and topography data: in an isostatic framework, they imply that the shell is thinnest at the south pole, medium thick at the north pole, and thickest at the equator.
Thus, the north-south asymmetries observed at the surface and inferred in the interior make it pressing to compute tidal dissipation in a body with a floating shell of variable thickness.
It is also likely that the shell is softer in the SPT, so that lateral variations of shell rheology should also be included.

In Paper~I, I developed the theory of non-uniform viscoelastic thin shells, coupled it to tides, and solved the resulting tidal thin shell equations in terms of the stress function $F$ and the radial displacement $w$.
The ultimate purpose was the prediction of tidal stresses within the shell.
Here, I show how to use the same $(F,w)$ solution to predict tidal dissipation in the core and shell.
The former must be internally spherically symmetric, but the latter can be fully non-uniform in thickness and rheology.
The shell must however satisfy the following requirements (approximately satisfied in Enceladus):
thickness less than 10 to 20\% of the surface radius, homogeneous shell density, negligible density contrast between the shell and the top layer of the ocean, uniform Poisson's ratio, and linear viscoelasticity.
The uniformity of Poisson's ratio deserves special attention as it implies a new dissipation constraint, called `no Poisson dissipation' in contrast with the more usual condition of `no bulk dissipation' related to the uniformity of the bulk modulus.
Dissipation predictions under the two constraints differ by a few percent, which is generally smaller than the error due to the common incompressible assumption, except for the thinnest shells (Fig.~\ref{Figk2Thick}).
While the uniformity of Poisson's ratio can be seen as a drawback of the thin shell approach, it is not certain either that bulk dissipation vanishes in tidal heating processes.

The thin shell approach is basically a 2D theory because its primary variables ($F$ and $w$) do not depend on depth.
Nevertheless, thin shell theory does not assume that all variables are constant with depth.
In particular, the strain varies quasi linearly across the shell thickness, while the stress varies in proportion to the local strain and local shear modulus.
Thus, the thin shell approach predicts 3D quantities such as the volumetric dissipation rate (Eqs.~(\ref{PowerThinShell})-(\ref{PowerThinShellComp})).
Integrating the dissipation rate then yields the shell dissipation flux at the surface and the total shell power (Eqs.~(\ref{SurfaceFluxThinShell})-(\ref{EdotShellMicro})).
In that `micro' framework, the shell dissipation rate, flux, and power are expressed as sums of membrane ($FF$), mixed ($Fw$), and bending ($ww$) contributions, the first one being by far the largest.

Regarding core dissipation, I compute core deformations by treating the non-uniform shell as a pressure load (effective tidal potential method), before computing the core dissipation rate, flux, and power, which are nearly insensitive to lateral variations of shell properties (Eqs.~(\ref{PowerStrainInv})-(\ref{EdotCoreMicro})).
If one is only interested by the total power produced in the core, it is easier to partition the total power produced in the body into core and shell contributions (Eqs.~(\ref{EdotCoreMacro})-(\ref{EdotShellMacro})): this is the `macro' approach to tidal dissipation, to be contrasted with the previous `micro' approach.
Comparing the core power and shell power obtained in the micro and macro approaches provides a self-consistency check on the correctness of numerical codes.

Benchmarking against a laterally uniform shell provides a good way to estimate errors due to the thin shell approximation.
The impact of a non-zero shell thickness on the shell power is less than 2 and 4\% (for a laterally uniform shell) if the shell thickness is less than 20 and $50\rm\,km$, respectively.
These errors are computed by comparing apples with apples, i.e.\ thin and thick shell models having the same uniform Poisson's ratio (Fig.~\ref{Figk2Thin}A).
If the thin shell is compared to the thick shell with no bulk dissipation, the difference is below 4\% if the shell thickness is between 20 and $50\rm\,km$, but can climb to 9\% for very thin shells (Fig.~\ref{Figk2Thin}B).
Spatial patterns of the shell dissipation rate and flux are well predicted by the thin shell approach.
The benchmarking of a laterally non-uniform thin shell against the finite-element method of \citet{soucek2019} is outside the scope of this paper, but preliminary comparisons with FEM results show good agreement \citep{behounkova2018}.

As an illustration, I compute tidal dissipation within Enceladus's shell assuming that it is conductive, in thermal equilibrium, and with thickness variations predicted by isostasy.
The conductive model is as realistic as possible regarding ice conductivity (dependent on temperature) and surface temperature (latitudinal variation).
Dissipation and heat transfer are solved as a coupled system, because viscosity depends strongly on temperature which in turn depends on internal heating, although rather weakly.
If dissipation is low, the temperature profile is mostly determined by the surface and melting temperatures, but high dissipation results in a large rheology feedback (about 35\%): the rheology of the bottom of the shell becomes softer because of locally higher temperatures.
The addition of the forced libration to eccentricity tides increases dissipation by about 30\%.
Variations in shell thickness only slightly increase the total shell power (here by 20\%), but they have a major effect on the shell dissipation pattern: dissipation is highest where Enceladus's shell is thinnest.
In particular, the shell dissipation flux at Enceladus's south pole is three times larger in the isostatic model ($d_{SP}=7\rm\,km$) than in the uniform thickness model ($d=23\rm\,km$).
If the shell is hard and the lateral variations of the shell properties are of long wavelength, one can actually predict the shell dissipation flux by scaling the flux for a laterally uniform shell (Eqs.~(\ref{scaling})-(\ref{amplifactor})).
The scaling factor depends on the inverse shell thickness (at least if rheology is nearly laterally uniform) and can be understood in the thin shell approach by the dominance of the membrane contribution.

The shell power for nominal values of viscoelastic parameters is only a few percent of the conductive power, in agreement with \citet{soucek2019}.
If the shell dissipation rate is ten times higher, as suggested by recent laboratory experiments, it contributes to nearly half the conductive power and accounts for most of the spatial variations of the conductive flux.
The same effect is obtained with an eccentricity three times larger than the present one.
It does not make sense to explore higher rates of dissipation within the shell, because the shell dissipation flux remains too small along the tidal axis to balance conductive cooling.
In a steady state model, dissipation within an unconsolidated core must contribute the rest.
Given that core or ocean dissipation cannot be avoided, it is logical to ask whether one could entirely dispense with shell dissipation.
Core dissipation, however, remains nearly north-south symmetric even in a model with an extremely thin and soft shell at the south pole.
Therefore, dissipation in a homogeneous core cannot be responsible for the flux asymmetry observed at the surface.

In conclusion, explaining Enceladus's heat anomaly in magnitude and localization requires pushing the envelope very far: dissipation must be simultaneously high in the shell and in the core.
To avoid this non-parsimonious solution, one should investigate whether dissipation within a inhomogeneous core is sufficient to maintain the non-uniform shell in thermal equilibrium.
Alternatively, non-steady state models -- which maybe solve the problem of the total power without recourse to core dissipation -- must demonstrate that they can predict the surface heat pattern and the variations of shell thickness.
Models in which Enceladus's shell is currently thinning face the same difficulties as thermal equilibrium models in explaining the origin of the emitted energy and the non-uniformity of tidal heating and shell thickness.
If Enceladus's shell is currently thickening \citep{luan2017}, the higher dissipation flux in the past could have provided enough heat to keep the ocean liquid, but present-day shell thickness variations are more difficult to explain: thickness variations generated in the past are not only smaller (because they are bounded by the smaller average thickness) but they also tend to be averaged out by faster thickening where the shell is thinner.

\small
\section*{Acknowledgments}
I thank Attilio Rivoldini for his help with the BVP solver, Antony Trinh for information about the libration of the core, and Alice Nadeau for explanations about equilibrium surface temperatures.
This work is financially supported by the Belgian Federal Science Policy Office through the Brain Pioneer contract BR/314/PI/LOTIDE and the PRODEX grant No. 4000120791.

\normalsize

\newpage

\begin{appendices}

\section{Spherical differential operators}
\label{SphericalOperators}
\renewcommand{\theequation}{A.\arabic{equation}} 
\setcounter{equation}{0}  

Table~\ref{TableOp1} gives the definitions of the spherical differential operators appearing in this paper.
Table~\ref{TableOp2} gives their expressions in terms of scalar operators.

\begin{table}[ht]\centering
\ra{1.3}
\small
\caption[Spherical differential operators: definitions]
{\small
Spherical differential operators: definitions.
See Appendices A and B of Paper~I and Appendix~B of \citet{beuthe2013}.
The null space is characterized by the harmonic degree of the functions belonging to it.
}
\begin{tabular}{@{}lll@{}}
\hline
Notation & Definition & Null space \\
\hline
Tensorial operators & & \\
$\bar {\cal O}_1={\cal O}_1-1$ & $\partial^2_{\theta}$ & $n=0$ \\
$\bar {\cal O}_2={\cal O}_2-1$ & $(\sin\theta)^{-2} \, \partial^2_{\varphi} + \cot \theta \, \partial_\theta$ & $n=0$ \\
$\bar {\cal O}_3={\cal O}_3$ & $(\sin\theta)^{-1} \, ( \partial_\theta \partial_\varphi- \cot \theta \, \partial_\varphi )$ & $n=0,1$ \\
\hline
Scalar operators & & \\
$\Delta$ & $\bar {\cal O}_1 + \bar {\cal O}_2$ & $n=0$ \\
${\cal D}_2(a\,;b)$ & $(\partial_\theta a) (\partial_\theta b) + (\sin\theta)^{-2} \, (\partial_\varphi a) (\partial_\varphi b)$ & $n=0$ \\
${\cal D}_4(a\,;b)$ & $(\bar{\cal O}_1 \, a)(\bar{\cal O}_1 \, b) +  (\bar{\cal O}_2 \, a)(\bar{\cal O}_2 \, b) + 2 \, (\bar{\cal O}_3 \, a)(\bar{\cal O}_3 \, b)$ & $n=0$ \\
$\Delta'$ &  ${\cal O}_1 + {\cal O}_2$ &  $n=1$ \\
${\cal D}_4'(a\,;b)$ & $({\cal O}_1 \, a)({\cal O}_1 \, b) +  ({\cal O}_2 \, a)({\cal O}_2 \, b) + 2 \, ({\cal O}_3 \, a)({\cal O}_3 \, b)$ &  $n=1$ \\
${\cal C}(a\,;b)$ & $\Delta' ( a \, \Delta' \,  b )$ & $n=1$ \\
${\cal A}(a\,;b)$ & $({\cal O}_1 \, a)({\cal O}_2 \, b) +  ({\cal O}_2 \, a)({\cal O}_1 \, b) - 2 \, ({\cal O}_3 \, a)({\cal O}_3 \, b)$ & $n=1$ \\
\hline
\end{tabular}
\label{TableOp1}
\end{table}%

\begin{table}[ht]\centering
\ra{1.3}
\small
\caption[Spherical differential operators: identities]
{\small
Spherical differential operators: identities.
See Appendices A and B of Paper~I and Appendix~B of \citet{beuthe2013}.
}
\begin{tabular}{@{}cllll@{}}
\hline
\multicolumn{4}{l}{Identity} & Remark \\
\hline
\multicolumn{5}{l}{$a$ and $b$ are arbitrary functions of $(\theta,\varphi)$, $a_{\rm uni}$ is uniform:}  \\
(a) & $\Delta'$ &=& $\Delta + 2$ & -- \\
(b) & ${\cal D}_2(a\,;b)$ &=& $\frac{1}{2} \left[ \Delta (a b) - (\Delta a) \, b - a \, (\Delta b) \right] $ & -- \\
(c) & ${\cal D}_4(a\,;b)$ &=& $ (\Delta a)(\Delta b) + (\Delta a) \, b + a \, (\Delta b) + 2 a b - {\cal A}(a\,;b)$ & -- \\
(d) & ${\cal D}_4'(a\,;b)$ &=& $ (\Delta'a)(\Delta'b)-{\cal A}(a\,;b)$ & -- \\
(e) & ${\cal A}(a\,;b)$ &=& $\frac{1}{4}  \left[ \, - \, \Delta'\Delta' (ab) - (\Delta'\Delta' \, a) \, b - a \, (\Delta'\Delta' \, b) \right.$ & -- \\
 & & & $ \hspace{4mm} + \, 2 \, (\Delta' \, a)(\Delta' \, b) + \, 2 \,  \Delta' \left( (\Delta' \, a) \, b +  a \, (\Delta' \, b) \right)  $ & \\
 & & & $\left. \hspace{4mm} - \, 2 \left( \Delta' (ab) + (\Delta' \, a) \, b + a \, ( \Delta' \, b) \right) + 8 \, ab \, \right]$ & \\
(f) & ${\cal A}(a_{\rm uni}\,;b)$ &=& $a_{\rm uni}\,\Delta'b$ & -- \\
\hline
\multicolumn{5}{l}{$a_n$ is a spherical harmonic of degree $n$:} \\
(g) & $\Delta \, a_n$ &=& $\delta_n \, a$ & $\delta_n=-n(n+1)$ \\
(h) & $\Delta' \, a_n$ &=& $\delta_n' \, a$ & $\delta_n'=-(n-1)(n+2)$ \\
(i) & ${\cal A}(a_n;a_n^*)$ &=& $\frac{1}{4} \left[ - \Delta \Delta + \left( 4\delta'_n -6 \right)\Delta + 4 \delta'_n \right] |a_n|^2$ &  \\
(j) & $\langle {\cal A}(a_n;a^*_n) \rangle$ &=& $\delta'_n \, \langle |a_n|^2 \rangle$ & $\langle x \rangle = (4\pi)^{-1} \int_S x \, d\Omega$ \\
\hline
\end{tabular}
\label{TableOp2}
\end{table}%

\section{Dissipation rate in terms of $F$ and $w$}
\label{DissipationRate}
\renewcommand{\theequation}{B.\arabic{equation}} 
\setcounter{equation}{0}  

The shell dissipation rate (Eq.~(\ref{PowerThinShellGeneral})) is a linear combination of the strain invariants ${\cal E}_2$ and ${\cal E}_{\rm tr}$ (Eqs.~(\ref{E2})-(\ref{Etr})).
The tangential strains are given by Eq.~(J.1) of Paper~I, here corrected for a typo:
\begin{eqnarray}
\epsilon_{\theta\theta} &=& \alpha \left( {\cal O}_2 - \nu {\cal O}_1 \right) F + \alpha  \left(1-\nu\right) \left( \Omega + \Omega_M \right) - \left(z/R\right) {\cal O}_1 w \, ,
\nonumber \\
\epsilon_{\varphi\varphi} &=& \alpha \left( {\cal O}_1 - \nu {\cal O}_2 \right) F + \alpha  \left(1-\nu\right) \left( \Omega + \Omega_M \right) - \left(z/R\right) {\cal O}_2 w \, ,
\nonumber \\
\epsilon_{\theta\varphi} &=& - \alpha \left( 1+\nu \right) {\cal O}_3 \, F - \left(z/R\right) {\cal O}_3 w \, ,
\label{strainsFw}
\end{eqnarray}
where $z$ and $\chi$ are defined in Table~\ref{TableVisco}.
The potentials for the tangential load $\Omega$ and its moment $\Omega_M$ (Eq.~(E.5) of Paper~I) vanish here ($\Omega=\Omega_M=0$) because tangential loads are neglected in the tidal coupling of the thin shell.
In Paper~I, the factor $(1-\nu)$ was missing in front of the terms $(\Omega + \Omega_M)$, but other equations of Paper~I are not affected by this typo.
Another typo without consequences is that the definitions of the elastic extensibility and bending rigidity are exchanged in Eq.~(4) of Paper~I.

Computing the trace invariant ${\cal E}_{\rm tr}$ is straightforward.
After substituting Eq.~(\ref{strainsFw}) into Eq.~(\ref{Etr}), I apply the identity ${\cal O}_1+{\cal O}_2=\Delta'$ (Table~\ref{TableOp1}).
I write the result as
\begin{equation}
{\cal E}_{\rm tr} = {\cal E}_{\rm tr} ^{FF^*} + {\cal E}_{\rm tr} ^{Fw^*} + {\cal E}_{\rm tr} ^{F^*w} + {\cal E}_{\rm tr} ^{ww^*} \, ,
\label{EtrFw}
\end{equation}
where
\begin{eqnarray}
{\cal E}_{\rm tr} ^{FF^*} &=& |\alpha|^2  \left(1-\nu \right)^2 \left| \Delta' F \right|^2 ,
\nonumber \\
{\cal E}_{\rm tr} ^{Fw^*} &=& - \frac{\alpha z^*}{R} \, (1-\nu)  \left( \Delta' F \right) \left( \Delta' w^* \right) ,
\nonumber  \\
{\cal E}_{\rm tr} ^{ww^*} &=& \frac{|z|^2}{R^2}\left| \Delta' w \right|^2 ,
\label{ETRcomponents}
\end{eqnarray}
and ${\cal E}_{\rm tr} ^{F^*w}=({\cal E}_{\rm tr} ^{Fw^*})^*$.

Computing the strain invariant ${\cal E}_2$ is a bit more involved.
After substituting Eq.~(\ref{strainsFw}) into Eq.~(\ref{E2}), I combine the operators $({\cal O}_1,{\cal O}_2,{\cal O}_3)$ into $\Delta'$, ${\cal A}$, and ${\cal D}_4'$ using the definitions of Table~\ref{TableOp1}.
Next, I express ${\cal D}_4'$ in terms of $\Delta'$ and ${\cal A}$ with the identity~(d) of Table~\ref{TableOp2}.
I write the result as
\begin{equation}
{\cal E}_2 = {\cal E}_2^{FF^*} + {\cal E}_2^{Fw^*} + {\cal E}_2^{F^*w} + {\cal E}_2^{ww^*} \, ,
\label{E2Fw}
\end{equation}
where
\begin{eqnarray}
{\cal E}_2^{FF^*} &=& |\alpha|^2 \left( \left(1+ \nu^2 \right) \left| \Delta' F \right|^2 - \left( 1+\nu \right)^2 {\cal A}(F \,; F^*) \right) ,
\nonumber \\
{\cal E}_2^{Fw^*} &=& \frac{\alpha z^*}{R} \left( \nu \left( \Delta' F \right) \left( \Delta' w^* \right) - \left( 1 + \nu \right) {\cal A}(F \,; w^*) \right) ,
\nonumber \\
{\cal E}_2^{ww} &=& \frac{|z|^2}{R^2} \left( \left| \Delta' w \right|^2 -{\cal A}(w \,; w^*) \right) ,
\label{E2components}
\end{eqnarray}
and ${\cal E}_2^{F^*w}=({\cal E}_2^{Fw^*})^*$.
Substituting Eqs.~(\ref{EtrFw}) to (\ref{E2components}) into Eq.~(\ref{PowerThinShellGeneral}), I can write the dissipation rate within the shell in the form of Eqs.~(\ref{PowerThinShell})-(\ref{PowerThinShellComp}).

\section{Surface flux in terms of $F$ and $w$}
\label{AppendixSurfaceFlux}
\renewcommand{\theequation}{C.\arabic{equation}} 
\setcounter{equation}{0}  

The shell surface flux (Eq.~(\ref{Fshell})) is given by
\begin{equation}
{\cal F}_{shell} = \omega \int_d  {\rm Im}(\mu) \left({\cal E}^{FF} + {\cal E}^{Fw} + {\cal E}^{ww} \right)  \left( 1 + \bar\zeta \, \right)^2 d\zeta \, ,
\label{FshellApp}
\end{equation}
where $({\cal E}^{FF},{\cal E}^{Fw},{\cal E}^{ww})$ are given by Eq.~(\ref{PowerThinShellComp}) and $\bar\zeta=\zeta/R$.

In more compact notation, I write
\begin{equation}
\left( \chi , \alpha \right) = \left(  \frac{a}{b} , \frac{1}{2\left(1+\nu \right) b \, d} \right) \, ,
\end{equation}
where
\begin{eqnarray}
a &=& \mu_0+\varepsilon \mu_1 \, ,
\nonumber \\
b &=& \mu_0 + 2 \varepsilon \mu_1 + \varepsilon^2 \mu_2 \, .
\end{eqnarray}

First, the integral over ${\cal E}^{FF}$ is proportional to
\begin{equation}
\int_d {\rm Im}(\mu) \left( 1 + \bar\zeta \,  \right)^2 d\zeta = d \, {\rm Im}(b) \, .
\end{equation}
Now,
\begin{eqnarray}
{\rm Im}(\alpha) &=&
\frac{1}{2\left(1+\nu \right)d} \, {\rm Im}\left(\frac{1}{b}\right)
\nonumber \\ 
&=& -2 \left(1+\nu\right) d \left|\alpha\right|^2 {\rm Im}(b) \, ,
\end{eqnarray}
so that
\begin{equation}
\left|\alpha\right|^2 \int_d {\rm Im}(\mu) \left( 1 + \bar\zeta \, \right)^2 d\zeta = - \frac{1}{2\left(1+\nu\right)} \, {\rm Im}(\alpha) \, .
\label{intFF}
\end{equation}

Second, the integral over ${\cal E}^{Fw}$ includes a term proportional to
\begin{eqnarray}
\int_d {\rm Im}(\mu) \, z \left( 1 + \bar\zeta \, \right)^2 d\zeta
&=& \int_d {\rm Im}(\mu) \left( -\left(1 + \bar\zeta \,  \right) + \chi \left( 1 + \bar\zeta \, \right)^2 \right) d\zeta
\nonumber \\
&=& d \left( - {\rm Im}(a) + \chi \, {\rm Im}(b) \right) .
\end{eqnarray}
Now,
\begin{eqnarray}
{\rm Im}(\chi) &=& \frac{1}{|b|^2} \left( {\rm Im}(a) \, b - a \, {\rm Im}(b) \right)
\nonumber \\
&=& 2 \left( 1+\nu\right) d \, \alpha^* \left({\rm Im}(a) - \chi \, {\rm Im}(b) \right) ,
\end{eqnarray}
so that
\begin{equation}
\alpha^* \int_d {\rm Im}(\mu) \, z \left( 1 + \bar\zeta \, \right)^2 d\zeta = - \frac{1}{2\left(1+\nu\right)} \, {\rm Im}(\chi) \, .
\label{intFw}
\end{equation}
The contribution of the other term (proportional to $\alpha{}z^*$) gives the same result.

Third, the integral over ${\cal E}^{ww}$ is proportional to
\begin{eqnarray}
\int_d {\rm Im}(\mu) \, |z|^2 \left( 1 + \bar\zeta \, \right)^2 d\zeta
&=& \int_d {\rm Im}(\mu) \left( 1 - 2 \, {\rm Re}(\chi) \left(1 + \bar\zeta \, \right) + \left|\chi\right|^2 \left(1 + \bar\zeta \,  \right)^2 \right) d\zeta
\nonumber \\
&=& d \left( {\rm Im}(\mu_0) - 2 \, {\rm Re}\Big(\frac{a}{b}\Big) {\rm Im}(a) + \frac{|a|^2}{|b|^2} \, {\rm Im}(b) \right) \, .
\end{eqnarray}
Now,
\begin{eqnarray}
{\rm Im}(D) &=&
\frac{2 \, d R^2}{1-\nu} \, {\rm Im} \left( \frac{ \mu_0 b - a^2}{b} \right)
\nonumber \\ 
&=& \frac{2 \, d R^2}{1-\nu} \left( {\rm Im}(\mu_0) - \frac{2\,{\rm Re}(ab^*) \, {\rm Im}(a) - |a|^2 \,  {\rm Im}(b)}{|b|^2} \right) ,
\end{eqnarray}
so that
\begin{equation}
\int_d {\rm Im}(\mu) \, |z|^2 \left( 1 + \bar\zeta \, \right)^2 d\zeta = \frac{1-\nu}{2R^2} \, {\rm Im}(D) \, .
\label{intww}
\end{equation}

Substituting Eqs.~(\ref{intFF}), (\ref{intFw}), and (\ref{intww}) into Eq.~(\ref{FshellApp}), I can write the shell surface flux as Eqs.~(\ref{SurfaceFluxThinShell})-(\ref{SurfaceFluxThinShellComp}).

\section{Deformation of the core}
\label{DeformationCore}
\renewcommand{\theequation}{D.\arabic{equation}} 
\setcounter{equation}{0}  

In the thin shell approach, the tidal deformation of the core is obtained by solving the viscoelastic-gravitational problem for the associated fluid-crust model forced by an effective tidal potential (Section~\ref{EffectiveTidalPotential}).
If the core is homogeneous and incompressible and the ocean is homogeneous, the propagator matrix method \citep{sabadini2004} yields analytical formulas for the Love numbers and the functions $y_{i n}^T$.

Non-dimensional parameters are defined by
\begin{equation}
\left( y, \xi, \hat\mu_c \right) = \left( \frac{R_c}{R} , \frac{\rho}{\rho_b} , \frac{\mu_c}{\rho_b{g}R} \right) .
\end{equation}
The fluid-crust tidal Love numbers are (Appendix C.2 of \citet{beuthe2015})
\begin{equation}
h_n^\circ = k_n^\circ + 1 = 
\frac{ A_n + \left(2n+1\right) y^4 \, \hat\mu_c }{ B_n  + \left(2n+1-3\,\xi \right) y^4 \, \hat\mu_c } \, ,
\label{hn0}
\end{equation}
where $(A_n,B_n)$ are polynomials in $(y,\xi)$ defined in Table~\ref{TableCore}.
If the core is very rigid ($\mu_c>0.1\rm\,GPa$), these Love numbers become independent of the core parameters:
\begin{equation}
h_n^\circ = k_n^\circ + 1 = \frac{ 1 }{1-\xi_n } \, .
\label{hn0R}
\end{equation}
The radial Love number at the core-ocean boundary reads
\begin{eqnarray}
h_n^{\circ c}
&=& g \, y_{1,n}^T(R_c)
\nonumber \\
&=&  \frac{{\rm f}_n \left(2n+1\right)^2 \left(1-\xi \right) y^{n+2}}{ B_n  +  \left( 2n+1-3\,\xi  \right) y^4 \, \hat\mu_c } \, .
\label{hn0core}
\end{eqnarray}
The three radial functions required for the dissipation rate read
\begin{equation}
\left( y_{1n}^T(r) \, , \, y_{3n}^T(r) \, , \, \frac{r y_{4n}^T(r)}{\mu_c} \right) = \frac{h_n^{\circ c}}{g} \Big( {\rm f}_{n1}(\hat{r}) \, , \, {\rm f}_{n3}(\hat{r}) \, , \, {\rm f}_{n4}(\hat{r}) \Big) \, ,
\label{yifun}
\end{equation}
where $\hat{r}=r/R_c$ and ${\rm f}_{ni}(\hat{r})$ are non-dimensional functions defined by Table~\ref{TableCore}.
This solution tends to the well-known solution for a homogeneous body if one takes the limits $y\rightarrow1$ and $\xi\rightarrow0$ (e.g.\ Eq.~(D.3) of \citet{beuthe2016a}).
Since Eq.~(\ref{yifun}) depends on $(y,\xi)$ through the common factor $h_n^{\circ c}$, the strains in the core of a 3-layer body are scaled down by a common factor from the strains in a homogeneous body.

\begin{table}[ht]\centering
\ra{1.3}
\small
\caption[Non-dimensional functions for core deformations]
{\small
Non-dimensional functions for core deformations. Variables are the harmonic degree $n$, the density ratio $\xi=\rho/\rho_b$, the relative core radius $y=R_c/R$, and the reduced radius $\hat r=r/R_c$.}
\begin{tabular}{@{}lll@{}}
 \hline
${\rm f}_n$ &=& $n \left[2 \left(n-1\right) \left(3+4n+2n^2\right) \right]^{-1}$ \\
$p_A$ &=& $\left( 2 \left( n-1 \right) + 3 y^{2n+1} \right) \left(1-\xi\right) + \left(2n+1\right) y^3 \, \xi$ \\
$p_B$ &=& $\left( 2n+1-3\xi \right) \left[ 2\left(n-1\right)\left(1-\xi\right)+\left(2n+1\right) y^3\,\xi \right] - 9 \left(1-\xi\right) y^{2n+1} \xi$ \\
$A_n$ &=& ${\rm f}_n \left(2n+1\right) \left(1-\xi \right) p_A$ \\
$B_n$ &=& ${\rm f}_n \left(1-\xi \right) p_B$ \\
${\rm f}_{n1}(\hat{r})$ &=& $\frac{1}{2n+1} \, \left( n(n+2) - (n^2-1) \, \hat{r}^2 \right) \hat{r}^{n-1}$ \\
${\rm f}_{n3}(\hat{r})$ &=& $\frac{1}{n(2n+1)} \left(n(n+2) - (n-1)(n+3) \, \hat{r}^2 \right) \hat{r}^{n-1}$ \\
${\rm f}_{n4}(\hat{r})$ &=& $\frac{2}{2n+1} \left(n-1\right) \left(n+2\right) \left(1 - \hat{r}^2 \right) \hat{r}^{n-1}$ \\
\hline
\end{tabular}
\label{TableCore}
\end{table}%

\section{Dissipation in a laterally uniform thin shell}
\label{PatternsUniformThinShell}
\renewcommand{\theequation}{E.\arabic{equation}} 
\setcounter{equation}{0}  

\subsection*{Dissipation rate}

If the shell is laterally uniform and the tidal potential is of degree $2$, it is possible to factorize the thin shell dissipation rate $P_{shell}$ (Eqs.~(\ref{PowerThinShell})-(\ref{PowerThinShellComp})) in radial and angular functions as in Eq.~(\ref{PowerSpherical}).
One just needs to know that:
\begin{enumerate}
\item The variables $(F,w)$ are proportional to the tidal potential $U_2$ (see Table~\ref{TableUniform}).
\item The operator $|\Delta'U_2|^2$ depends on the angular function $\Psi_A$:
\begin{eqnarray}
|\Delta'U_2|^2 &=& |-4U_2|^2
\nonumber \\
&=& 16 \, (\omega R)^4 \, \Psi_A \, .
\end{eqnarray}
\item The operator ${\cal A}(U_2;U_2^*)$ is a linear combination of the angular functions $(\Psi_A,\Psi_C)$:
\begin{eqnarray}
{\cal A}(U_2;U_2^*) &=& - \frac{1}{4} \left( \Delta\Delta + 22 \Delta + 16 \right) |U_2|^2
\nonumber \\
&=& (\omega R)^4 \left( 8 \, \Psi_A - 12 \, \Psi_C \right) ,
\label{relationAPsi}
\end{eqnarray}
where the first equality results from the identity~(i) of Table~\ref{TableOp2} and the second one from Eq.~(22) of \citet{beuthe2013}.
\end{enumerate}
The three components of $P_{shell}$ are thus equal to
\begin{eqnarray}
{\cal E}^{FF} &=& f_2 \, \frac{1+\nu}{\left(5+\nu\right)^2} \, \left| \chi \right|^2 \Big( 8 \left(1-\nu\right) \Psi_A + 12 \left(1+\nu\right) \Psi_C \Big) \, ,
\nonumber \\
{\cal E}^{Fw} &=& 2 \, f_2 \, \frac{1+\nu}{5+\nu} \,\, {\rm Re}(\chi  z^*) \, \Big( 8 \, \Psi_A - 12 \, \Psi_C \Big) \, ,
\nonumber \\
{\cal E}^{ww} &=& f_2 \, \frac{1}{1-\nu} \, |z|^2 \, \Big( 8 \left(1+\nu\right) \Psi_A + 12 \left(1-\nu \right) \Psi_C \Big) \, ,
\label{PowerUniformComp}
\end{eqnarray}
where $f_2=(R|h_2|/g)^2\omega^4$.
Equating $P_{shell}$ to the factorized dissipation rate (Eq.~(\ref{PowerSpherical})), I obtain the dissipation weight functions for the thin shell:
\begin{eqnarray}
f_A + f_K &=& 16 \, \frac{r^2}{R^2} \, \frac{|h_2|^2}{g^2} \left( \frac{1-\nu^2}{(5+\nu)^2} \, |\chi|^2 + 2 \,\frac{1+\nu}{5+\nu}  \, {\rm Re}(\chi  z^*) + \frac{1+\nu}{1-\nu} \left| z \right|^2 \right) ,
\nonumber \\
f_B &=& 0 \, ,
\nonumber \\
f_C &=& 24 \, \frac{r^2}{R^2} \, \frac{|h_2|^2}{g^2} \left( \frac{(1+\nu)^2}{(5+\nu)^2} \, |\chi|^2 - 2 \, \frac{1+\nu}{5+\nu} \, {\rm Re}(\chi  z^*) + \left| z \right|^2 \right) ,
\label{WeightsThinShell}
\end{eqnarray}
where $r=R+\zeta$.
The three terms within the brackets correspond to membrane, mixed, and bending contributions, respectively.

\subsection*{Surface flux}

Similarly to the dissipation rate, the surface flux of a laterally uniform thin shell can be written as a weighted sum of the angular functions $\Psi_J$.
Equating ${\cal F}_{shell}$ (Eqs.~(\ref{SurfaceFluxThinShell})-(\ref{SurfaceFluxThinShellComp})) to the factorized surface flux (Eq.~(\ref{SurfaceFluxABC})), one can show that 
\begin{eqnarray}
{\cal F}_{mem} &=& f_2 \, \omega \, \frac{1}{(5+\nu)^2} \, {\rm Im}\bigg(\frac{|\chi|^2}{\alpha}\bigg) \Big( 4 \left(1-\nu \right) \Psi_A + 6 \left( 1+\nu \right) \Psi_C \Big) ,
\nonumber \\
{\cal F}_{mix} &=& -2 f_2 \, \omega \, \frac{1}{5+\nu} \, {\rm Im}(\chi) \, {\rm Re}\bigg(\frac{\chi}{\alpha}\bigg) \Big(  4 \, \Psi_A - 6 \, \Psi_C  \Big) ,
\nonumber \\
{\cal F}_{bend} &=& f_2 \, \omega  \, \frac{{\rm Im}(D)}{R^2}  \Big( 4 \left( 1+\nu \right) \Psi_A + 6 \left(1-\nu \right) \Psi_C \Big) ,
\label{SurfaceFluxThinShellCompU}
\end{eqnarray}
in which $f_2=(R|h_2|/g)^2\omega^4$ as above.
The surface flux weights (Eq.~(\ref{FluxWeight})) thus read
\begin{eqnarray}
\hspace{-7mm}
{\cal F}_A + {\cal F}_K &=& 4 f_2 \, \omega \left( \frac{1-\nu}{(5+\nu)^2}  \, {\rm Im}\bigg(\frac{|\chi|^2}{\alpha}\bigg) - \frac{2}{5+\nu} \, {\rm Im}(\chi) \, {\rm Re}\bigg(\frac{\chi}{\alpha}\bigg) + \frac{1+\nu}{R^2} \, {\rm Im}(D) \right) ,
\nonumber \\
{\cal F}_B &=& 0 \, ,
\nonumber \\
{\cal F}_C &=& 6 f_2 \, \omega \left( \frac{1+\nu}{(5+\nu)^2} \, {\rm Im}\bigg(\frac{|\chi|^2}{\alpha}\bigg) + \frac{2}{5+\nu} \, {\rm Im}(\chi) \, {\rm Re}\bigg(\frac{\chi}{\alpha}\bigg) + \frac{1-\nu}{R^2} \, {\rm Im}(D) \right) .
\label{SurfaceFluxWeightsThinShell}
\end{eqnarray}

In the membrane limit, the contribution of Pattern~C to the average surface flux is
\begin{equation}
\lim_{d\rightarrow0} \, \frac{ {\cal F}_C }{{\cal F}_T} =
\left\{
\begin{tabular}{ll}
$3 \, (1+\nu)/(5+\nu)$ & if $\,\nu=\nu_e$ \, ,
\\
$9/(11+2\bar\kappa)$ & if $K=K_e$ .
\end{tabular}
\right.
\label{FCratio}
\end{equation}
The first line can be deduced from Eq.~(\ref{SurfaceFluxWeightsThinShell}) while the second line results from Eq.~(94) of \citet{beuthe2014}.
The parameter $\bar\kappa$ quantifies the effective bulk dissipation of the thin shell and varies between 0 (incompressible limit) and 0.5 for a conductive shell with $\nu_e=0.33$.
If the membrane is incompressible, the ${\cal F}_C /{\cal F}_T$ ratio is equal to $9/11$ (whatever the dissipation condition) and reproduces (if there is no forced libration) the pattern obtained by \citet{ojakangas1989a} (see Eq.~(61) of \citet{beuthe2013}).
If the membrane is compressible, the ${\cal F}_C /{\cal F}_T$ ratio is always equal to $3/4$ ($\nu_e=1/3$) if there is no Poisson dissipation but varies between $3/4$ and $9/11$ in there is no bulk dissipation, the precise value depending on the bottom viscosity (see Table~\ref{TableRatios}).

\subsection*{Thin shell power}

In the micro approach, the total power dissipated in the shell is obtained by integrating the surface flux (Eq.~(\ref{EdotShellMicro})).
If the shell is laterally uniform, the integral can be done analytically by substituting the degree-$n$ solution of Table~\ref{TableUniform} into the surface flux equations (Eqs.~(\ref{SurfaceFluxThinShell})-(\ref{SurfaceFluxThinShellComp})), replacing the operator $\Delta'$ by its eigenvalue $\delta_n'$ and integrating the operator ${\cal A}$ with the identity (j) of Appendix~\ref{SphericalOperators}.
The result reads
\begin{equation}
\dot{E}_{shell} = \dot{E}_{mem} + \dot{E}_{mix} + \dot{E}_{bend} \, ,
\label{EshelldecompApp}
\end{equation}
where the membrane, mixed, and bending terms are given by
\begin{eqnarray}
\dot{E}_{mem} &=& c_n \, {\rm Im}\Big( (\chi^*/\psi) \Lambda_n^M \Big) \, ,
\nonumber \\
\dot{E}_{mix} &=& c_n \, 2 \, {\rm Im}(\chi) \, {\rm Re}\Big( \Lambda_n^M/\psi \Big) \, ,
\nonumber \\
\dot{E}_{bend} &=& c_n \, {\rm Im}\Big( \chi\psi \, \Lambda_n^B \Big) \, ,
\label{EshelldecompU}
\end{eqnarray}
in which $c_n = (2\pi\omega R^2\rho/g) |h_n|^2 \langle |U_n^T|^2 \rangle$.
For the reference conductive shell ($d=23\rm\,km$, $\eta_{\rm m}=10^{13}\rm\,Pa.s$), the membrane/mixed/bending contributions to the degree-2 total power are
\begin{equation}
\left( \dot{E}_{mem} \, , \dot{E}_{mix} \, , \dot{E}_{bend} \right) /\dot{E}_{shell}
\approx \left( 86.2 \, , 8.2 \, , 5.5  \right) \% \, .
\label{PowerRatios1}
\end{equation}
The membrane and bending terms can be summed with the identity ${\rm Im}(a^*b)+2\,{\rm Im}(a){\rm Re}(b) = {\rm Im} (ab)$ (valid for any complex numbers $a$ and $b$), where $a=\chi$ and $b=\Lambda_n^M/\psi$.
The total power can thus be written as
\begin{eqnarray}
\dot{E}_{shell}
&=& c_n \, {\rm Im}\Big( \Lambda_n^M + \Lambda_n^B + \Lambda_n^{corr} \Big)
\nonumber \\
&=&  \frac{\omega R}{G} \, \frac{2n+1}{2} \, \xi_n \, |h_n|^2 \, {\rm Im}(\Lambda_n) \, \langle | U_n^T |^2 \rangle \, ,
\label{EshellLambda}
\end{eqnarray}
where $\Lambda_n^{corr}$ includes the next-to-leading contributions (see Table~\ref{TableUniform}).
For the same reference conductive shell as above, the membrane/bending/next-to-leading contributions to the degree-2 total power are
\begin{equation}
\Big( {\rm Im}( \Lambda_2^M) \, , {\rm Im}( \Lambda_2^B) \, , {\rm Im}( \Lambda_2^{corr}) \Big) /{\rm Im}( \Lambda_2)
\approx \left( 94.7 \, , 4.9 \, , 0.4  \right) \% \, .
\label{PowerRatios2}
\end{equation}
The mixing contribution is now included in the membrane term, whereas it was separate in Eq.~(\ref{PowerRatios1}).
Both equations predict the same bending contribution of 5\% (differences are of second order in the thin shell approximation; $\Lambda_2^{corr}$ is also of second order, see Appendix~I of Paper~I).

\subsection*{Core-shell partition}

The partition of the total power (Eq.~(\ref{EdotTotalMacro})) into core and shell contributions is equivalent to the decomposition of
${\rm Im}(k_n)$ into a term proportional to ${\rm Im}(k_n^\circ)$, associated with the interior below the shell, and a term proportional to ${\rm Im}(\Lambda_n)$, associated with the shell itself:
\begin{equation}
{\rm Im}( k_n ) =  \left| \frac{k_n+1}{k_n^\circ+1} \right|^2 {\rm Im}( k_n^\circ ) - \, \xi_n \left| h_n \right|^2 {\rm Im}( \Lambda_n ) \,.
\label{IdentityMicroMacro}
\end{equation}
This identity was already proved for a membrane (see Appendix~H of \citet{beuthe2014}) and is also valid for a thin shell.

\end{appendices}

\scriptsize

\end{document}